\documentclass[reprint,amssymb,superscriptaddress,showpacs, nofootinbib,prx]{revtex4-2}

\usepackage{multirow}
\usepackage{array}
\usepackage{makecell}
\usepackage{amsmath}
\usepackage{placeins}
\usepackage{graphicx}
\usepackage{dcolumn}
\usepackage{bm}
\usepackage{hyperref}

\usepackage{lipsum}
\usepackage[free-standing-units=true]{siunitx}
\usepackage{xspace} 
\usepackage{graphicx}
\usepackage{dsfont}
\usepackage{epstopdf}
\usepackage{bm,  amssymb, bbold}
\usepackage{float}
\usepackage{placeins}
\usepackage{xcolor}
\usepackage[T1]{fontenc}
\usepackage[separate-uncertainty=true]{siunitx}
\usepackage{boldline,multirow}

\newcommand{\Eref}[1]{Eq.~\eqref{#1}}

\newcommand{\Fref}[1]{Fig.~\ref{#1}}
\newcommand{\Tref}[1]{Tab.~\ref{#1}}

\begin{document}

\title{Loss mechanisms in high-coherence multimode mechanical resonators coupled to superconducting circuits}

\author{Raquel Garcia-Belles}
\altaffiliation{Equal contribution}
\affiliation{Department of Physics, ETH Zurich, CH-8093 Zurich, Switzerland}
\affiliation{Quantum Center, ETH Zurich, CH-8093 Zurich, Switzerland}
\author{Alexander Anferov}
\altaffiliation{Equal contribution}
\affiliation{Department of Physics, ETH Zurich, CH-8093 Zurich, Switzerland}
\affiliation{Quantum Center, ETH Zurich, CH-8093 Zurich, Switzerland}
\author{Lukas F. Deeg}
\affiliation{University of Innsbruck, Institute for Experimental Physics, A-6020 Innsbruck, Austria}
\affiliation{Institute for Quantum Optics and Quantum Information, Austrian Academy of Sciences, A-6020 Innsbruck, Austria}
\author{Loris Colicchio}
\affiliation{Department of Physics, ETH Zurich, CH-8093 Zurich, Switzerland}
\affiliation{Quantum Center, ETH Zurich, CH-8093 Zurich, Switzerland}
\author{Arianne Brooks}
\affiliation{Department of Physics, ETH Zurich, CH-8093 Zurich, Switzerland}
\affiliation{Quantum Center, ETH Zurich, CH-8093 Zurich, Switzerland}
\author{Tom Schatteburg}
\affiliation{Department of Physics, ETH Zurich, CH-8093 Zurich, Switzerland}
\affiliation{Quantum Center, ETH Zurich, CH-8093 Zurich, Switzerland}
\author{Maxwell Drimmer}
\affiliation{Department of Physics, ETH Zurich, CH-8093 Zurich, Switzerland}
\affiliation{Quantum Center, ETH Zurich, CH-8093 Zurich, Switzerland}
\author{Ines C. Rodrigues}
\affiliation{Department of Physics, ETH Zurich, CH-8093 Zurich, Switzerland}
\affiliation{Quantum Center, ETH Zurich, CH-8093 Zurich, Switzerland}
\author{Rodrigo Benevides}
\altaffiliation[Currently at: ]{Institute of Physics, University of Sao Paulo, Sao Paulo 05508-090, Brazil}
\affiliation{Department of Physics, ETH Zurich, CH-8093 Zurich, Switzerland}
\affiliation{Quantum Center, ETH Zurich, CH-8093 Zurich, Switzerland}
\author{Marco Liffredo}
\affiliation{Advanced NEMS Laboratory, EPFL,  CH-1015 Lausanne, Switzerland}
\author{Jyotish Patidar}
\affiliation{EMPA, Swiss Federal Laboratories for Materials Science and Technology,  CH-8600 Dubendorf, Switzerland}
\author{Oleksandr Pshyk}
\affiliation{EMPA, Swiss Federal Laboratories for Materials Science and Technology,  CH-8600 Dubendorf, Switzerland}
\author{Matteo Fadel}
\affiliation{Department of Physics, ETH Zurich, CH-8093 Zurich, Switzerland}
\affiliation{Quantum Center, ETH Zurich, CH-8093 Zurich, Switzerland}
\author{Luis Guillermo Villanueva}
\affiliation{Advanced NEMS Laboratory, EPFL, CH-1015 Lausanne, Switzerland}
\author{Sebastian Siol}
\affiliation{EMPA, Swiss Federal Laboratories for Materials Science and Technology, CH-8600 Dubendorf, Switzerland}
\author{Gerhard Kirchmair}
\affiliation{University of Innsbruck, Institute for Experimental Physics, A-6020 Innsbruck, Austria}
\affiliation{Institute for Quantum Optics and Quantum Information, Austrian Academy of Sciences, A-6020 Innsbruck, Austria}
\author{Yiwen Chu}
\affiliation{Department of Physics, ETH Zurich, CH-8093 Zurich, Switzerland}
\affiliation{Quantum Center, ETH Zurich, CH-8093 Zurich, Switzerland}

\date{\today}
\begin{abstract}
Circuit quantum acoustodynamics (cQAD) devices have a wide range of applications in quantum science, all of which depend crucially on the quantum coherence of the mechanical subsystem. In this context, high-overtone bulk acoustic-wave resonators (HBARs) are particularly promising, since they have shown very high quality factors with negligible dephasing. However, the introduction of piezoelectric films, which are necessary for coupling to a superconducting circuit, can lead to additional loss channels, such as surface scattering and two-level systems (TLS). Here, we study the acoustic dissipation of HBAR resonators in cQAD systems and find that the defect density of the piezoelectric material and its interface with the bulk are limiting factors for the coherence. 
We measure acoustic modes with phonon lifetimes up to $400~\mu$s and lifetime-limited coherence times approaching one millisecond in the quantum regime.
When coupled to a superconducting qubit, this leads to a hybrid system with a large quantum coherence cooperativity of $C_{T_2}=1.1\times10^5$. These results represent a new milestone for the performance of cQAD devices and offer concrete paths forward for further improvements.
\end{abstract}

\maketitle

\section*{Introduction}

Mechanical resonators cooled to their quantum mechanical ground state are promising for a wide range of applications, including high-precision sensing \cite{linehan2025listening, mason2019continuous}, tests of fundamental physics \cite{pikovski2012probing, schrinski2023macroscopic}, and serving as memories \cite{hann2019hardware} or transducers \cite{van2025optical} in quantum information processing architectures.
A particularly versatile example are cQAD systems composed of GHz-frequency mechanical resonators electromechanically coupled to microwave superconducting qubits \cite{oconnell_quantum_2010, satzinger_quantum_2018, yang_mechanical_2024, chou_deterministic_2025, bozkurt_mechanical_2025, wollack_quantum_2022}.
These systems are compact, hardware efficient, and feature strong nonlinearities required for full quantum control and measurement. Furthermore, their microwave frequencies allow for ground-state operation when passively cooled below 100~mK.

An important performance metric for cQAD systems is the cooperativity between the different components, which quantifies the ratio of coherent coupling to decoherence \cite{clerk_hybrid_2020}.
Reaching the strong coupling regime with cooperativities well above unity is challenging, as it requires the combination of high mechanical coherence times at the single-phonon level and sufficiently large electromechanical coupling strengths, while also preserving the superconducting qubit's coherence. 
Various cQAD systems have achieved strong coupling, including those based on phononic crystal (PNC) resonators \cite{arrangoiz-arriola_resolving_2019, wollack_quantum_2022, lee_strong_2023, bozkurt_mechanical_2025}, surface acoustic wave (SAW) resonators \cite{satzinger_quantum_2018, chou_deterministic_2025, kitzman_phononic_2023} and HBARs \cite{chu_quantum_2017, von_lupke_parity_2022, kervinen_sideband_2020}. 
Nevertheless, further identifying and eliminating dissipation and decoherence channels is crucial for future applications and continued progress in the field.

Dissipation in mechanical resonators has been addressed by using high-quality materials \cite{gokhale_epitaxial_2020, maccabe_nano-acoustic_2020}, reducing surface roughness and impurities \cite{vorobiev_effect_2011, luo_lifetime-limited_2025}, better acoustic confinement techniques \cite{galliou_extremely_2013, maccabe_nano-acoustic_2020}, introducing dissipation dilution \cite{tsaturyan_ultracoherent_2017, ghadimi_elastic_2018}, or simply operating at cryogenic temperatures \cite{galliou_extremely_2013}.
However, at low temperatures and low phonon populations, coupling to TLS becomes increasingly relevant and is a major source of loss and dephasing \cite{cleland_studying_2024, bozkurt_mechanical_2025, behunin_dimensional_2016, maksymowych_frequency_2025}.
This proves particularly problematic for PNC resonators with small mode volumes and SAW resonators, which concentrate energy on the surface \cite{gruenke-freudenstein_surface_2025}.
HBARs, on the other hand, store a much larger fraction of their mechanical energy in the bulk of a high-quality crystal. This results in a highly multimode system that consistently demonstrates high quality factors ($Q$) and stable frequencies \cite{gokhale_epitaxial_2020, luo_lifetime-limited_2025, galliou_extremely_2013}. 
In fact, optomechanically actuated quartz HBARs have recently demonstrated record-level $f\times Q$ products and lifetime-limited coherence in the high-phonon-population limit \cite{luo_lifetime-limited_2025}.
However, HBARs in cQAD devices require thin piezoelectric films, which introduce additional defects and interfaces.

In this work, we study dissipation in composite AlN-sapphire HBARs in the low-power and low-temperature context of quantum devices, uncovering the coherence limits imposed by the piezoelectric layer and its interface with the bulk crystal. We probe the mechanical coherence using both a superconducting qubit and microwave spectroscopy. At around 10~mK, the thermal phonon population is largely suppressed, revealing dominant dissipation from two-level systems and either mechanical absorption in the AlN or scattering at interfaces, depending on the growth method of the piezoelectric thin film. 
Nearly all of 
our mechanical modes have single-phonon Q's above $10^6$ with negligible dephasing, and we measure single-phonon lifetimes as high as $T_{1,p}=396\pm11~\mu$s corresponding to $Q=1.3\times10^7$, and a lifetime-limited mechanical coherence time of $T_{2,p}^*=806\pm24~\mu$s. 
Together with a qubit-phonon coupling rate of $g_0/2\pi= 298\pm3$ kHz and the ability to preserve good qubit coherence properties, these results demonstrate a single-quantum coherence cooperativity of $C_{T_2}=g_0^2T_{2,q}^*T_{2,p}^*=1.13(2)\times10^5$. This is more than one order of magnitude higher than in other cQAD platforms, to the best of our knowledge, and also a significant improvement on our previous devices \cite{yang_mechanical_2024, von_lupke_parity_2022, chu_creation_2018}. 

\section*{Results}
\subsection*{Device description}\label{sec:devices}

\begin{figure}[t]
    \centering
    \includegraphics[width=0.95\linewidth]{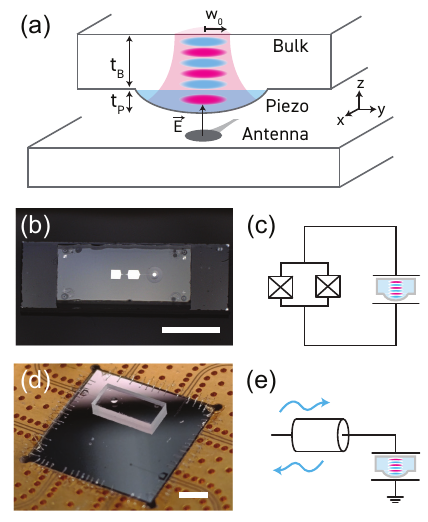}
    \caption{\textbf{Device geometry.} (a) Schematic illustration of a longitudinal overtone mode in an HBAR, where the dome-shaped piezoelectric layer mediates the coupling between the mechanical motion and the antenna's out-of-plane electric field $\vec{E}$. The diagram also indicates the bulk thickness $t_B$, piezoelectric layer thickness $t_P$, and mode waist $w_0$. (b) Photograph of the HBAR resonator flip-chip bonded to a flux-tunable superconducting qubit. (c) Circuit model of the system in (b). (d) Photograph of the HBAR resonator flip-chip bonded to a coplanar waveguide microwave antenna. (e) Driven microwave circuit modelling the system in (d). The scale-bars in (b) and (d) correspond to $2~$mm. }
    \label{fig:device}
\end{figure}

The HBARs studied in this work consist of a high-quality sapphire crystal substrate with a thin film of piezoelectric AlN grown on its surface, which transduces electric signals into longitudinally-polarized acoustic waves, as illustrated in Figure \ref{fig:device}(a).
This coupling effect is strongest when the thickness of the piezo layer $t_P$ matches an odd multiple of $\lambda/2$, where $\lambda$ is the acoustic wavelength.
For $t_P\sim1~\mu$m in AlN, this corresponds to optimal coupling near 5~GHz.
The vacuum boundary conditions on either end of the material stack result in acoustic resonances with approximate frequencies $f_n\approx n v_B/2t_B$ for integer overtone number $n$, which are primarily determined by the thickness $t_B\sim450~\mu$m and the longitudinal speed of sound $v_B\sim11000~$m/s of the bulk sapphire layer.

A key aspect of this design is the dome-shaped transducer, realized with a lithographic process that transfers a reflowed resist geometry onto the device surface through dry etching \cite{kharel_ultra-high-_2018}. The radial symmetry of this geometry supports phonon modes with Laguerre-Gaussian (LG) mode profiles. In this work, we primarily study the LG$_{00}$ overtone modes, which have a Gaussian profile with a mode waist $w_0\sim30~\mu$m determined by the radius of curvature of the dome, the total thickness of the HBAR, and the acoustic wavelength.
Since $w_0$ is much smaller than the diameter of the dome ($\sim500~\mu$m), diffraction and clamping losses are strongly suppressed (see Supplementary Information I C).
The active region of the mechanical resonator is effectively suspended with vacuum on both sides, which suppresses gas damping, and the absence of electrodes inside or on the resonator surface mitigates ohmic and quasiparticle loss \cite{wollack_loss_2021, valimaa_electrode_2018}.
Therefore, this device geometry offers a compelling platform for investigating intrinsic mechanical losses linked to material quality. To explore this, we characterize multiple HBAR samples with AlN films grown through different techniques, each providing various degrees of microstructure and crystallinity. These techniques include pulsed direct-current (DC) sputtering \cite{dubois_stress_2001} (sample A), hydride vapor-phase epitaxy (HVPE) \cite{lee_nanovoid-driven_2020} (samples B.1-4, C, and D), and high-power impulse magnetron sputtering (HiPIMS) \cite{patidar_low_2025} (sample E). Samples A, B.1-2, and C are characterized in the main text, and the results for samples B.3-4, D and E are presented in the Supplementary.

In the following, we use two methods to probe the HBAR modes. First, in a standard cQAD geometry \cite{von_lupke_parity_2022}, the mechanical resonator is coupled to an aluminum superconducting transmon qubit via the out-of-plane electric field passing through the piezoelectric transducer, as shown in Figure \ref{fig:device}(b), enabling the manipulation of single-phonon states in the HBAR.
The qubit frequency can be tuned on-resonance with a particular mechanical mode by controlling the external magnetic flux passing through the superconducting quantum interference device (SQUID), or with finer control via the AC Stark shift from an external microwave drive.
A second, simpler approach useful for rapid characterization is to replace the transmon qubit with a $50~\Omega$ superconducting coplanar waveguide (CPW) antenna and drive the piezoelectric transducer directly with a Vector Network Analyzer (VNA) \cite{zhang_resonant_2003, gruenke-freudenstein_surface_2025, gokhale_epitaxial_2020}, as shown in Figure \ref{fig:device}(d).
This purely classical measurement method can probe the phonon modes over a broader range of frequencies, powers, and temperatures.
For both methods, we assemble the HBAR in a flip-chip geometry, with the superconducting circuit elements located on the bottom chip and the HBAR chip on top oriented so the piezo faces downwards (Fig. \ref{fig:device}(a)).
The coupling is proportional to the overlap between the out-of-plane electric field and the strain field in the transducer, so for optimal coupling, we match the antenna radius to $w_0$.
Since the electric field strength decreases with the out-of-plane distance, we keep a small vacuum gap ($\sim1~\mu$m) between the transducer and the bottom antenna, controlled by the thickness of spacers on the edges of the HBAR chip.

\begin{figure*}[ht]
    \centering
    \includegraphics[width=0.96\linewidth]{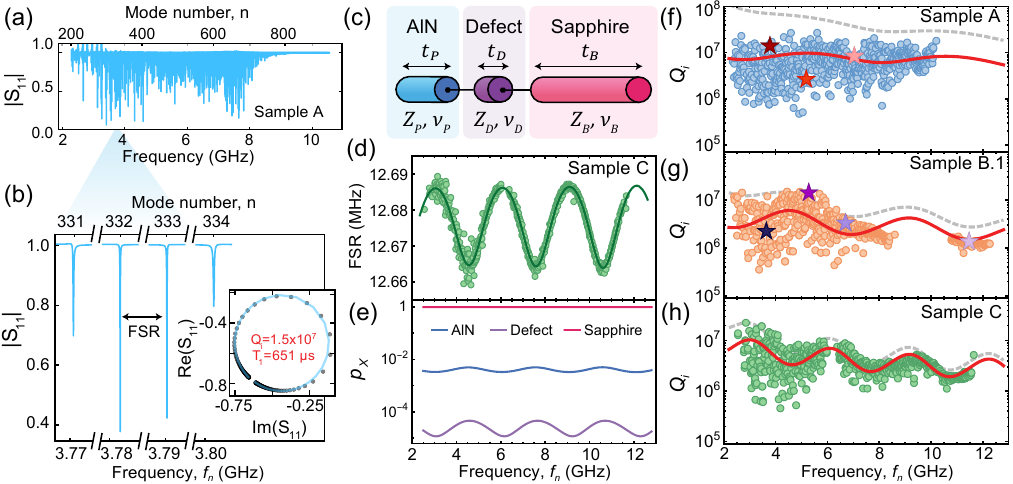}
    \caption{\textbf{Spectroscopy mode characterization.} (a) Typical broadband reflection spectrum measured through the microwave antenna, with a narrower span (b) highlighting the spectral features of each mode. The inset shows the complex reflection response of mode $n=332$, with the corresponding fit shown as a solid line. (c) Schematic illustrating the material stack used to model the resonator. The defect layer is only included to model the HVPE-deposited HBARs (Samples B.1 and C) (d) Frequency spacing between successive modes as a function of mode frequency. The solid line is a fit to the multi-material model. (e) Energy participation ratio for each layer in the model. (f-h) Internal quality factor of each resonant mode as a function of mode frequency. The error bars, corresponding to one standard deviation of the $Q_{\text{i}}$ fit uncertainty, are smaller than the marker size. The quality factor predicted by the energy participation model including surface scattering and absorption is shown as a solid red line. The dashed gray line shows only the surface scattering contribution. Stars indicate modes used for the power and temperature measurements below. }
    \label{fig:antennameas}
\end{figure*}


\subsection*{Spectroscopy Measurements}\label{sec:antenna}
Using the CPW antenna, we probe the cryogenic frequency response of the HBAR modes at 10 mK. The modes are driven coherently to mean phonon occupancies of $\bar{n}_p\sim 10^6-10^{10}$ on resonance.
The measured reflection spectrum is shown in Figs. \ref{fig:antennameas}(a) and (b), in which we observe a series of dips in reflection magnitude $|S_{11}|$ (normalized to the slowly varying background).
Each of these corresponds to a mechanical mode resonance, and fitting the complex response to an asymmetric reflection model (see Methods\ref{sec:S11fit}) yields the internal $Q_{\text{i}}$ and external $Q_{\text{e}}$ quality factors as well as the resonance frequency $f_n$ for each mode.

For our devices, the free spectral range (FSR) $f_{n+1}-f_n$ is approximately $13$ MHz, as shown in \Fref{fig:antennameas}(d).
To first order, this is consistent with the substrate thickness and longitudinal sound velocity in sapphire. 
Closer inspection reveals that the FSR is not constant over the whole frequency range, with a small modulation in frequency.
This behavior can be explained by the discontinuity in acoustic impedance caused by the piezoelectric layer.
Modeling the resonator as coupled sections of media \cite{mason_electromechanical_1948} with unique acoustic impedance $Z_X$, sound velocity $v_X$, and thickness $t_X$ (see \Fref{fig:antennameas}(c)), and imposing matching boundary conditions for the stress and velocity at the interface, we arrive at a transcendental equation that can be solved for the mode frequencies $f_n$ (see Methods).
For two layers (piezoelectric $P$ and bulk $B$, with thicknesses $t_P$ and $t_B$ respectively) the equation takes the form:
\begin{equation}\label{eq:frequencies}
    \tan\left(\frac{2\pi f_n}{v_P}t_P\right)=-\frac{Z_B}{Z_P}\tan\left(\frac{2\pi f_n}{v_B}t_B\right).
\end{equation}
Using this model to fit the experimental resonance frequencies yields precise estimates of the material parameters ($Z_X$,$v_X$,$t_X$) under certain assumptions (see Methods\ref{sec:fitting}), with extracted values summarized in \Tref{tab:parameters}.
The resulting solutions for $f_n$ are in good agreement with the measured mode frequencies for devices with sputter-deposited AlN (A and E. See Fig. \ref{fig:fig11particips}).
For samples with HVPE grown AlN (B.1 and C in \Fref{fig:antennameas}), we also include a third interface layer between the piezo and bulk layers (see Methods\ref{sec:multilayer} for a detailed derivation).
This is motivated by transmission electron microscopy (TEM) images of the piezo-bulk interface of these samples, which reveal a $t_D\sim10~$nm deep layer of damage extending into the substrate (see \Fref{fig:TEM}(a)), whose origin is likely related to a surface treatment to promote the high-crystallinity growth of AlN \cite{lee_nanovoid-driven_2020,lee_realization_2023}. 

\subsection*{Mechanical Quality Factors}
With the material properties of the acoustic resonator determined from its mode frequencies, we can make deductions about the dominant loss channels.
We show the measured high-power internal quality factors as a function of mode frequency for three representative resonators in \Fref{fig:antennameas}(f-h).
We find internal quality factors up to $2\times10^7$ in the measured frequency range, on par with state-of-the-art composite HBARs \cite{gokhale_epitaxial_2020}. Across all measured modes, we also find $Q_{\text{e}}\gtrsim10^7$, placing the system in the critically coupled or undercoupled regime, where the internal losses dominate the linewidth.

We observe significant mode-to-mode fluctuations in $Q_{\text{i}}$, in addition to a slower periodic modulation with mode frequency in the HVPE samples B.1 and C. The modulation period is different in these two samples due to their different AlN thicknesses (see Tab. \ref{tab:parameters}). We first focus on the periodic modulation. To understand this behavior, we use the mode profiles calculated when solving the resonator mode frequencies to determine the energy participation ratio $p_X$ of each layer $X$ (see Methods\ref{sec:participation}).
We plot the total energy participation of each layer for sample C in Fig. \ref{fig:antennameas}(e), observing oscillations similar to those in the FSR as well as $Q_{\text{i}}$.
Motivated by electrical analogues \cite{wang_surface_2015,minev_energy-participation_2021}, we model the total acoustic dissipation in the high phonon occupation limit as a participation-weighted sum of each layer's loss:
\begin{equation}\label{eq:classicalloss}
    \frac{1}{Q_{\text{i}}(\bar{n}_p\gg1)} = \sum_X p_X\left(\frac{1}{Q_{X,\text{mech}}}+\frac{1}{Q_{X,\sigma}}\right)
\end{equation}
Here, $Q_{X,\sigma}^{-1}$ corresponds to roughness-induced scattering loss in each layer, calculated from the root-mean-square (RMS) surface roughnesses from AFM measurements (Eq. \eqref{Eq:surfaceroughnessmultilayer} in Methods). The total surface scattering loss contribution $\sum_Xp_XQ_{X,\sigma}^{-1}$ is shown as a dashed gray line in \Fref{fig:antennameas}(f-h).
$Q_{X,\text{mech}}^{-1}$ is the intrinsic mechanical absorption loss for each layer, which could originate from internal friction and scattering at grain boundaries and defects as the material vibrates \cite{braginsky_systems_1986, hutchison_ultrasonic_1960, nowick_chapter_1972-3}. 
We note that at the low temperatures of the experiment ($10~$mK), loss due to anharmonic phonon scattering is negligible \cite{akhiezer_absorption_1939, ghaffari_quantum_2013}.

With $p_X$ and $Q^{-1}_{X,\sigma}$ independently determined, we fit Eq. \eqref{eq:classicalloss} to the data to obtain the average mechanical losses in each layer, with the total dissipation shown as red lines in \Fref{fig:antennameas}(f-h).
The high cross-correlation between layer participations limits the fit accuracy, so we instead quote lower bounds for the quality factors in each layer in Table \ref{tab:parameters}. 
Nevertheless, the vast differences in energy participation between layers shed light on the limiting quality factor of each layer.
We find consistently high lower bounds for the bulk quality factor $Q_{B,\text{mech}}\gtrsim10^7$ across all devices, consistent with expectations for a crystalline material like sapphire, for which single-crystal mechanical resonators have been reported to exhibit quality factors above $10^8$ \cite{braginsky_systems_1986}.

By contrast, the piezoelectric AlN layer is grown by different methods with varying crystallinity and defect density.
For the sputter-deposited HBAR (sample A),
we estimate a lower bound $Q_{P,\text{mech}}\sim2\times10^4$. 
With a smoother piezo-bulk interface, surface scattering loss in this sample is substantially lower, suggesting that its loss is dominated by mechanical absorption in the AlN, with a nearly flat frequency dependence. The low surface scattering loss also results in an $f\times Q$ product of $1.8\times10^{17}$~Hz, indicating a higher limit for sputter-deposited HBARs than previously expected \cite{gokhale_epitaxial_2020}. Sample E with HiPIMS-grown AlN, characterized in the Supplementary section \ref{sec:hipims}, shows a similar performance.

For the HVPE HBARs (samples B.1 and C), we obtain lower bound estimates $Q_{P,\text{mech}}>3-10\times10^4$, consistent with the higher crystalline quality observed in X-ray diffraction (XRD) and TEM analysis (see Figs. \ref{fig:XRD} and \ref{fig:TEM}) compared to the sputter-deposited HBARs.
However, these HBARs present a thin interface defect layer with higher dissipation by orders of magnitude ($Q_{D,\text{mech}}\sim10^2$).
Furthermore, the strong surface scattering in this layer dominates over the low mechanical absorption of the HVPE-deposited AlN, resulting in a clear frequency modulation together with a $1/f$ trend in $Q_\text{i}$, as shown in Fig. \ref{fig:antennameas}(g,h).
Given the favorable estimates for dissipation in the HVPE AlN film, optimizing this interface could be a promising path for future low-loss devices.

Regarding the mode-to-mode uncorrelated fluctuations in the quality factors, which are comparable for samples A, B.1 and C at low frequencies, we find that this behavior appears to be independent of the AlN growth method. In addition, we occasionally observe split modes, a feature previously reported in single-crystal sapphire resonators \cite{braginsky_systems_1986}. These effects might originate in the bulk, for example, due to polarization-conversion loss, where the longitudinal phonon modes leak into nearby shear or transverse modes, or due to variations in the sampling of defects by different HBAR modes, leading to an inhomogeneous mechanical absorption constant.


\subsection*{TLS loss}
\begin{figure}[t]
    \centering
    \includegraphics[width=0.96\linewidth]{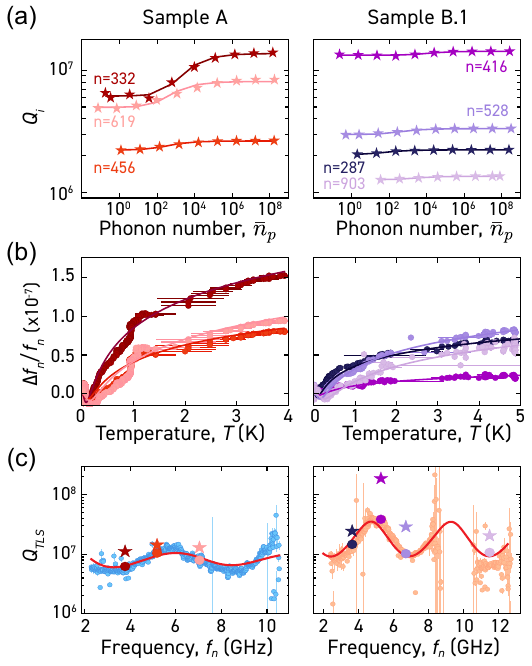}
    \caption{\textbf{TLS loss characterization.} The color scheme identifies corresponding phonon modes across panels. The left panels show measurements of Sample A, and the right panels of Sample B.1. (a) Internal quality factors $Q_{\text{i}}$ as a function of the average phonon population for selected mechanical overtones in Samples A and B.1. The solid lines are fits to the resonant TLS loss model in Eq. \eqref{eq:TLSpower}. (b) Fractional frequency shift of the phonon modes with temperature. The solid lines are fits to Eq. \eqref{eq:TLStemp}. (c) Inverse of the TLS loss tangent ($Q_{\text{TLS}}$) for all phonon modes calculated from the fractional frequency shifts of the modes between $10~$mK and $4~$K, shown as dots. The color-coded stars and circles are the $Q_{\text{TLS}}$ extracted from the fits in (a) and (b) respectively. The solid red line is a fit to the energy participation model in Eq. \eqref{eq:TLSparticipation} used to extract the TLS loss tangent of AlN in each sample. In all the panels, error bars represent one standard deviation in the variable fit uncertainty (or propagated uncertainty), except the horizontal error bars in panel (b), which indicate the temperature variation within $\pm1~$minute of data acquisition.}
    \label{fig:TLS}
\end{figure}

To investigate loss channels relevant to the single-phonon regime,
we measure the frequency response of a subset of phonon modes in samples A and B.1 as a function of drive power and temperature $T$. We select three and four modes per device, respectively, highlighted with stars in Fig. \ref{fig:antennameas}(f, g). 
In all the selected modes, we observe an improvement of the internal quality factors as the average phonon number $\bar{n}_p$ increases, see Fig. \ref{fig:TLS}(a). This improvement is more evident in the highest $Q_{\text{i}}$ modes of sample A (sputter-deposited AlN), and is consistent with the saturation of near-resonant TLS \cite{phillips_two-level_1987, hunklinger_3_1976}:
\begin{equation}\label{eq:TLSpower}
    \frac{1}{Q_{\text{i}}}=\frac{1}{Q_{\text{TLS}}}\frac{\tanh\left(\frac{hf_n}{2k_BT}\right)}{\sqrt{1+\bar{n}_p/n_c}}+\frac{1}{Q_{\text{i}}(\bar{n}_p\gg1)}.
\end{equation}
 Here, $Q_{\text{TLS}}^{-1}$ is the effective TLS loss tangent at zero temperature, $n_c$ is the critical phonon number for TLS saturation, and $Q_{\text{i}}(\bar{n}_p\gg1)^{-1}$ is the power-independent loss associated with other loss mechanisms, as defined in the previous section. 
 In addition, with increasing temperature, we observe a positive fractional frequency shift $\Delta f_n/f_n$ of the HBAR modes, shown in Fig. \ref{fig:TLS}(b). This shift is opposite to that expected from thermal expansion and is explained by the dispersive interaction of the phonon modes with a TLS bath \cite{phillips_two-level_1987, hunklinger_3_1976}: 
\begin{equation}\label{eq:TLStemp}
    \frac{\Delta f_n}{f_n}=\frac{Q_{\text{TLS}}^{-1}}{\pi}\left(\text{Re}\left\{\Psi\left(\frac{1}{2}+\frac{h f_n}{ 2\pi i  k_BT}\right)\right\}-\ln\left(\frac{hf_n}{ 2\pi k_BT}\right)\right)
\end{equation}
where $\Psi$ is the digamma function. 

To quantify the TLS loss in both devices, we fit the inverse quality factors for each mode to Eq. \eqref{eq:TLSpower}. 
These fits are shown in Fig. \ref{fig:TLS}(a), and the extracted values of $Q_{\text{TLS}}$ are plotted as stars in Fig. \ref{fig:TLS}(c). Similarly, the fits of the fractional frequency shifts $\Delta f_n/f_n$ to Eq. \eqref{eq:TLStemp} are shown in Fig. \ref{fig:TLS}(b), and the corresponding $Q_{\text{TLS}}$ values extracted from the fit are plotted as circles in Fig. \ref{fig:TLS}(c). 
\begin{figure*}[ht!]
    \centering
    \includegraphics[width=0.96\linewidth]{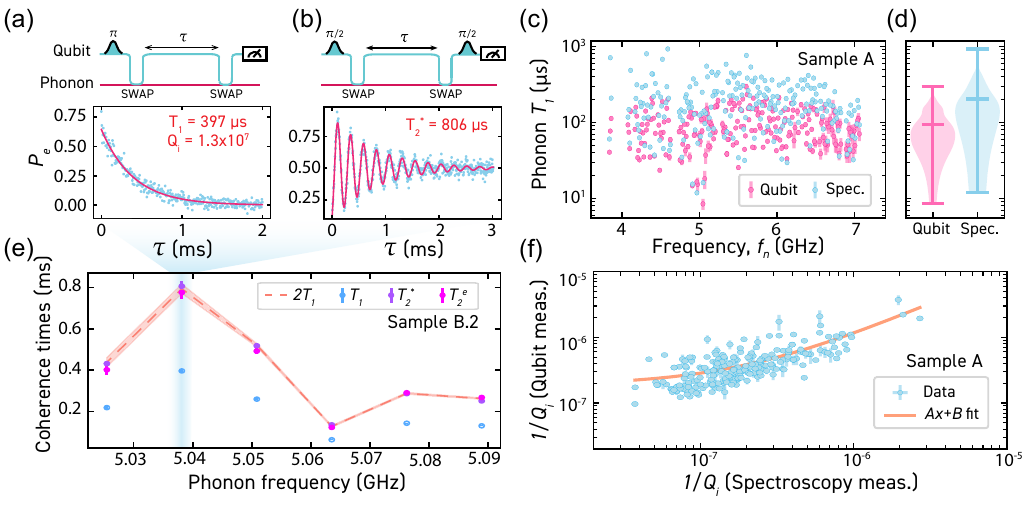}
    \caption{\textbf{Phonon coherence times in the quantum regime.} (a-b) Pulse sequences to measure phonon $T_1$ and $T_2^*$ through the qubit, and corresponding phonon population as a function of wait time $\tau$ in both measurements. The red lines are fits to a decaying exponential (a) or exponentially decaying sinusoid (b). (c) Phonon mode relaxation time ($T_1$) values in Sample A, obtained either by swapping a single excitation with the qubit and observing its decay (pink), or calculated from the internal quality factors extracted by microwave spectroscopy at high powers (blue). (d) Violin plots comparing the $T_1$ distributions of both measurement methods. (e) Phonon decay $T_1$, and coherence times $T_2^*$ and $T_2^e$ for six phonon modes in Sample B.2, with $T_2^*$ times approaching the lifetime limit $2T_1$. The band around $2T_1$ represents the uncertainty. (f) Correlation between inverse internal quality factors measured with the qubit and with microwave spectroscopy in sample A, with a linear fit. In all the panels, error bars represent one standard deviation in the variable fit uncertainty, or the propagated uncertainty.}
    \label{fig:qubitmeas}
\end{figure*}

Since most of the TLS-related frequency shifts occur between 10 mK and 4 K, we can estimate the corresponding $Q_{\text{TLS}}$ over the entire frequency range by using Eq. \eqref{eq:TLStemp} and measuring the fractional frequency shift of all the visible phonon modes at these two temperatures. 
To account for any global frequency drifts that may occur in the time span between these two measurements, we use the continuous frequency shifts measured for the selected modes (Fig. \ref{fig:TLS}(b)) to calibrate for a possible offset. The inferred $Q_{\text{TLS}}$ are shown in Figure \ref{fig:TLS}(c). These measurements reveal oscillations in the effective TLS loss tangent of the modes at different frequencies, similar to the oscillations we observe in the mechanical quality factors at high phonon-populations (Fig. \ref{fig:antennameas}).  
To explain this effect, we fit the TLS loss tangent to a weighted sum of different contributions
\begin{equation}\label{eq:TLSparticipation}
    Q_{\text{TLS}}^{-1}=\sum_Xp^{\text{pot}}_{X}Q_{X,\text{TLS}}^{-1},
\end{equation}
where $p_{X}^{\text{pot}}$ is the potential energy participation ratio of the $X$th layer (see Methods), and $Q_{X,\text{TLS}}^{-1}$ its TLS loss tangent. 
To reduce the number of free parameters, we assume an ideal sapphire crystal with $Q_{B,\text{TLS}}^{-1}=0$, and obtain  $Q_{P,\text{TLS}}^{-1}=5.9(1)\times10^{-5}$ for sample A, and $Q_{P,\text{TLS}}^{-1}=1.27(2)\times10^{-5}$, and $Q_{D,\text{TLS}}^{-1}=7.5(3)\times10^{-4}$ for sample B.1. The fits are displayed as a solid line in Fig. \ref{fig:TLS}(c), in good agreement with the data. Note that, analogously to dielectric loss, we use the potential (rather than the total) energy participation ratio, since TLS couple to the strain field \cite{behunin_dimensional_2016, emser_thin-film_2024, phillips_two-level_1987} (see Methods and Supplementary Information I D). In this case, $p_D^{\text{pot}}\not\propto p_P^{\text{pot}}$ because its thickness is sub-wavelength, which allows us to independently fit the TLS loss tangents. Importantly, $Q_{\text{TLS}}$ sets an upper bound on the quality factors achievable by HBARs in the quantum regime. The nearly five times lower TLS loss tangent of HVPE-grown AlN compared to sputter-deposited AlN is consistent with the narrower XRD rocking curve measured for this sample (see Methods), and suggests that this material may be better suited for quantum applications due to its lower defect density.

It is worth mentioning that the two methods for finding $Q_{\text{TLS}}$ that we use might yield different results \cite{phillips_two-level_1987, emser_thin-film_2024}. 
Namely, $Q_{\text{TLS}}$ obtained from the phonon-number dependent measurements and the fit to Eq. \eqref{eq:TLSpower} depends on the TLS distribution in frequency around the phonon mode and their coupling strength, which might be inhomogeneous. 
In contrast, $Q_{\text{TLS}}$ obtained from the fractional frequency shifts is also affected by far-off-resonant TLS, therefore, it yields a better measurement of the average TLS ensemble distribution and a lower variance \cite{emser_thin-film_2024}. 
This could explain the discrepancy between the stars and the circles in Fig. \ref{fig:TLS}(c) (See Supplementary Information I D for a detailed discussion).

\subsection*{Qubit measurements}\label{sec:qubit}
\begin{figure*}[ht!]
    \centering
    \includegraphics[width=0.93\linewidth]{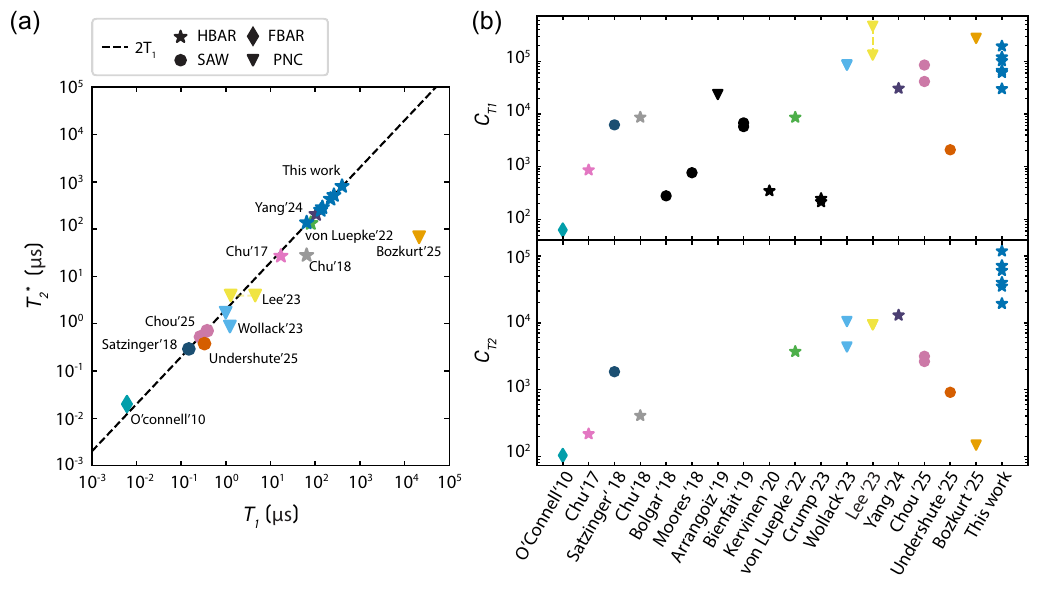}
    \caption{\textbf{Dephasing limit and strong coupling in cQAD.} (a) Decay ($T_1$) and coherence times ($T_2^*$) of mechanical modes reported in works that achieve strong coupling between a mechanical resonator in the quantum ground state and a superconducting qubit \cite{bozkurt_mechanical_2025, undershute_decoherence_2025, yang_mechanical_2024, lee_strong_2023, wollack_quantum_2022, von_lupke_parity_2022, chu_creation_2018, satzinger_quantum_2018, chu_quantum_2017, oconnell_quantum_2010}. The modes characterized in this work for sample B.2 are displayed as blue stars. The dashed black line represents the lifetime limit of coherence with negligible dephasing. (b) Overview of the $T_1$ and $T_2^*$ cooperativities in the same works and additional references \cite{bolgar_quantum_2018, moores_cavity_2018, arrangoiz-arriola_resolving_2019, bienfait_phonon-mediated_2019, kervinen_sideband_2020, crump_coupling_2023} demonstrating resonant strong coupling with only $T_1$ data provided (black markers).}
    \label{fig:discussion}
\end{figure*}

To directly probe phonon dissipation in a truly quantum regime, we measure the energy relaxation ($T_1$) and coherence ($T_2^*$) times of the phonon modes by swapping single excitations from a superconducting flux-tunable qubit into the modes, see Fig. \ref{fig:qubitmeas}(a,b). Importantly, thanks to the modularity of our device architecture, the HBAR characterized in the previous section, Sample A, is the same HBAR used here. This allows us to compare the phonon lifetimes of each mode with the quality factors measured previously using the relation $T_1=Q_{\text{i}}\cdot(2\pi f_n)^{-1}$.  To focus only on the effects of TLS loss, we subtract the inverse Purcell decay induced by the qubit on the phonon modes (see Supplementary Information I E). We observe a degradation of the average phonon lifetimes measured with the qubit with respect to the high-power $T_1$s measured via microwave spectroscopy; see Fig. \ref{fig:qubitmeas}(c,d). We compare the total loss $Q_{\text{i}}^{-1}$ of each phonon mode in both measurements in Fig. \ref{fig:qubitmeas}(f), and observe that there is a linear correlation. Performing a linear fit, we find $Q_{\text{i,(qubit)}}^{-1}=1.9(2)\times10^{-7}+0.98(5)Q_{\text{i,(spec.)}}^{-1}$ with $R^2=0.6$. This result is consistent with the power-dependent TLS loss model $Q_{\text{i}}^{-1}=Q_{\text{TLS}}^{-1}+Q_{\text{i}}^{-1}(\bar{n}_p\gg1)$ from Eq. \eqref{eq:TLSpower}, where $\tanh(hf_n/2k_BT)/\sqrt{1+\bar{n}_p/n_c}\approx1$ at low temperatures and drive powers. We identify the offset of the linear fit with $Q_{\text{TLS}}^{-1}=1.9(2)\times10^{-7}$, which is consistent with the average TLS loss of Sample A characterized in Fig. \ref{fig:TLS} of the previous section. The relatively low value of $R^2$ is explained by the spread of the data, which is expected for resonant TLS loss \cite{emser_thin-film_2024} (see Supplementary Information I D). 

In a separate experiment, we employ a single Josephson junction superconducting qubit with frequency tunability enabled via the AC Stark shift and idle frequency $f_q=5.096~$GHz, $T_{1,q}=33.4~\mu$s, and $T_{2,q}^*=40.0~\mu$s to characterize the energy relaxation times ($T_{1,p}$), and Ramsey ($T_{2,p}^*$) and echo ($T_{2,p}^e$) coherence times of six phonon modes from an HVPE-deposited HBAR (Sample B.2) \cite{chu_quantum_2017}. Note that this sample was fabricated on the same wafer as Sample B.1 previously characterized. As shown in Fig. \ref{fig:qubitmeas}(e), all modes exhibit $T_{2,p}^*\approx 2T_{1,p}$, indicating dephasing rates well below the energy decay rate and consistent with lifetime-limited coherence \cite{luo_lifetime-limited_2025}. Among the modes, we identify one with $T_{1,p}=(397\pm11)~\mu$s and $T_{2,p}^*=(806\pm24)~\mu$s, corresponding to a single-phonon internal quality factor $Q_{\text{i}}=1.3(3)\times10^7$. This value is comparable to the best high-power quality factors obtained via the microwave spectroscopy measurements on Sample B.1, suggesting that TLS-induced dissipation is not yet the dominant loss mechanism in these high crystallinity HBARs in the quantum regime. This observation is consistent with the TLS loss tangent measured for Sample B.1, shown in Fig. \ref{fig:TLS}(c) of the previous section.
\section*{Discussion}\label{sec:conclusion}
Most quantum applications require not only long lifetimes $T_1$, but also long coherence times $T_2^*$, which quantify how robustly a system can maintain superposition and entanglement, essential resources for quantum computing, sensing, and communication. In this work, we demonstrate lifetime-limited coherence times for HBARs in the quantum regime, 
see Fig. \ref{fig:discussion}(a). The negligible dephasing observed can be attributed to the strong spatial confinement of the phonon mode within the bulk of the substrate, which shields it from surface-related noise and environmental perturbations. Furthermore, the large volume of the acoustic mode significantly dilutes the coupling to individual TLS, which avoids the dephasing observed in more compact resonators \cite{bozkurt_mechanical_2025}. 

The cooperativities $C_{T_1}=4g_0^2T_{1,q}T_{1,p}$ and $C_{T_2}=g_0^2T_{2,q}^*T_{2,p}^*$ serve as key figures of merit for quantifying the strength and utility of qubit-phonon coupling in hybrid quantum systems.
Since the pioneering work by O'Connell \cite{oconnell_quantum_2010}, Chu \cite{chu_quantum_2017} and Satzinger \cite{satzinger_quantum_2018}, the cooperativities in cQAD systems have substantially increased, see Fig. \ref{fig:discussion}(b). In this work, we have established HBARs as the cQAD platform with the highest coherence cooperativity $C_{T_2}=1.13(2)\times10^5$ to date, leading by more than one order of magnitude. Moreover, our study also shows that this result can still be improved. Even without better material quality, the low TLS density measured in the current HBARs with highly crystalline AlN results in $Q_{\text{TLS}}$ up to $4\times10^7$. Based on this, HBARs optimized for lower  frequencies near $3~$GHz, where surface scattering loss is reduced, could reach phonon lifetimes and coherence times beyond 1 ms. Assuming the same qubit loss, the qubit lifetime could also benefit from going towards this low-frequency regime, while the EM coupling rate $g_0$ and energy participations could be optimized by adjusting the thickness of the piezoelectric transducer and reducing the mode volume. 

The limiting loss mechanism for the HBARs based on HVPE-grown AlN measured in the quantum regime remains surface scattering and absorption at the damaged interface. Achieving $\sim1~\mu$m single-crystal AlN films without this defect layer could substantially boost the quality factors of the HBAR modes, which highlights the relevance and challenges of high-quality material growth \cite{gokhale_epitaxial_2020, lee_nanovoid-driven_2020, patidar_low_2025, dubois_stress_2001}. In this context, the methodology and participation model that we developed in this work provide a new framework for assessing the quality of piezoelectric and other thin films intended for acoustic applications \cite{galliou_new_2016, zhang_high-tone_2006}. Our model can be extended to the characterization of HBARs with other material stacks \cite{gokhale_epitaxial_2020, kervinen_sideband_2020, tian_hybrid_2020}, provided that diffraction and other loss mechanisms are properly accounted for or mitigated. Even more broadly speaking, the multimode nature of the HBARs also makes them a promising platform for probing TLS distributions across a broad range of frequencies in various materials used for quantum devices.


\section*{Methods}
    
\subsection*{Additional measurements}
Additional measurements and control experiments are presented in Supplementary Information \ref{sec:additionalmeasurements} to validate the theoretical models and rule out alternative loss mechanisms. These include temperature-dependent measurements up to $\sim80~$K to examine frequency shifts and quality-factor scaling, measurements of a flat HBAR without a dome (Sample F) to assess the diffraction loss limit, and an HBAR incorporating a larger-radius antenna to quantify the influence of antenna geometry on dissipation and to access higher-order Laguerre-Gaussian modes (Samples B.3-4). We further characterize a thinner HBAR (Sample D, thickness $\sim170~\mu$m) to test the participation-ratio model, and an HBAR fabricated with HiPIMS-grown AlN (Sample E), which exhibits performance comparable to that of Sample A. 

\subsection*{Material growth and sample fabrication}
\textbf{AlN growth:} HBAR samples were fabricated from 2-inch AlN-on-sapphire wafers with AlN film thicknesses between 1 to 1.8 $\mu$m. All films were grown on double-side-polished, c-plane oriented sapphire substrates. 

Sample A used AlN grown by pulsed DC sputtering \cite{dubois_stress_2001} on a $\sim$500 $\mu$m-thick sapphire substrate, following a 10 min piranha clean and deposition at $300^o$C. Samples B.1-4, C and D were fabricated from commercially available HVPE-grown AlN (LumiGNTech) on $\sim435~\mu$m sapphire substrates ($\sim170~\mu$m for sample D). In particular, samples B.1-4 were fabricated on the same wafer. Prior to growth, the sapphire substrates underwent a surface treatment to promote the highly crystalline growth of AlN \cite{lee_nanovoid-driven_2020, lee_realization_2023}, which introduces pits and defects at the AlN/sapphire interface. An additional sample (Sample E), discussed in the Supplementary Information, employed AlN grown by HiPIMS \cite{patidar_low_2025} on a $\sim430~\mu$m sapphire substrate with no prior cleaning or surface treatment. 

\textbf{HBAR fabrication:} After AlN growth, the HBARs are fabricated in a series of photolithography steps. We start by depositing 100~nm-thick aluminum markers for alignment during the subsequent steps. These are defined by exposing a bilayer stack of 400~nm LOR 5B and $1~\mu$m AZ1505 resists using an EVG 620 NT, and developing in a 1:4 AZ400k:water solution and water. After aluminum deposition and liftoff in NMP at $75~^o$C we clean the wafers with sonication in acetone and IPA for 5 minutes each and blow dry with nitrogen. The next step is patterning and etching the domes, which are crucial for the confinement and stability of the acoustic modes. We follow the same recipe as in Ref. \cite{chu_creation_2018} with very minor adaptations. Before spinning the resist, we first bake the wafer at $110~^o$C for 2 minutes to dehydrate the surface. We spin a $\sim400~$nm layer of PMMA 950k e-beam resist and bake it at $180~^o$C for 5 minutes. The function of this e-beam resist layer is simply to protect the surface of the AlN from the photoresist developer. We then spin a $5.5~\mu$m layer of AZP 4620 and bake it at $100~^o$C for 5 minutes. We then perform UV exposure using the same EVG tool and develop in 1:4 AZ400k:water for 3 minutes and rinse in water for 2 minutes to define $\sim250~\mu$m radius photoresist disks. To start the reflow, we expose the wafer to HMDS vapor for 10~minutes at room temperature. We then tape the wafer to a hotplate at $60~^o$C and flip it upside down on a beaker with PGMEA heated to $55~^o$C. After 2-3 hours the disks of resist will reflow into domes. To stabilize the domes, these are first baked on the hotplate at $100~^o$C for $10~$minutes, followed by a hard-bake at $140~^o$C for 2 minutes. In both the heating and cooling of the wafer we avoid sudden temperature changes. The resist domes are used as a mask to transfer the dome-shape onto the AlN during etching. The etching is performed in an Oxford PlasmaPro 100 RIE/ICP etcher using a BCl$_3$/Ar plasma at 40/2~sccm, 7~mTorr, 120~W RF Power (375 V DC Bias), and 1000~W ICP power, as described in \cite{yang_mechanical_2024}. The etch rate of the resist is $\sim70$~nm/min while the etch rate of the AlN and the sapphire is $\sim25~$nm/min. We overetch into the sapphire to ensure that the AlN remains only on top of the dome with a thickness $\sim\lambda/2$, where $\lambda$ is the acoustic wavelength with optimal coupling. After the etch, we deposit $100$~nm of aluminum to pattern the dicing marks and labels on the chips using the same process as for the alignment markers. The last step is fabricating four pillar spacers at the edges of each HBAR chip which will set the separation between the HBAR and the bottom chip during the flip-chip bonding. This is done by spinning a $4-6~\mu$m layer of SU8 resist baked at $110~^o$C for 1~min, which is then exposed with UV light and baked at $100~^o$C (post exposure bake). After developing in mr-Dev 600 for 1~minute and IPA for 10~s, the pillars will remain in the areas that were exposed with the UV light. To make sure that the pillars remain attached during subsequent dicing and cleaning steps, the wafer is hard-baked at $180~^o$C for at least 5~minutes. 

A summary of growth and design parameters of the HBARs characterized in this work is provided in Table \ref{tab:parameters}.

\textbf{Qubit and CPW antenna fabrication:} Transmon qubits were fabricated on sapphire substrates by aluminum evaporation using e-beam lithography and the Dolan bridge technique to define the Josephson junctions, following standard recipes \cite{chu_creation_2018, wang_surface_2015}. Flip-chip bonding was performed as described in \cite{von_lupke_parity_2022, von_lupke_quantum_2023}. The bonded devices were packaged in aluminum cavities and mechanically secured using small pieces of indium to ensure thermal contact. 

For spectroscopy measurements, HBARs were flip-chip bonded to $50~\Omega$ CPW antenna chips. The antennas were fabricated by evaporating 100~nm of aluminum onto $430~\mu$m-thick sapphire substrates patterned by standard photolithography (same process as for the HBAR alignment markers). For measurement, the antenna chip was mounted in a copper enclosure using GE varnish for thermalization and wire-bonded to a printed circuit board with SMP connectors.

\subsection*{Material characterization}
\subsubsection*{XRD}
\begin{figure}[t!]
    \centering
    \includegraphics[width=\linewidth]{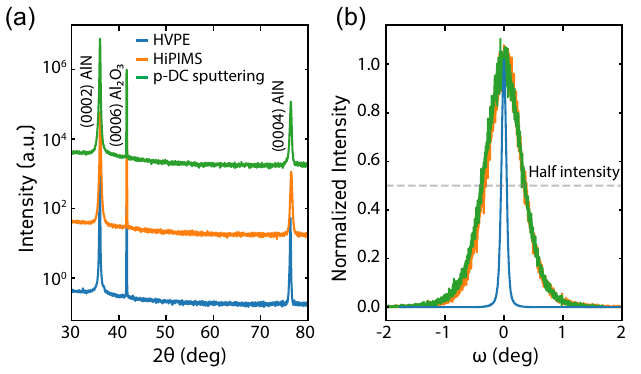}
    \caption{\textbf{X-ray diffraction (XRD) analysis of AlN-sapphire wafers with AlN films grown by different methods.} (a) XRD $2\theta-\omega$ angle scan demonstrating a (0002) out-of-plane texture for all the AlN films and c-plane oriented sapphire for the substrates. For visual clarity, the orange and green lines have been offset by a factor $\times10^2$ and $\times10^4$, respectively. (b) Rocking curves ($\omega$-scans) around the (0002) AlN peaks.}
    \label{fig:XRD}
\end{figure}

To evaluate the crystalline quality and orientation of the AlN films, we perform XRD measurements on three wafers representative of those used to fabricate the HBARs characterized in this work. The scans are performed using a Bruker D8 equipped with an EIGER2 R 500K detector in Bragg-Brentano geometry and Cu K$\alpha$ radiation ($\lambda=1.5406$ \r{A}). As shown in Figure \ref{fig:XRD}(a), the $2\theta-\omega$ scans exhibit prominent peaks at 36.0$^{\circ}$, 76.4$^{\circ}$ and 41.7$^{\circ}$, corresponding to the (0002) and (0004) AlN reflections and the third-order reflection (0006) of the c-plane sapphire substrate, respectively. To measure the out-of-plane crystalline alignment of the AlN films, we perform rocking curve scans centered on the (0002) AlN peaks, shown in Figure \ref{fig:XRD}(b).  The (0002) rocking curve full width at half maximum (FWHM) is 350 arc-sec for the HVPE-grown AlN film, indicating excellent out-of-plane alignment and high crystalline quality. The FWHM of the pulsed-DC sputtered film is 2640 arc-sec, and 2472 arc-sec for the HiPIMS film. The larger FWHM of the rocking curve suggests a higher density of structural imperfections in the sputtered films. In the context of acoustic devices, these defects can contribute to increased mechanical damping and may correlate with a higher density of TLS.
\subsubsection*{STEM}
To investigate the microstructure of the different AlN growths and the quality of the AlN/sapphire interface, we obtain scanning transmission electron microscopy (STEM) images from three different HBAR samples. In each sample, the lamella is obtained from the top of the HBAR dome and thinned down in a Thermo Fisher Scientific Helius 5 UX focused ion beam (FIB) system. The lamella is lifted out with a 30 kV ion beam, and sequentially thinned at 30 kV, then 5 kV, and finally 2 kV to minimize surface damage.

\begin{figure}
    \centering
    \includegraphics[width=\linewidth]{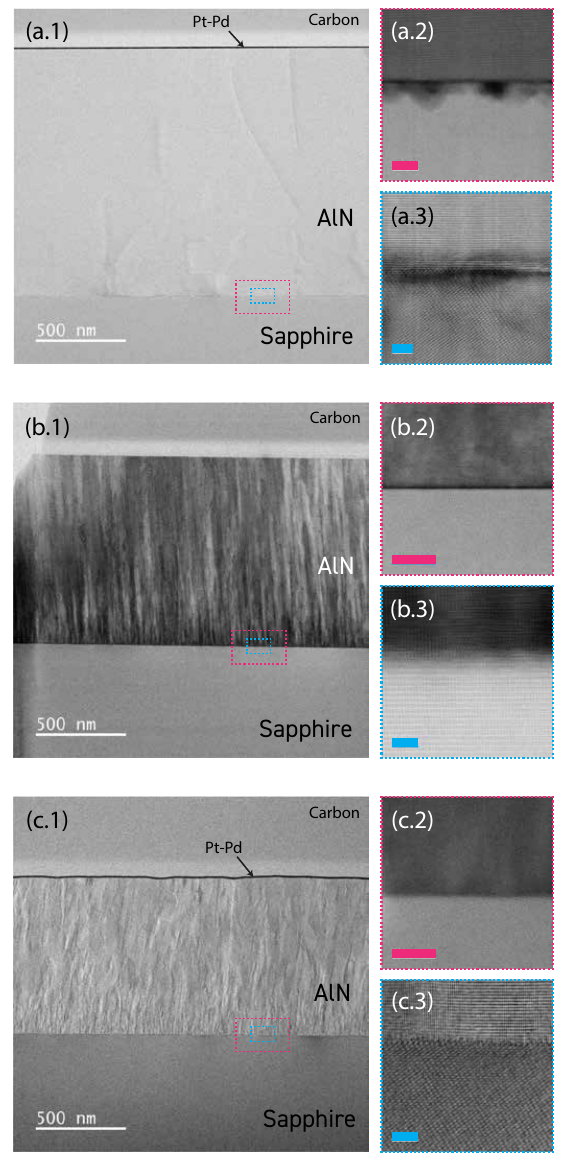}
    \caption{\textbf{Electron microscopy analysis of the AlN/sapphire interface.} Pink and blue scale-bars correspond to 10~nm and 2~nm, respectively. (a) HVPE-grown AlN on sapphire: (a.1) STEM-BF image; (a.2) STEM-HAADF image of the AlN/sapphire interface showing $\sim10~$nm-deep pits at the sapphire surface; (a.3) HRSTEM-BF image of the interface. (b) pulsed DC-sputtered AlN on sapphire: (b.1) STEM-BF image showing a polycrystalline, columnar grain structure of the AlN; (b.2) HRSTEM-HAADF image of the interface; (b.3) HRSTEM-BF close-up of the same region. (c) HiPIMS-grown AlN on sapphire: (c.1) STEM-BF image; (c.2) STEM-HAADF image of the interface. (c.3) STEM-BF image at higher magnification of the interface region.}
    \label{fig:TEM}
\end{figure}

STEM and high-resolution STEM (HRSTEM) images of the three samples are acquired with a JEOL Grand ARM microscope and presented in Figure \ref{fig:TEM}. The HVPE grown AlN sample, sourced from an HBAR fabricated on the same wafer as samples B.1-4, is shown in Figures \ref{fig:TEM}(a.1)-(a.3). This material exhibits a nearly single-crystal structure with a low density of dislocations and other structural defects. However, the magnified high-angle annular dark-field (HAADF) STEM images of the AlN/sapphire interface in this sample show clear signs of damage, with pits on the sapphire surface extending $\sim10$ nm deep in the substrate. Energy-dispersive X-ray spectroscopy (EDS) analysis of this interface also revealed a reduced concentration of aluminum in the damaged region, suggesting the presence of a lower-density interfacial layer (see Supplementary section \ref{sec:EDS}). 

Figures \ref{fig:TEM}(b.1) and (c.1) show STEM images of AlN films deposited by pulsed DC sputtering and HiPIMS, respectively. Both films exhibit a polycrystalline microstructure with columnar grains predominantly oriented normal to the substrate surface. The grain boundaries are clearly visible in both samples. The corresponding magnified STEM images (Fig. \ref{fig:TEM}(b.2) and (c.2)) and HRSTEM images (Fig. \ref{fig:TEM}(b.3) and (c.3)) reveal sharp, continuous interfaces with sapphire, without visible substrate damage. The two imaged samples were obtained from HBARs fabricated on the same wafers as Samples A and E, respectively. 
\subsubsection*{AFM}
\label{sec:AFM}
\begin{figure}[t!]
    \centering
    \includegraphics[width=\linewidth]{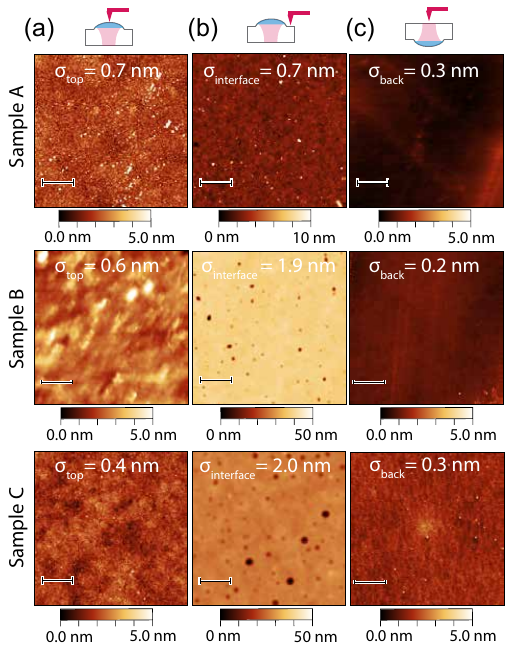}
    \caption{\textbf{Atomic force microscopy (AFM) scans of the HBARs.} All scans are acquired on a $10\times10~\mu$m$^2$ area, with the scale-bar corresponding to $2~\mu$m.  (a) Surface morfology at the top of the AlN dome. (b) Planar region adjacent to the dome, where the AlN has been fully etched. (c) Backside surface of the HBAR. The RMS surface roughness $\sigma$ extracted from each scan is indicated on the corresponding image. }
    \label{fig:AFM}
\end{figure}

To assess the surface morphology of the fabricated HBARs, we perform atomic force microscopy (AFM) scans using a Bruker Dimension FastScan operating in tapping mode. The three samples analyzed are representative of Samples A, B.1-4 and C, respectively, as they were fabricated from the same respective wafers. Scans are conducted at three distinct locations on each HBAR: on top of the AlN dome (Fig. \ref{fig:AFM}(a), left-column), on a flat region adjacent to the dome where the AlN has been completely etched (Fig. \ref{fig:AFM}(b), center-column), and on the backside of the HBAR (Fig. \ref{fig:AFM}(c), right-column). Scans of the AlN dome indicate that the etching process preserves a RMS surface roughness below 0.8~nm. The backside exhibits RMS roughness below 0.5 nm, expected for polished sapphire. Given the comparable etch rates of AlN and sapphire ($\sim25$~nm/min for both), regions where  AlN has been fully etched provide an upper bound estimate of the roughness at the AlN/sapphire interface. Samples B and C, with HVPE grown AlN, show localized surface pits and elevated RMS surface roughness ($\sim 2$~nm) at the interface, consistent with the damage observed in the STEM images. In contrast, Sample A, with sputtered AlN, displays a smoother interface with an RMS roughness of 0.7~nm.

\subsection*{Experimental setup}
\begin{figure*}[t!]
    \centering
    \includegraphics[width =0.9 \linewidth]{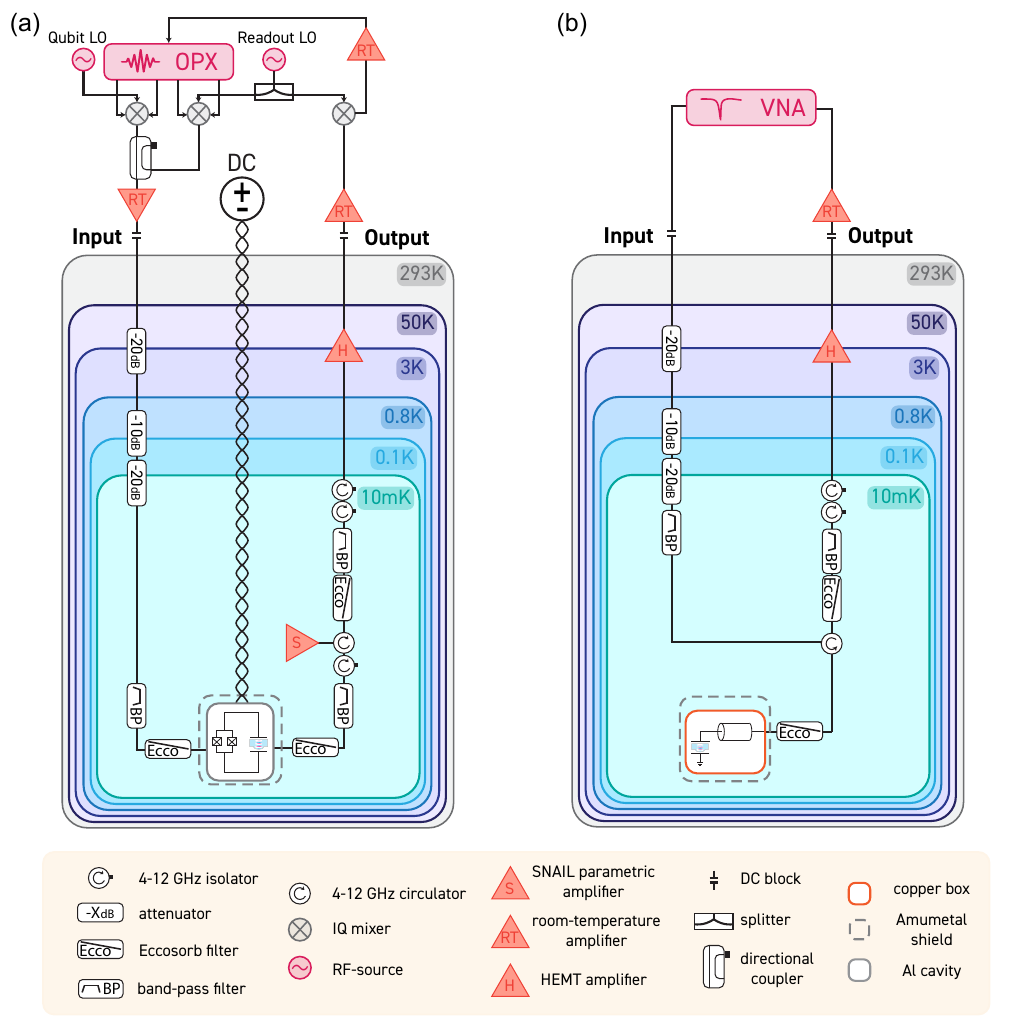}
    \caption{\textbf{Cryogenic experimental setup.} (a) Schematic of the wiring and control electronics used in the qubit measurements. (b) Schematic of the measurement setup used in the spectroscopy measurements.}
    \label{fig:experimentalsetup}
\end{figure*}
The HBARs are measured in a dilution refrigerator (Bluefors LD400) with a base temperature of approximately 10~mK. The measurements with a qubit are performed in the experimental setup depicted in Fig. \ref{fig:experimentalsetup}(a). The qubit bonded to the HBAR is housed inside a three-dimensional aluminum microwave cavity, used for qubit readout and control. For measurements involving the flux-tunable qubit (Qubit-1), a flux-hose cavity \cite{gargiulo_fast_2021} is employed and connected to a DC line providing a continuous current from a low-noise DC source (Yokogawa, GS200). For measurements using the single Josephson junction qubit (Qubit-2), a conventional aluminum cavity is used without a DC line. The cavites are shielded from external magnetic fields by an Amumetal shield. Qubit drive and readout pulses are generated using a Quantum Machines OPX+ arbitrary waveform generator (AWG), and up-converted to GHz frequencies using IQ mixers driven by a continuous LO tone from a microwave source (R\&S, SGS100A). The drive and readout signals are combined at a directional coupler, pre-amplified, and routed into the dilution refrigerator through an input line incorporating multiple stages of attenuation and filtering to suppress thermal and microwave noise. Upon leaving the cavity, the readout signal is amplified by a SNAIL parametric amplifier \cite{frattini2018optimizing} (used only with Qubit-2), a high-electron mobility transistor (HEMT) amplifier (LNF-LNC4-8G), and additional room-temperature amplifiers. The amplified signals are then down-converted using an IQ mixer and digitized at the OPX+ for demodulation and analysis. 

For the HBAR spectroscopy measurements, the packaged antenna chip is mounted on the base plate of the dilution refrigerator, as shown in Fig. \ref{fig:experimentalsetup}(b). Reflection measurements are performed using a circulator, which routes the reflected signal from the sample to the output line. The signal is then amplified by a HEMT (LNF-LNC4-16C) and a room-temperature amplifier. In this configuration, the VNA (Keysight, P5004A) is connected to the input and output lines of the refrigerator and we measure the transmission through the setup. The bandwidth of the cryogenic setup is limited to 4-12 GHz by the isolators and circulators. Outside of this bandwidth, some phonon modes can still be detected, but the resonances suffer substantial Fano interference \cite{rieger_fano_2023}. This effect is accounted for in the fitting procedure described in Methods. Data acquisition is automated, with each resonance measured individually using a 100~kHz frequency span, 3001~points, 10~averages, and an IF bandwidth of 1~kHz. The microwave power applied at the sample input for high-power measurements is around -70~dBm. For drive power sweeps on samples A and B.1, measurements are performed with a reduced frequency span of $\sim20~$kHz, 501 points, and an IF bandwidth of 100~Hz. Spectra at different powers are acquired sequentially with a single average per power, repeated multiple times, and averaged in post-processing. This ensures identical effective acquisition times for each power and comparable broadening due to frequency drifts across all data points. In all the measurements, the VNA is locked to an external Rb atomic clock. 

\subsection*{Modeling of the frequency response and phonon number estimates}
\label{sec:S11fit}
The internal quality factor $Q_{\text{i}}$, external quality factor $Q_{\text{e}}$, and resonance frequency $f_n$ of the phonon modes are obtained by fitting the coherent frequency response around a resonance to the reflection model
\begin{equation}\label{eq:S11}
    S_{11}=1-\frac{2e^{i\phi}}{1+\frac{Q_e}{Q_{\text{i}}}+2iQ_{\text{e}}(\frac{f}{f_n}-1)},
\end{equation}
derived from input-output theory. $\phi$ is a phase that accounts for interference with background signals \cite{probst_efficient_2015,rieger_fano_2023}. Since we do not calibrate the microwave transmission through the entire experimental setup, when processing the data we first remove the background from the measured signal. This is achieved by fitting the background using the function $(A+Bf+Cf^2)\cdot(e^{i (a+bf+cf^2)})$, where the quadratic pre-factor accounts for distortions arising from the frequency-dependent response of the different microwave components in the measurement chain, while the complex exponential accounts for its electrical phase delay. The fits are performed using the Python package \texttt{stlab} \cite{noauthor_steelelab-delftstlab_2025}.

To estimate the average phonon occupation number $\bar{n}_p$ on resonance in the HBAR modes for a given input power $P_{\text{in}}$, we use the expression
\begin{equation}\label{eq:phononnumber}
    \bar{n}_p = \frac{4\frac{\omega_n}{Q_{\text{e}}}}{(\frac{\omega_n}{Q_{\text{e}}}+\frac{\omega_n}{Q_{\text{i}}})^2\hbar\omega_n}P_{\text{in}},
\end{equation}
where $\omega_n=2\pi f_n$, and $P_{\text{in}}$ is the input power referenced to the input port of the antenna chip, expressed in Watts. To estimate $P_{\text{in}}$, we consider the power sent by the source (e.g. from the VNA) and the attenuation of the cables and microwave components in the input line of the refrigerator leading to the sample.


\subsection*{Qubit parameters}

\begin{table}[h!]
    \centering
    \renewcommand{\arraystretch}{1.5}
    \caption{\textbf{List of qubit parameters.} The transmon qubit properties are measured at the qubit's idle frequency, far detuned from the phonon modes.}
    \label{tab:qubitparameters}
    \begin{tabular}{l l c c}
    \hline
    &&\multicolumn{2}{c}{Value}\\
    Quantity&Symbol&Qubit-1&Qubit-2\\
    \hline
    \hline
        frequency& $\omega_q/2\pi$ (GHz)&3.8 - 7.0&5.096\\
        relaxation time& $T_1$ ($\mu$s)&3.3 - 0.5&33.4\\
        coherence time (Ramsey)& $T_{2,q}^*$ ($\mu$s)&0.5 - 0.2&40.0\\
        coherence time (Echo)& $T_{2,q}^e$ ($\mu$s)&1.0 - 0.4&43.1\\
        qubit-phonon coupling&$g_0/2\pi$ (kHz) &380 - 480&298\\
        \hline
    \end{tabular}
\end{table} 

\begin{figure}
    \centering
    \includegraphics[width=\linewidth]{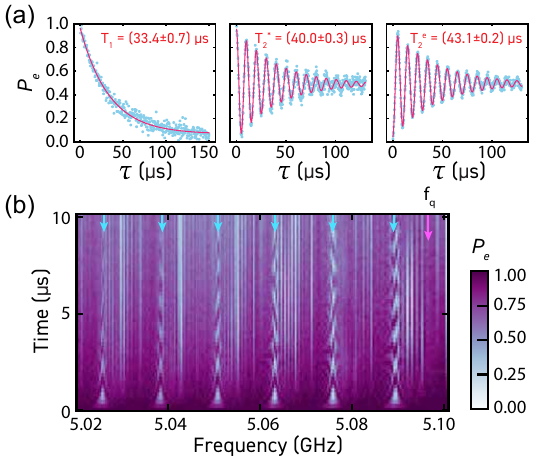}
    \caption{\textbf{Qubit properties and phonon spectra.} (a) Measurements of Qubit-2 $T_1$, $T_2^*$ and $T_2^e$ times at its idling frequency ($f_q=5.096$ GHz). (b) Qubit excited state population measured after exciting the qubit at the idle frequency (indicated by the pink arrow) and waiting for a variable time as the qubit frequency is shifted via the AC Stark shift. The clear oscillations correspond to vacuum Rabi oscillations between the qubit and the LG$_{00}$ overtone modes (marked with blue arrows). The additional features correspond to high order modes with weaker coupling to the qubit.}
    \label{fig:qubitspec}
\end{figure}

In Table \ref{tab:qubitparameters}, we list the relevant parameters of the flux-tunable qubit (Qubit-1) and single Josephson junction qubit (Qubit-2), used in the characterization of Samples A and B.2, respectively. The coherence times of Qubit-2, shown in Figure \ref{fig:qubitspec}(a), are measured with the qubit at an idle frequency far detuned from phonon modes to avoid introducing additional decay channels on the qubit. To probe the interaction between Qubit-2 and the phonon modes, we send a microwave pulse detuned by 80~MHz from the readout cavity frequency, which induces an AC Stark shift and tunes the frequency of the qubit up to $\sim 100$ MHz. After preparing the qubit in the excited state at the idle frequency, we tune its frequency for a variable duration and then read out its state at the idle frequency. We observe coherent population swaps when the qubit is on resonance with a phonon mode, as shown in Figure \ref{fig:qubitmeas}(b). In the main text, we report the coherence times of the six phonon modes closest to the qubit. From the rate of the vacuum Rabi oscillations, we extract the phonon-qubit coupling rate $g_0/2\pi$, which is consistently measured to be $\sim298\pm3~$kHz for all LG$_{00}$ overtone modes interacting with Qubit-2. In addition to these fundamental modes, we observe features corresponding to higher-order transverse phonon modes, with weaker coupling to the qubit. Qubit-1 is measured in a similar way, but introducing an additional flux offset to tune the qubit frequency over a substantially broader range. 

Note that the qubit induces an additional decay channel for the phonon modes via the inverse Purcell effect. We quantitatively characterize this loss mechanism by determining the qubit decay rate, the qubit-phonon coupling strength, and the detuning between the qubit idle frequency and the phonon modes (see Supplementary Information I E).


\subsection*{Multilayer Participation Model}\label{sec:multilayer}

\begin{table*}[htb!]
    \centering
    \caption{\textbf{Summary of fitted acoustic properties for individual layers $X$}. For piezo and bulk layers, material density and sound velocity are constrained to be consistent between different samples. For the defect layer, only its thickness is constrained between samples. For mechanical Q factors of each layer, we quote a lower bound, calculated by assuming no mechanical loss in any other layers.} 
    \label{tab:parameters}
    \renewcommand{\arraystretch}{1.5}
    \setlength{\tabcolsep}{6.5pt}
    \begin{tabular}{ c c c c c c c c c }
    \hline
    \hline
         Sample &AlN growth& RoC (mm) & Layer & $t_X$ ($\mu$m)& $v_X$ (km/s)  & $\rho_X$ $\left(\text{g}/\text{cm}^3\right)$& min $Q_{X,\text{mech}}$ & $Q_{X,\text{TLS}}$\\
         \hline
         \multirow{2}{*}{A}&\multirow{2}{*}{\shortstack{pulsed DC \\ sputtering}}&\multirow{2}{*}{6.4}
         &P&1.0078&10.920&3.306&$1.89(11)\times10^4$&$1.71(3)\times10^4$\\
         &&&B&484.45&11.059&3.98&$8.99(54)\times10^6$&\\
         \hline
         \multirow{3}{*}{B.1}&\multirow{3}{*}{HVPE}&\multirow{3}{*}{6.9}
         &P&1.1778&10.920&3.306&$3.02(20)\times10^4$&$7.9(1)\times10^4$\\
         &&&D&0.010&11.767&1.823&$244(164)$&$1.33(5)\times10^3$\\
         &&&B&434.57&11.059&3.98&$1.14(76)\times10^7$&\\
         \hline
         \multirow{3}{*}{B.4}&\multirow{3}{*}{HVPE}&\multirow{3}{*}{6.9}
         &P&1.219&10.920&3.306&$4.46(29)\times10^4$&\\
         &&&D&0.010&10.530&2.12&$452(309)$&\\
         &&&B&434.59&11.059&3.98&$1.57(99)\times10^7$&\\
         \hline
         \multirow{3}{*}{C}&\multirow{3}{*}{HVPE}&\multirow{3}{*}{16.4}
         &P&1.7676&10.920&3.306&$1.05(89)\times10^5$&\\
         &&&D&0.010&10.450&2.12&$822(474)$&\\
         &&&B&434.40&11.059&3.98&$1.48(97)\times10^7$&\\
         \hline
         \multirow{3}{*}{D}&\multirow{3}{*}{HVPE}&\multirow{3}{*}{7.6}
         &P&1.5192&10.920&3.306&$1.04(74)\times10^5$&\\
         &&&D&0.010&12.0135&2.90&$649(464)$&\\
         &&&B&171.175&11.059&3.98&$1.29(92)\times10^7$&\\
         \hline
         \multirow{2}{*}{E}&\multirow{2}{*}{HiPIMS}&\multirow{2}{*}{11.0}
         &P&0.9239&10.785&3.306&$2.14(11)\times10^4$&$1.13(3)\times10^4$\\
         &&&B&422.74&11.059&3.98&$9.82(51)\times10^6$&\\
    \hline
    \hline
    \end{tabular}
\end{table*}

By analogy to voltage and current in an electrical transmission line, longitudinal acoustic waves in a homogeneous solid can be described by the local stress $T$ and velocity $\dot{u}$ \cite{mason_electromechanical_1948}. A resonant acoustic mode formed in a given length of material can be expressed as a standing wave of longitudinal plane waves traveling in opposite directions.
The general solution for the profile of a standing longitudinal plane wave in layer $X$ can then be expressed in terms of a position offset $\delta_{X}$:
\begin{equation}
\begin{aligned}
    T_X(z) &= T_X^+e^{-i k_X z}+T_X^-e^{+i k_X z}\\&= 2T_X \sin\left(k_X (z+\delta_X)\right)\\
    \dot{u}_X(z) &= \dot{u}_X^+e^{-i k_X z}+\dot{u}_X^-e^{+i k_X z}\\&=2\frac{T_X}{Z_{X}} \cos\left(k_X (z+\delta_X)\right)
\end{aligned}
\label{eq:genericsols}
\end{equation}
For these solutions, the wavevectors $k_X = 2\pi f /v_X$ satisfy the dispersion relation of the material. For each wave, stress and velocity are related by the acoustic impedance $Z_X$, expressed in terms of the density $\rho_X$ and the sound velocity $v_X$ of the solid, in the same manner as electrical impedance \cite{pozar2021microwave}: 
\begin{equation}
    Z_X =\rho_X v_X= \frac{T_X^+}{\dot{u}_X^+} =  -\frac{T_X^-}{\dot{u}_X^-} 
    \label{eq:impedance}
\end{equation}
The resonant longitudinal modes of a particular material stack can then be calculated by matching force and velocity between each layer and applying the appropriate boundary conditions.
In the case of the piezoelectric (P) - defect (D) - bulk (B) stack representing the HBAR resonator (illustrated in \Fref{fig:antennameas}(c)) that is unclamped on both sides, the boundary conditions enforce zero stress on the outer boundaries, which leads to $\delta_B = -t_B$ and $\delta_P = t_P+t_D$ (using the coordinate system from \Fref{fig:device} and placing the origin $z=0$ at the D-B interface).
The interface boundary conditions then dictate:
\begin{equation}
\begin{aligned}
    T_P \sin (k_P t_P) = T_D \sin \left(k_D (\delta_D -t_D)\right),\\
    \frac{T_P}{Z_P} \cos (k_P t_P) = \frac{T_D}{Z_D} \cos \left(k_D (\delta_D -t_D)\right)
    \label{eq:boundaryconds}
\end{aligned}
\end{equation}
for the P-D boundary, and 
\begin{equation}
\begin{aligned}
    T_D \sin (k_D\delta_D) = T_B \sin (- k_B t_B),\\
    \frac{T_D}{Z_D} \cos (k_D \delta_D) = \frac{T_B}{Z_B}\cos(-k_B t_B)
    \label{eq:boundaryconds2}
\end{aligned}
\end{equation}
for D-B interface.
\Eref{eq:boundaryconds2} can be solved directly for the position offset and the amplitude of the mode in the defect layer:
\begin{align}
\delta_D &= \frac{v_D}{2\pi f}\arctan\left(\frac{Z_B}{ -Z_D }\tan\left(\frac{2\pi f}{v_B}t_B\right)\right)\label{eq:phi},\\
T_D &=  \frac{T_B}{Z_B}\sqrt{Z_D^2 \cos^2 \frac{2\pi f}{v_B}t_B+Z_B^2 \sin^2 \frac{2\pi f}{v_B}t_B}=\eta T_B.\end{align}
Here, we introduce the factor $\eta$ to simplify the notation. Combined with the P-D boundary equations, this yields a self-consistent equation similar to the two-layer case discussed in the main text:
\begin{equation}
    \frac{T_B}{T_P} = \frac{1}{\eta}\frac{\sin \frac{2\pi f}{v_P}t_P}{\sin \frac{2\pi f}{v_D}(\delta_D - t_D)}=
    \frac{Z_D}{\eta Z_P}\frac{\cos \frac{2\pi f}{v_P} t_P}{\cos \frac{2\pi f}{v_D}(\delta_D - t_D)}
    \label{eq:3mode}
\end{equation}
In the limit $Z_D\to Z_B$, $\delta_D= -t_B$ and $\eta = 1$, recovering the two-layer case (\Eref{eq:frequencies}).

\subsubsection*{Fitting Procedure}
\label{sec:fitting}

In principle, frequency solutions $f'_m$ to \Eref{eq:3mode} can be matched against the measured mode frequencies until a good match is found.
However, this is rather cumbersome as each material layer $X$ introduces three separate variables $t_X$, $v_X$ and $Z_X$.
Closer observation of \Eref{eq:phi}-\eqref{eq:3mode} reveals that the variables always appear in pairs, making it significantly easier to simplify the expressions in terms of accumulated phases $\beta_X = t_X/v_X$ for each layer, as well as impedance ratios $Z_P/Z_D$ and $Z_B/Z_D$, yielding only five independent model parameters.

Direct mode matching between numerical solutions and the frequency data is further complicated by the inability to experimentally measure every single mode, whether from bandwidth or transducer limitations.
To remedy this, we can first assign a mode index $n$ to each measured mode frequency $f_n$ through integer division by the average FSR $\Delta_0$, and similarly for the solved roots $f'_m$:
\begin{equation}
    n \equiv \Bigl\lfloor\frac{f_n + \Delta_0/2}{\Delta_0}\Bigr\rfloor,\qquad
    m \equiv \Bigl\lfloor\frac{f'_m + \Delta_0/2}{\Delta_0}\Bigr\rfloor
\end{equation}
To address potential gaps in either $f_n$ or $f'_m$, the root solver is first seeded with measured mode frequencies $f_n$, and the error $\varepsilon$ is computed using only overlapping indices $\{m\}\cap\{n\}$:
\begin{equation}
\varepsilon = \sum_{n=m}\left((f_n-n\Delta_0) - \psi - (f'_m-m\Delta_0)\right)^2 
\end{equation}
Here $\psi$ is a final fit parameter corresponding to the frequency shift from the Gouy phase \cite{saleh_fundamentals_1991} arising from the curved dome surface.
With index mismatches addressed, the error can then be minimized as a function of the reduced model parameters ($\beta_P$,$\beta_D$,$\beta_P$, $Z_P/Z_D$,$Z_B/Z_D$) in a standard fashion.
Compared to simply fitting the FSR (shown in \Fref{fig:antennameas}(e)), this fitting method proves far more accurate, as shown in \Fref{fig:fig11particips}(a-c).
We further observe clear evidence for the presence of the defect layer as the oscillations in frequency shift grow in amplitude with increasing mode frequency, which could not be explained with the simpler two-layer model.
Constraining material density and sound velocity for the bulk and piezo layers to be consistent across samples yields unique solutions for the individual parameters of each layer, summarized in Table \ref{tab:parameters}.

\subsubsection*{Energy Participation}
\label{sec:participation}
\begin{figure*}[t!]
\centering
\includegraphics[width=6.67in]{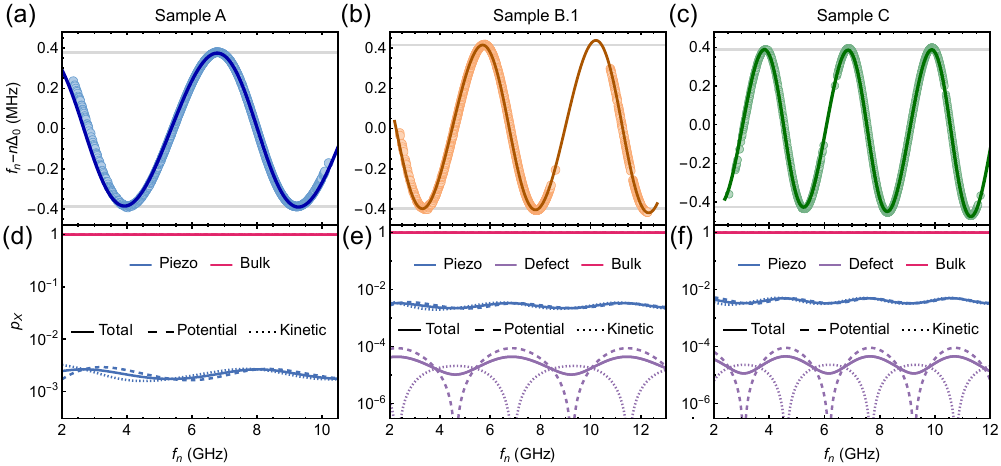}
\caption{
\textbf{Multilayer participation model.} (a-c) Frequency deviations for samples A, B.1 and C respectively. The solid lines are optimized solutions to \Eref{eq:3mode}. To reveal higher-order behavior, the average FSR $\Delta_0$ is subtracted from both the data and the model.
Grey lines mark the minimum and maximum of a simple two-layer fit as a guide to the eye, highlighting the need for a third layer for samples B.1 and C.
(d-f) Energy participations calculated from the optimized multilayer model for samples A-C respectively.
The dashed and dotted lines correspond to potential and kinetic participation ratios for each layer.
\label{fig:fig11particips}}
\end{figure*}

Having determined the relevant acoustic properties of each layer, we can use the general solutions for force and velocity in \Eref{eq:genericsols} to determine the mode profile, and thus the energy stored in each layer.
By analogy to an electrical transmission line \cite{mason_electromechanical_1948}, the kinetic, potential and total energy in a longitudinal acoustic wave can be expressed as the integral of the squared stress and velocity fields:
\begin{equation}\label{eq:energy}
    \mathcal{E}_{\text{tot}} = \mathcal{E}_{\text{pot}} + \mathcal{E}_{\text{kin}} = \underbrace{\frac{1}{2}\int\frac{T(z)^2}{v Z}dz}_{\text{Potential}}+\underbrace{\frac{1}{2}\int\frac{Z}{v}\dot{u}(z)^2dz}_{\text{Kinetic}} 
\end{equation}
Integrating over each layer, we arrive at expressions for the kinetic and potential energy in each layer in terms of the piezo layer maximum strain $T_P$:
\begin{equation}\label{eq:potentialkineticenergies}
\begin{aligned}
\mathcal{E}^{\text{pot},\text{kin}}_P &= 
\frac{T_P^2 t_P}{v_PZ_P} \mp  \frac{T_P^2}{Z_P 4\pi f}\sin\frac{4\pi f t_P}{v_P}\\
\mathcal{E}^{\text{pot},\text{kin}}_D &= 
\frac{T_P^2 t_D \xi^2}{v_D Z_D}\mp \\&\frac{T_P^2\xi^2}{Z_D 4\pi f}\left(\sin\frac{ \delta_D 4\pi f}{v_D}-\sin\frac{4\pi f (\delta_D - t_D)}{v_D}\right)\\
\mathcal{E}^{\text{pot},\text{kin}}_B &= 
\frac{T_P^2 t_B}{v_B Z_B}\frac{\xi^2}{\eta^2} \mp \frac{T_P^2}{Z_B4\pi f}\frac{\xi^2}{\eta^2}\sin \frac{4\pi f t_B}{v_B}
\end{aligned}
\end{equation}
Where the potential energies have a minus sign, and the kinetic energies a plus, and we have defined $\xi = \sin\left(\frac{2\pi f t_P}{v_P}\right)/\sin\left(\frac{2\pi f}{v_D} (\delta_D - t_D)\right)$ for convenience.
Since the second terms are of opposite sign for the kinetic and potential energies, the total acoustic energies for each layer ($\mathcal{E}^{\text{tot}}_X=\mathcal{E}^{\text{pot}}_X+\mathcal{E}^{\text{kin}}_X$) are just twice the first term of each expression.

We are particularly interested in the energy distribution for each mode, which is quantified by the unit-less energy participation ratio \cite{minev_energy-participation_2021,wang_surface_2015} in each layer $X$:
\begin{equation}\label{eq:participation}
    p^{\text{pot}}_{X} = \frac{\mathcal{E}^{\text{pot}}_X}{\sum_Y\mathcal{E}^{\text{pot}}_Y}, \quad p^{\text{kin}}_{X} = \frac{\mathcal{E}^{\text{kin}}_X}{\sum_Y\mathcal{E}^{\text{kin}}_Y}, \quad p_X=\frac{\mathcal{E}^{\text{tot}}_X}{\sum_Y\mathcal{E}^{\text{tot}}_Y}.
\end{equation}
These participation ratios are summarized for each layer and each energy type in \Fref{fig:fig11particips}(d-f).
In all cases, the energy  is contained predominantly in the bulk, which is orders of magnitude thicker than the other layers. At the same time, we observe clear oscillations in energy participation with respect to mode frequency.
Since the piezoelectric and defect layers are sub-wavelength, these oscillations have different phases when considering only the potential or kinetic energy, providing a method for further distinguishing loss mechanisms.
These oscillations are most pronounced in the defect layer, where the mode profile effectively shifts between stress and velocity nodes as a function of mode frequency.

\subsection*{Multilayer Scattering}
\begin{figure}[htb]
\centering
\includegraphics[width=\linewidth]{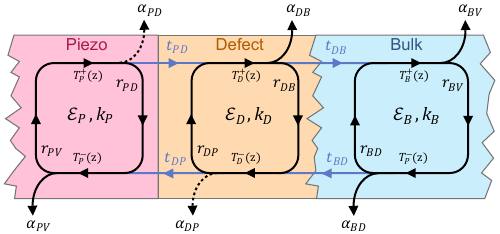}
\caption{\textbf{Multilayer resonance power flow diagram.} The one-dimensional resonator is modeled as three coupled cavities, with losses $
\alpha$ arising from diffuse reflection $r$ and transmission $t$.
As discussed in the text, the piezo-defect boundary is modeled as smooth and lossless, as indicated by the dashed arrow for the power loss $\alpha_{PD}$}
\label{fig:fig12scatter}
\end{figure}

To determine the dissipation from scattering on the boundaries of each layer, we consider the power flow for three coupled layers in terms of the reflection $r$ and transmission $t$ parameters for each boundary, summarized in \Fref{fig:fig12scatter}. These coefficients satisfy

\begin{equation}
    t^2 + r^2 =p^2(t_0 ^2 +r_0^2)= p^2 = 1- \alpha^2,
\end{equation}
where $p$ is the specularity parameter that depends on the surface roughness (see Supplementary Information I B), $r_0$ and $t_0$ are the coherent reflection and transmission coefficients, and we have combined the total power lost into one term $\alpha$.
The top (piezo-vacuum) and bottom (bulk-vacuum) boundary roughnesses can be measured directly.
With insight from cross-sectional images (\Fref{fig:TEM}(a.2)), we make the assumption that the D-B interface is far rougher than the P-D interface, which appears relatively smooth.
Thus, we neglect the scattering from the P-D interface (dashed arrows in \Fref{fig:fig12scatter}), and assign the measured interface roughness from \Fref{fig:AFM}(b) to the D-B boundary.

We can use the previous solutions (\Eref{eq:3mode}) 
to determine the total energy participation in each layer.
Without changing the power losses for each boundary, by setting $t=0$ and $r=p$ we can now consider the system as split into three individual cavities, provided the standing wave circulating in each split cavity satisfies the original coupled boundary conditions.
The total energy lost by the system can be rewritten as the product of the energy participation $p_X$ in each cavity with its individual loss, derived in Supplementary section I B for a single layer Fabry-Perot cavity
\begin{equation}\label{Eq:surfaceroughnessmultilayer}
    Q_{X,\sigma}=\frac{t_X v_X}{4\pi f_n(\sigma_{X,1}^2+\sigma_{X,2}^2)},
\end{equation}
where $\sigma_{X,1}$ and $\sigma_{X,2}$ are the RMS roughness of the top and bottom surfaces of layer $X$, respectively.
Following the convention used for electrical systems \cite{wang_surface_2015, minev_energy-participation_2021}, the total scattering loss of the system can be written as a sum of the individual scattering losses $Q_{X,\sigma}$ weighted by the total energy participation of each layer $p_X$:
\begin{equation}
    \frac{1}{Q_\sigma} = \frac{p_P}{Q_{\sigma,P}}+\frac{p_D}{Q_{\sigma,D}}+\frac{p_B}{Q_{\sigma,B}}
\end{equation}
This loss inherits both the inverse frequency dependence of the scattering loss in Eq. \eqref{Eq:surfaceroughnessmultilayer} and the oscillatory behavior arising from oscillations in the energy participation.


\section*{Acknowledgements}
We thank Simon Storz, Yu Yang, Hugo Doeleman, Uwe von Luepke and Pietro Borghi for valuable discussions, ScopeM and specifically Zeng Peng for the FIB preparation of the lamellas and acquisition of STEM images. Y. C. and other authors affiliated with ETH Zürich acknowledge funding from Projects CRSII--222812, 200021\_204073, and Ambizione Grant no. 208886 from the Swiss National Science Foundation (SNSF), ETH Grant 23-2 ETH-041, and the European Research Council (ERC) under the European Union’s Horizon 2020 research and innovation programme (Grant agreement No. 948047). 
This research was funded in whole or in part by the Austrian Science Fund (FWF) DOI 10.5776/W1259 as well as 10.557766/F71. For open access purposes, the author has applied a CC BY public copyright license to any author accepted manuscript version arising from this submission.
J.P., O.P. and S.S. acknowledge funding by SNSF under Projects 200021\_196980 and 10004403. M.L. and L.G.V. acknowledge funding from the SNSF under Projects CRSII--222812 and 200020\_184935.  

\section*{Data availability}
The data that support the findings of this study are available from the corresponding
authors upon reasonable request.

\section*{Author contributions}
R.G.B., A.B. and Y.C. conceived and planned the experiments. R.G.B., L.F.D. and L.C. acquired the data. R.G.B., A.A., L.F.D. and L.C. analyzed the data and derived theoretical models. R.G.B., A.B., M.D., I.C.R., R.B. and M.F. fabricated the devices. T.S. assisted with cryogenic measurements. M.L. and L.G.V. supplied the DC-sputtered AlN films, J.P. and S.S. supplied the HiPIMS AlN films, and G.K. provided the flux-hose cavity. A.B., R.B., M.L. and O.P. performed XRD measurements. Y.C. supervised the work. R.G.B., A.A. and Y.C. wrote the manuscript with input from all the authors.


\begin{thebibliography}{83}%
\makeatletter
\providecommand \@ifxundefined [1]{%
 \@ifx{#1\undefined}
}%
\providecommand \@ifnum [1]{%
 \ifnum #1\expandafter \@firstoftwo
 \else \expandafter \@secondoftwo
 \fi
}%
\providecommand \@ifx [1]{%
 \ifx #1\expandafter \@firstoftwo
 \else \expandafter \@secondoftwo
 \fi
}%
\providecommand \natexlab [1]{#1}%
\providecommand \enquote  [1]{``#1''}%
\providecommand \bibnamefont  [1]{#1}%
\providecommand \bibfnamefont [1]{#1}%
\providecommand \citenamefont [1]{#1}%
\providecommand \href@noop [0]{\@secondoftwo}%
\providecommand \href [0]{\begingroup \@sanitize@url \@href}%
\providecommand \@href[1]{\@@startlink{#1}\@@href}%
\providecommand \@@href[1]{\endgroup#1\@@endlink}%
\providecommand \@sanitize@url [0]{\catcode `\\12\catcode `\$12\catcode `\&12\catcode `\#12\catcode `\^12\catcode `\_12\catcode `\%12\relax}%
\providecommand \@@startlink[1]{}%
\providecommand \@@endlink[0]{}%
\providecommand \url  [0]{\begingroup\@sanitize@url \@url }%
\providecommand \@url [1]{\endgroup\@href {#1}{\urlprefix }}%
\providecommand \urlprefix  [0]{URL }%
\providecommand \Eprint [0]{\href }%
\providecommand \doibase [0]{https://doi.org/}%
\providecommand \selectlanguage [0]{\@gobble}%
\providecommand \bibinfo  [0]{\@secondoftwo}%
\providecommand \bibfield  [0]{\@secondoftwo}%
\providecommand \translation [1]{[#1]}%
\providecommand \BibitemOpen [0]{}%
\providecommand \bibitemStop [0]{}%
\providecommand \bibitemNoStop [0]{.\EOS\space}%
\providecommand \EOS [0]{\spacefactor3000\relax}%
\providecommand \BibitemShut  [1]{\csname bibitem#1\endcsname}%
\let\auto@bib@innerbib\@empty
\bibitem [{\citenamefont {Linehan}\ \emph {et~al.}(2025)Linehan, Ryan and Trickle, Tanner and Conner, Christopher R and Ghosh, Sohitri and Lin, Tongyan and Sholapurkar, Mukul and Cleland, Andrew N}]{linehan2025listening}%
  \BibitemOpen
  \bibfield  {author} {\bibinfo {author} {Linehan, R.}, \bibinfo {author} {Trickle, T.}, \bibinfo {author} {Conner, C.~R.}, \emph {et~al.},\ }\bibfield  {title} {\bibinfo {title} {Listening for new physics with quantum acoustics},\ }\href {https://journals.aps.org/prd/abstract/10.1103/63zj-d8z4} {\bibfield  {journal} {\bibinfo  {journal} {Physical Review D}\ }\textbf {\bibinfo {volume} {112}},\ \bibinfo {pages} {115005} (\bibinfo {year} {2025})}\BibitemShut {NoStop}%
\bibitem [{\citenamefont {Mason}\ \emph {et~al.}(2019)Mason, David and Chen, Junxin and Rossi, Massimiliano and Tsaturyan, Yeghishe and Schliesser, Albert}]{mason2019continuous}%
  \BibitemOpen
  \bibfield  {author} {\bibinfo {author} {Mason, D.}, \bibinfo {author} {Chen, J.}, \bibinfo {author} {Rossi, M.}, \emph {et~al.},\ }\bibfield  {title} {\bibinfo {title} {Continuous force and displacement measurement below the standard quantum limit},\ }\href {https://www.nature.com/articles/s41567-019-0533-5} {\bibfield  {journal} {\bibinfo  {journal} {Nature Physics}\ }\textbf {\bibinfo {volume} {15}},\ \bibinfo {pages} {745} (\bibinfo {year} {2019})}\BibitemShut {NoStop}%
\bibitem [{\citenamefont {Pikovski}\ \emph {et~al.}(2012)Pikovski, Igor and Vanner, Michael R and Aspelmeyer, Markus and Kim, MS and Brukner, {\v{C}}aslav}]{pikovski2012probing}%
  \BibitemOpen
  \bibfield  {author} {\bibinfo {author} {Pikovski, I.}, \bibinfo {author} {Vanner, M.~R.}, \bibinfo {author} {Aspelmeyer, M.}, \emph {et~al.},\ }\bibfield  {title} {\bibinfo {title} {Probing planck-scale physics with quantum optics},\ }\href {https://www.nature.com/articles/nphys2262} {\bibfield  {journal} {\bibinfo  {journal} {Nature Physics}\ }\textbf {\bibinfo {volume} {8}},\ \bibinfo {pages} {393} (\bibinfo {year} {2012})}\BibitemShut {NoStop}%
\bibitem [{\citenamefont {Schrinski}\ \emph {et~al.}(2023)Schrinski, Bj{\"o}rn and Yang, Yu and Von L{\"u}pke, Uwe and Bild, Marius and Chu, Yiwen and Hornberger, Klaus and Nimmrichter, Stefan and Fadel, Matteo}]{schrinski2023macroscopic}%
  \BibitemOpen
  \bibfield  {author} {\bibinfo {author} {Schrinski, B.}, \bibinfo {author} {Yang, Y.}, \bibinfo {author} {Von~L{\"u}pke, U.}, \emph {et~al.},\ }\bibfield  {title} {\bibinfo {title} {Macroscopic quantum test with bulk acoustic wave resonators},\ }\href {https://journals.aps.org/prl/abstract/10.1103/PhysRevLett.130.133604} {\bibfield  {journal} {\bibinfo  {journal} {Physical review letters}\ }\textbf {\bibinfo {volume} {130}},\ \bibinfo {pages} {133604} (\bibinfo {year} {2023})}\BibitemShut {NoStop}%
\bibitem [{\citenamefont {Hann}\ \emph {et~al.}(2019)Hann, Connor T and Zou, Chang-Ling and Zhang, Yaxing and Chu, Yiwen and Schoelkopf, Robert J and Girvin, Steven M and Jiang, Liang}]{hann2019hardware}%
  \BibitemOpen
  \bibfield  {author} {\bibinfo {author} {Hann, C.~T.}, \bibinfo {author} {Zou, C.-L.}, \bibinfo {author} {Zhang, Y.}, \emph {et~al.},\ }\bibfield  {title} {\bibinfo {title} {Hardware-efficient quantum random access memory with hybrid quantum acoustic systems},\ }\href {https://journals.aps.org/prl/abstract/10.1103/PhysRevLett.123.250501} {\bibfield  {journal} {\bibinfo  {journal} {Physical review letters}\ }\textbf {\bibinfo {volume} {123}},\ \bibinfo {pages} {250501} (\bibinfo {year} {2019})}\BibitemShut {NoStop}%
\bibitem [{\citenamefont {van Thiel}\ \emph {et~al.}(2025)van Thiel, Thierry C and Weaver, MJ and Berto, Federico and Duivestein, Pim and Lemang, Mathilde and Schuurman, Kiki Louise and {\v{Z}}emli{\v{c}}ka, M and Hijazi, Frederick and Bernasconi, Alexandra C and Ferrer, Cristobal and others}]{van2025optical}%
  \BibitemOpen
  \bibfield  {author} {\bibinfo {author} {van Thiel, T.~C.}, \bibinfo {author} {Weaver, M.}, \bibinfo {author} {Berto, F.}, \emph {et~al.},\ }\bibfield  {title} {\bibinfo {title} {Optical readout of a superconducting qubit using a piezo-optomechanical transducer},\ }\href {https://www.nature.com/articles/s41567-024-02742-3} {\bibfield  {journal} {\bibinfo  {journal} {Nature Physics}\ }\textbf {\bibinfo {volume} {21}},\ \bibinfo {pages} {401} (\bibinfo {year} {2025})}\BibitemShut {NoStop}%
\bibitem [{\citenamefont {O{\textquoteright}Connell}\ \emph {et~al.}(2010)O{\textquoteright}Connell, A. D. and Hofheinz, M. and Ansmann, M. and Bialczak, Radoslaw C. and Lenander, M. and Lucero, Erik and Neeley, M. and Sank, D. and Wang, H. and Weides, M. and Wenner, J. and Martinis, John M. and Cleland, A. N.}]{oconnell_quantum_2010}%
  \BibitemOpen
  \bibfield  {author} {\bibinfo {author} {O{\textquoteright}Connell, A.~D.}, \bibinfo {author} {Hofheinz, M.}, \bibinfo {author} {Ansmann, M.}, \emph {et~al.},\ }\bibfield  {title} {\bibinfo {title} {Quantum ground state and single-phonon control of a mechanical resonator},\ }\href {https://doi.org/10.1038/nature08967} {\bibfield  {journal} {\bibinfo  {journal} {Nature}\ }\textbf {\bibinfo {volume} {464}},\ \bibinfo {pages} {697} (\bibinfo {year} {2010})}\BibitemShut {NoStop}%
\bibitem [{\citenamefont {Satzinger}\ \emph {et~al.}(2018)Satzinger, K. J. and Zhong, Y. P. and Chang, H.-S. and Peairs, G. A. and Bienfait, A. and Chou, Ming-Han and Cleland, A. Y. and Conner, C. R. and Dumur, {\'E} and Grebel, J. and Gutierrez, I. and November, B. H. and Povey, R. G. and Whiteley, S. J. and Awschalom, D. D. and Schuster, D. I. and Cleland, A. N.}]{satzinger_quantum_2018}%
  \BibitemOpen
  \bibfield  {author} {\bibinfo {author} {Satzinger, K.~J.}, \bibinfo {author} {Zhong, Y.~P.}, \bibinfo {author} {Chang, H.-S.}, \emph {et~al.},\ }\bibfield  {title} {\bibinfo {title} {Quantum control of surface acoustic-wave phonons},\ }\href {https://doi.org/10.1038/s41586-018-0719-5} {\bibfield  {journal} {\bibinfo  {journal} {Nature}\ }\textbf {\bibinfo {volume} {563}},\ \bibinfo {pages} {661} (\bibinfo {year} {2018})}\BibitemShut {NoStop}%
\bibitem [{\citenamefont {Yang}\ \emph {et~al.}(2024)Yang, Yu and Kladari{\'c}, Igor and Drimmer, Maxwell and von L{\"u}pke, Uwe and Lenterman, Daan and Bus, Joost and Marti, Stefano and Fadel, Matteo and Chu, Yiwen}]{yang_mechanical_2024}%
  \BibitemOpen
  \bibfield  {author} {\bibinfo {author} {Yang, Y.}, \bibinfo {author} {Kladari{\'c}, I.}, \bibinfo {author} {Drimmer, M.}, \emph {et~al.},\ }\bibfield  {title} {\bibinfo {title} {A mechanical qubit},\ }\href {https://doi.org/10.1126/science.adr2464} {\bibfield  {journal} {\bibinfo  {journal} {Science}\ }\textbf {\bibinfo {volume} {386}},\ \bibinfo {pages} {783} (\bibinfo {year} {2024})}\BibitemShut {NoStop}%
\bibitem [{\citenamefont {Chou}\ \emph {et~al.}(2025)Chou, Ming-Han and Qiao, Hong and Yan, Haoxiong and Andersson, Gustav and Conner, Christopher R. and Grebel, Joel and Joshi, Yash J. and Miller, Jacob M. and Povey, Rhys G. and Wu, Xuntao and Cleland, Andrew N.}]{chou_deterministic_2025}%
  \BibitemOpen
  \bibfield  {author} {\bibinfo {author} {Chou, M.-H.}, \bibinfo {author} {Qiao, H.}, \bibinfo {author} {Yan, H.}, \emph {et~al.},\ }\bibfield  {title} {\bibinfo {title} {Deterministic multi-phonon entanglement between two mechanical resonators on separate substrates},\ }\href {https://doi.org/10.1038/s41467-025-56454-0} {\bibfield  {journal} {\bibinfo  {journal} {Nature Communications}\ }\textbf {\bibinfo {volume} {16}},\ \bibinfo {pages} {1450} (\bibinfo {year} {2025})}\BibitemShut {NoStop}%
\bibitem [{\citenamefont {Bozkurt}\ \emph {et~al.}(2025)Bozkurt, Alk{\i}m B. and Golami, Omid and Yu, Yue and Tian, Hao and Mirhosseini, Mohammad}]{bozkurt_mechanical_2025}%
  \BibitemOpen
  \bibfield  {author} {\bibinfo {author} {Bozkurt, A.~B.}, \bibinfo {author} {Golami, O.}, \bibinfo {author} {Yu, Y.}, \emph {et~al.},\ }\bibfield  {title} {\bibinfo {title} {A mechanical quantum memory for microwave photons},\ }\href {https://doi.org/10.1038/s41567-025-02975-w} {\bibfield  {journal} {\bibinfo  {journal} {Nature Physics}\ }\textbf {\bibinfo {volume} {21}},\ \bibinfo {pages} {1469} (\bibinfo {year} {2025})}\BibitemShut {NoStop}%
\bibitem [{\citenamefont {Wollack}\ \emph {et~al.}(2022)Wollack, E. Alex and Cleland, Agnetta Y. and Gruenke, Rachel G. and Wang, Zhaoyou and Arrangoiz-Arriola, Patricio and Safavi-Naeini, Amir H.}]{wollack_quantum_2022}%
  \BibitemOpen
  \bibfield  {author} {\bibinfo {author} {Wollack, E.~A.}, \bibinfo {author} {Cleland, A.~Y.}, \bibinfo {author} {Gruenke, R.~G.}, \emph {et~al.},\ }\bibfield  {title} {\bibinfo {title} {Quantum state preparation and tomography of entangled mechanical resonators},\ }\href {https://doi.org/10.1038/s41586-022-04500-y} {\bibfield  {journal} {\bibinfo  {journal} {Nature}\ }\textbf {\bibinfo {volume} {604}},\ \bibinfo {pages} {463} (\bibinfo {year} {2022})}\BibitemShut {NoStop}%
\bibitem [{\citenamefont {Clerk}\ \emph {et~al.}(2020)Clerk, A. A. and Lehnert, K. W. and Bertet, P. and Petta, J. R. and Nakamura, Y.}]{clerk_hybrid_2020}%
  \BibitemOpen
  \bibfield  {author} {\bibinfo {author} {Clerk, A.~A.}, \bibinfo {author} {Lehnert, K.~W.}, \bibinfo {author} {Bertet, P.}, \emph {et~al.},\ }\bibfield  {title} {\bibinfo {title} {Hybrid quantum systems with circuit quantum electrodynamics},\ }\href {https://doi.org/10.1038/s41567-020-0797-9} {\bibfield  {journal} {\bibinfo  {journal} {Nature Physics}\ }\textbf {\bibinfo {volume} {16}},\ \bibinfo {pages} {257} (\bibinfo {year} {2020})}\BibitemShut {NoStop}%
\bibitem [{\citenamefont {Arrangoiz-Arriola}\ \emph {et~al.}(2019)Arrangoiz-Arriola, Patricio and Wollack, E. Alex and Wang, Zhaoyou and Pechal, Marek and Jiang, Wentao and McKenna, Timothy P. and Witmer, Jeremy D. and Van Laer, Rapha{\"e}l and Safavi-Naeini, Amir H.}]{arrangoiz-arriola_resolving_2019}%
  \BibitemOpen
  \bibfield  {author} {\bibinfo {author} {Arrangoiz-Arriola, P.}, \bibinfo {author} {Wollack, E.~A.}, \bibinfo {author} {Wang, Z.}, \emph {et~al.},\ }\bibfield  {title} {\bibinfo {title} {Resolving the energy levels of a nanomechanical oscillator},\ }\href {https://doi.org/10.1038/s41586-019-1386-x} {\bibfield  {journal} {\bibinfo  {journal} {Nature}\ }\textbf {\bibinfo {volume} {571}},\ \bibinfo {pages} {537} (\bibinfo {year} {2019})}\BibitemShut {NoStop}%
\bibitem [{\citenamefont {Lee}\ \emph {et~al.}(2023{\natexlab{a}})Lee, Nathan R.A. and Guo, Yudan and Cleland, Agnetta Y. and Wollack, E. Alex and Gruenke, Rachel G. and Makihara, Takuma and Wang, Zhaoyou and Rajabzadeh, Taha and Jiang, Wentao and Mayor, Felix M. and Arrangoiz-Arriola, Patricio and Sarabalis, Christopher J. and Safavi-Naeini, Amir H.}]{lee_strong_2023}%
  \BibitemOpen
  \bibfield  {author} {\bibinfo {author} {Lee, N.~R.}, \bibinfo {author} {Guo, Y.}, \bibinfo {author} {Cleland, A.~Y.}, \emph {et~al.},\ }\bibfield  {title} {\bibinfo {title} {Strong dispersive coupling between a mechanical resonator and a fluxonium superconducting qubit},\ }\href {https://doi.org/10.1103/PRXQuantum.4.040342} {\bibfield  {journal} {\bibinfo  {journal} {PRX Quantum}\ }\textbf {\bibinfo {volume} {4}},\ \bibinfo {pages} {040342} (\bibinfo {year} {2023}{\natexlab{a}})}\BibitemShut {NoStop}%
\bibitem [{\citenamefont {Kitzman}\ \emph {et~al.}(2023)Kitzman, J. M. and Lane, J. R. and Undershute, C. and Harrington, P. M. and Beysengulov, N. R. and Mikolas, C. A. and Murch, K. W. and Pollanen, J.}]{kitzman_phononic_2023}%
  \BibitemOpen
  \bibfield  {author} {\bibinfo {author} {Kitzman, J.~M.}, \bibinfo {author} {Lane, J.~R.}, \bibinfo {author} {Undershute, C.}, \emph {et~al.},\ }\bibfield  {title} {\bibinfo {title} {Phononic bath engineering of a superconducting qubit},\ }\href {https://doi.org/10.1038/s41467-023-39682-0} {\bibfield  {journal} {\bibinfo  {journal} {Nature Communications}\ }\textbf {\bibinfo {volume} {14}},\ \bibinfo {pages} {3910} (\bibinfo {year} {2023})}\BibitemShut {NoStop}%
\bibitem [{\citenamefont {Chu}\ \emph {et~al.}(2017)Chu, Yiwen and Kharel, Prashanta and Renninger, William H. and Burkhart, Luke D. and Frunzio, Luigi and Rakich, Peter T. and Schoelkopf, Robert J.}]{chu_quantum_2017}%
  \BibitemOpen
  \bibfield  {author} {\bibinfo {author} {Chu, Y.}, \bibinfo {author} {Kharel, P.}, \bibinfo {author} {Renninger, W.~H.}, \emph {et~al.},\ }\bibfield  {title} {\bibinfo {title} {Quantum acoustics with superconducting qubits},\ }\href {https://doi.org/10.1126/science.aao1511} {\bibfield  {journal} {\bibinfo  {journal} {Science}\ }\textbf {\bibinfo {volume} {358}},\ \bibinfo {pages} {199} (\bibinfo {year} {2017})}\BibitemShut {NoStop}%
\bibitem [{\citenamefont {von L{\"u}pke}\ \emph {et~al.}(2022)von L{\"u}pke, Uwe and Yang, Yu and Bild, Marius and Michaud, Laurent and Fadel, Matteo and Chu, Yiwen}]{von_lupke_parity_2022}%
  \BibitemOpen
  \bibfield  {author} {\bibinfo {author} {von L{\"u}pke, U.}, \bibinfo {author} {Yang, Y.}, \bibinfo {author} {Bild, M.}, \emph {et~al.},\ }\bibfield  {title} {\bibinfo {title} {Parity measurement in the strong dispersive regime of circuit quantum acoustodynamics},\ }\href {https://doi.org/10.1038/s41567-022-01591-2} {\bibfield  {journal} {\bibinfo  {journal} {Nature Physics}\ }\textbf {\bibinfo {volume} {18}},\ \bibinfo {pages} {794} (\bibinfo {year} {2022})}\BibitemShut {NoStop}%
\bibitem [{\citenamefont {Kervinen}\ \emph {et~al.}(2020)Kervinen, Mikael and V{\"a}limaa, Alpo and Ram{\'i}rez-Mu{\~n}oz, Jhon E. and Sillanp{\"a}{\"a}, Mika A.}]{kervinen_sideband_2020}%
  \BibitemOpen
  \bibfield  {author} {\bibinfo {author} {Kervinen, M.}, \bibinfo {author} {V{\"a}limaa, A.}, \bibinfo {author} {Ram{\'i}rez-Mu{\~n}oz, J.~E.},\ and\ \bibinfo {author} {Sillanp{\"a}{\"a}, M.~A.},\ }\bibfield  {title} {\bibinfo {title} {Sideband control of a multimode quantum bulk acoustic system},\ }\href {https://doi.org/10.1103/PhysRevApplied.14.054023} {\bibfield  {journal} {\bibinfo  {journal} {Physical Review Applied}\ }\textbf {\bibinfo {volume} {14}},\ \bibinfo {pages} {054023} (\bibinfo {year} {2020})}\BibitemShut {NoStop}%
\bibitem [{\citenamefont {Gokhale}\ \emph {et~al.}(2020{\natexlab{a}})Gokhale, Vikrant J. and Downey, Brian P. and Katzer, D. Scott and Nepal, Neeraj and Lang, Andrew C. and Stroud, Rhonda M. and Meyer, David J.}]{gokhale_epitaxial_2020}%
  \BibitemOpen
  \bibfield  {author} {\bibinfo {author} {Gokhale, V.~J.}, \bibinfo {author} {Downey, B.~P.}, \bibinfo {author} {Katzer, D.~S.}, \emph {et~al.},\ }\bibfield  {title} {\bibinfo {title} {Epitaxial bulk acoustic wave resonators as highly coherent multi-phonon sources for quantum acoustodynamics},\ }\href {https://doi.org/10.1038/s41467-020-15472-w} {\bibfield  {journal} {\bibinfo  {journal} {Nature Communications}\ }\textbf {\bibinfo {volume} {11}},\ \bibinfo {pages} {2314} (\bibinfo {year} {2020}{\natexlab{a}})}\BibitemShut {NoStop}%
\bibitem [{\citenamefont {MacCabe}\ \emph {et~al.}(2020)MacCabe, Gregory S. and Ren, Hengjiang and Luo, Jie and Cohen, Justin D. and Zhou, Hengyun and Sipahigil, Alp and Mirhosseini, Mohammad and Painter, Oskar}]{maccabe_nano-acoustic_2020}%
  \BibitemOpen
  \bibfield  {author} {\bibinfo {author} {MacCabe, G.~S.}, \bibinfo {author} {Ren, H.}, \bibinfo {author} {Luo, J.}, \emph {et~al.},\ }\bibfield  {title} {\bibinfo {title} {Nano-acoustic resonator with ultralong phonon lifetime},\ }\href {https://doi.org/10.1126/science.abc7312} {\bibfield  {journal} {\bibinfo  {journal} {Science}\ }\textbf {\bibinfo {volume} {370}},\ \bibinfo {pages} {840} (\bibinfo {year} {2020})}\BibitemShut {NoStop}%
\bibitem [{\citenamefont {Vorobiev}\ \emph {et~al.}(2011)Vorobiev, A. and Berge, J. and Gevorgian, S. and L{\"o}ffler, M. and Olsson, E.}]{vorobiev_effect_2011}%
  \BibitemOpen
  \bibfield  {author} {\bibinfo {author} {Vorobiev, A.}, \bibinfo {author} {Berge, J.}, \bibinfo {author} {Gevorgian, S.}, \emph {et~al.},\ }\bibfield  {title} {\bibinfo {title} {Effect of interface roughness on acoustic loss in tunable thin film bulk acoustic wave resonators},\ }\href {https://doi.org/10.1063/1.3610513} {\bibfield  {journal} {\bibinfo  {journal} {Journal of Applied Physics}\ }\textbf {\bibinfo {volume} {110}},\ \bibinfo {pages} {024116} (\bibinfo {year} {2011})}\BibitemShut {NoStop}%
\bibitem [{\citenamefont {Luo}\ \emph {et~al.}(2025)Luo, Yizhi and Diamandi, Hilel Hagai and Li, Hanshi and Bi, Runjiang and Mason, David and Yoon, Taekwan and Guo, Xinghan and Tang, Hanlin and Behunin, Ryan O. and Walker, Frederick J. and Ahn, Charles and Rakich, Peter T.}]{luo_lifetime-limited_2025}%
  \BibitemOpen
  \bibfield  {author} {\bibinfo {author} {Luo, Y.}, \bibinfo {author} {Diamandi, H.~H.}, \bibinfo {author} {Li, H.}, \emph {et~al.},\ }\href {https://doi.org/10.48550/arXiv.2504.07523} {\bibinfo {title} {Lifetime-limited gigahertz-frequency mechanical oscillators with millisecond coherence times}} (\bibinfo {year} {2025}),\ \bibinfo {note} {arXiv:2504.07523 [quant-ph]}\BibitemShut {NoStop}%
\bibitem [{\citenamefont {Galliou}\ \emph {et~al.}(2013)Galliou, Serge and Goryachev, Maxim and Bourquin, Roger and Abb{\'e}, Philippe and Aubry, Jean Pierre and Tobar, Michael E.}]{galliou_extremely_2013}%
  \BibitemOpen
  \bibfield  {author} {\bibinfo {author} {Galliou, S.}, \bibinfo {author} {Goryachev, M.}, \bibinfo {author} {Bourquin, R.}, \emph {et~al.},\ }\bibfield  {title} {\bibinfo {title} {Extremely low loss phonon-trapping cryogenic acoustic cavities for future physical experiments},\ }\href {https://doi.org/10.1038/srep02132} {\bibfield  {journal} {\bibinfo  {journal} {Scientific Reports}\ }\textbf {\bibinfo {volume} {3}},\ \bibinfo {pages} {2132} (\bibinfo {year} {2013})}\BibitemShut {NoStop}%
\bibitem [{\citenamefont {Tsaturyan}\ \emph {et~al.}(2017)Tsaturyan, Y. and Barg, A. and Polzik, E. S. and Schliesser, A.}]{tsaturyan_ultracoherent_2017}%
  \BibitemOpen
  \bibfield  {author} {\bibinfo {author} {Tsaturyan, Y.}, \bibinfo {author} {Barg, A.}, \bibinfo {author} {Polzik, E.~S.},\ and\ \bibinfo {author} {Schliesser, A.},\ }\bibfield  {title} {\bibinfo {title} {Ultracoherent nanomechanical resonators via soft clamping and dissipation dilution},\ }\href {https://doi.org/10.1038/nnano.2017.101} {\bibfield  {journal} {\bibinfo  {journal} {Nature Nanotechnology}\ }\textbf {\bibinfo {volume} {12}},\ \bibinfo {pages} {776} (\bibinfo {year} {2017})}\BibitemShut {NoStop}%
\bibitem [{\citenamefont {Ghadimi}\ \emph {et~al.}(2018)Ghadimi, A. H. and Fedorov, S. A. and Engelsen, N. J. and Bereyhi, M. J. and Schilling, R. and Wilson, D. J. and Kippenberg, T. J.}]{ghadimi_elastic_2018}%
  \BibitemOpen
  \bibfield  {author} {\bibinfo {author} {Ghadimi, A.~H.}, \bibinfo {author} {Fedorov, S.~A.}, \bibinfo {author} {Engelsen, N.~J.}, \emph {et~al.},\ }\bibfield  {title} {\bibinfo {title} {Elastic strain engineering for ultralow mechanical dissipation},\ }\href {https://doi.org/10.1126/science.aar6939} {\bibfield  {journal} {\bibinfo  {journal} {Science}\ }\textbf {\bibinfo {volume} {360}},\ \bibinfo {pages} {764} (\bibinfo {year} {2018})}\BibitemShut {NoStop}%
\bibitem [{\citenamefont {Cleland}\ \emph {et~al.}(2024)Cleland, Agnetta Y. and Wollack, E. Alex and Safavi-Naeini, Amir H.}]{cleland_studying_2024}%
  \BibitemOpen
  \bibfield  {author} {\bibinfo {author} {Cleland, A.~Y.}, \bibinfo {author} {Wollack, E.~A.},\ and\ \bibinfo {author} {Safavi-Naeini, A.~H.},\ }\bibfield  {title} {\bibinfo {title} {Studying phonon coherence with a quantum sensor},\ }\href {https://doi.org/10.1038/s41467-024-48306-0} {\bibfield  {journal} {\bibinfo  {journal} {Nature Communications}\ }\textbf {\bibinfo {volume} {15}},\ \bibinfo {pages} {4979} (\bibinfo {year} {2024})}\BibitemShut {NoStop}%
\bibitem [{\citenamefont {Behunin}\ \emph {et~al.}(2016)Behunin, R. O. and Intravaia, F. and Rakich, P. T.}]{behunin_dimensional_2016}%
  \BibitemOpen
  \bibfield  {author} {\bibinfo {author} {Behunin, R.~O.}, \bibinfo {author} {Intravaia, F.},\ and\ \bibinfo {author} {Rakich, P.~T.},\ }\bibfield  {title} {\bibinfo {title} {Dimensional transformation of defect-induced noise, dissipation, and nonlinearity},\ }\href {https://doi.org/10.1103/PhysRevB.93.224110} {\bibfield  {journal} {\bibinfo  {journal} {Physical Review B}\ }\textbf {\bibinfo {volume} {93}},\ \bibinfo {pages} {224110} (\bibinfo {year} {2016})}\BibitemShut {NoStop}%
\bibitem [{\citenamefont {Maksymowych}\ \emph {et~al.}(2025)Maksymowych, MP and Yuksel, M and Hitchcock, OA and Lee, NR and Mayor, FM and Jiang, W and Roukes, ML and Safavi-Naeini, AH}]{maksymowych_frequency_2025}%
  \BibitemOpen
  \bibfield  {author} {\bibinfo {author} {Maksymowych, M.}, \bibinfo {author} {Yuksel, M.}, \bibinfo {author} {Hitchcock, O.}, \emph {et~al.},\ }\bibfield  {title} {\bibinfo {title} {Spectral diffusion of nanomechanical resonators due to single quantum defects},\ }\href {https://journals.aps.org/prapplied/abstract/10.1103/22rx-v855} {\bibfield  {journal} {\bibinfo  {journal} {Physical Review Applied}\ }\textbf {\bibinfo {volume} {24}},\ \bibinfo {pages} {044066} (\bibinfo {year} {2025})}\BibitemShut {NoStop}%
\bibitem [{\citenamefont {Gruenke-Freudenstein}\ \emph {et~al.}(2025)Gruenke-Freudenstein, Rachel G. and Szakiel, Erik and Multani, Gitanjali P. and Makihara, Takuma and Hayden, Akasha G. and Khalatpour, Ali and Wollack, E. Alex and Akoto-Yeboah, Antonia and Salmani-Rezaie, Salva and Safavi-Naeini, Amir H.}]{gruenke-freudenstein_surface_2025}%
  \BibitemOpen
  \bibfield  {author} {\bibinfo {author} {Gruenke-Freudenstein, R.~G.}, \bibinfo {author} {Szakiel, E.}, \bibinfo {author} {Multani, G.~P.}, \emph {et~al.},\ }\bibfield  {title} {\bibinfo {title} {Surface and bulk two-level-system losses in lithium niobate acoustic resonators},\ }\href {https://doi.org/10.1103/tgs3-kkw8} {\bibfield  {journal} {\bibinfo  {journal} {Physical Review Applied}\ }\textbf {\bibinfo {volume} {23}},\ \bibinfo {pages} {064055} (\bibinfo {year} {2025})}\BibitemShut {NoStop}%
\bibitem [{\citenamefont {Chu}\ \emph {et~al.}(2018)Chu, Yiwen and Kharel, Prashanta and Yoon, Taekwan and Frunzio, Luigi and Rakich, Peter T. and Schoelkopf, Robert J.}]{chu_creation_2018}%
  \BibitemOpen
  \bibfield  {author} {\bibinfo {author} {Chu, Y.}, \bibinfo {author} {Kharel, P.}, \bibinfo {author} {Yoon, T.}, \emph {et~al.},\ }\bibfield  {title} {\bibinfo {title} {Creation and control of multi-phonon {Fock} states in a bulk acoustic-wave resonator},\ }\href {https://doi.org/10.1038/s41586-018-0717-7} {\bibfield  {journal} {\bibinfo  {journal} {Nature}\ }\textbf {\bibinfo {volume} {563}},\ \bibinfo {pages} {666} (\bibinfo {year} {2018})}\BibitemShut {NoStop}%
\bibitem [{\citenamefont {Kharel}\ \emph {et~al.}(2018)Kharel, Prashanta and Chu, Yiwen and Power, Michael and Renninger, William H. and Schoelkopf, Robert J. and Rakich, Peter T.}]{kharel_ultra-high-_2018}%
  \BibitemOpen
  \bibfield  {author} {\bibinfo {author} {Kharel, P.}, \bibinfo {author} {Chu, Y.}, \bibinfo {author} {Power, M.}, \emph {et~al.},\ }\bibfield  {title} {\bibinfo {title} {Ultra-high- \textit{{Q}} phononic resonators on-chip at cryogenic temperatures},\ }\href {https://doi.org/10.1063/1.5026798} {\bibfield  {journal} {\bibinfo  {journal} {APL Photonics}\ }\textbf {\bibinfo {volume} {3}},\ \bibinfo {pages} {066101} (\bibinfo {year} {2018})}\BibitemShut {NoStop}%
\bibitem [{\citenamefont {Wollack}\ \emph {et~al.}(2021)Wollack, E. Alex and Cleland, Agnetta Y. and Arrangoiz-Arriola, Patricio and McKenna, Timothy P. and Gruenke, Rachel G. and Patel, Rishi N. and Jiang, Wentao and Sarabalis, Christopher J. and Safavi-Naeini, Amir H.}]{wollack_loss_2021}%
  \BibitemOpen
  \bibfield  {author} {\bibinfo {author} {Wollack, E.~A.}, \bibinfo {author} {Cleland, A.~Y.}, \bibinfo {author} {Arrangoiz-Arriola, P.}, \emph {et~al.},\ }\bibfield  {title} {\bibinfo {title} {Loss channels affecting lithium niobate phononic crystal resonators at cryogenic temperature},\ }\href {https://doi.org/10.1063/5.0034909} {\bibfield  {journal} {\bibinfo  {journal} {Applied Physics Letters}\ }\textbf {\bibinfo {volume} {118}},\ \bibinfo {pages} {123501} (\bibinfo {year} {2021})}\BibitemShut {NoStop}%
\bibitem [{\citenamefont {V{\"a}limaa}\ \emph {et~al.}(2018)V{\"a}limaa, A. J. and Santos, J. T. and Ockeloen-Korppi, C. F. and Sillanp{\"a}{\"a}, M. A.}]{valimaa_electrode_2018}%
  \BibitemOpen
  \bibfield  {author} {\bibinfo {author} {V{\"a}limaa, A.~J.}, \bibinfo {author} {Santos, J.~T.}, \bibinfo {author} {Ockeloen-Korppi, C.~F.},\ and\ \bibinfo {author} {Sillanp{\"a}{\"a}, M.~A.},\ }\bibfield  {title} {\bibinfo {title} {Electrode configuration and electrical dissipation of mechanical energy in quartz crystal resonators},\ }\href {https://doi.org/10.1088/1361-6439/aac781} {\bibfield  {journal} {\bibinfo  {journal} {Journal of Micromechanics and Microengineering}\ }\textbf {\bibinfo {volume} {28}},\ \bibinfo {pages} {095014} (\bibinfo {year} {2018})}\BibitemShut {NoStop}%
\bibitem [{\citenamefont {Dubois}\ and\ \citenamefont {Muralt}(2001)Dubois, Marc-Alexandre and Muralt, Paul}]{dubois_stress_2001}%
  \BibitemOpen
  \bibfield  {author} {\bibinfo {author} {Dubois, M.-A.}\ and\ \bibinfo {author} {Muralt, P.},\ }\bibfield  {title} {\bibinfo {title} {Stress and piezoelectric properties of aluminum nitride thin films deposited onto metal electrodes by pulsed direct current reactive sputtering},\ }\href {https://doi.org/10.1063/1.1359162} {\bibfield  {journal} {\bibinfo  {journal} {Journal of Applied Physics}\ }\textbf {\bibinfo {volume} {89}},\ \bibinfo {pages} {6389} (\bibinfo {year} {2001})}\BibitemShut {NoStop}%
\bibitem [{\citenamefont {Lee}\ \emph {et~al.}(2020)Lee, Moonsang and Son, Hyungbin and Lee, Hae-Yong and Moon, Joonhee and Kim, Heejin and Park, Ji-In and Liu, Zheng and Hahm, Myung Gwan and Yang, Mino and Kim, Un Jeong}]{lee_nanovoid-driven_2020}%
  \BibitemOpen
  \bibfield  {author} {\bibinfo {author} {Lee, M.}, \bibinfo {author} {Son, H.}, \bibinfo {author} {Lee, H.-Y.}, \emph {et~al.},\ }\bibfield  {title} {\bibinfo {title} {Nanovoid-driven highly crystalline aluminum nitride and its application in solar-blind {UV} photodetectors},\ }\href {https://doi.org/10.1039/D0TC03208E} {\bibfield  {journal} {\bibinfo  {journal} {Journal of Materials Chemistry C}\ }\textbf {\bibinfo {volume} {8}},\ \bibinfo {pages} {14431} (\bibinfo {year} {2020})}\BibitemShut {NoStop}%
\bibitem [{\citenamefont {Patidar}\ \emph {et~al.}(2025)Patidar, Jyotish and Pshyk, Oleksandr and Thorwarth, Kerstin and Sommerh{\"a}user, Lars and Siol, Sebastian}]{patidar_low_2025}%
  \BibitemOpen
  \bibfield  {author} {\bibinfo {author} {Patidar, J.}, \bibinfo {author} {Pshyk, O.}, \bibinfo {author} {Thorwarth, K.}, \emph {et~al.},\ }\bibfield  {title} {\bibinfo {title} {Low temperature deposition of functional thin films on insulating substrates enabled by selective ion acceleration using synchronized floating potential {HiPIMS}},\ }\href {https://doi.org/10.1038/s41467-025-59911-y} {\bibfield  {journal} {\bibinfo  {journal} {Nature Communications}\ }\textbf {\bibinfo {volume} {16}},\ \bibinfo {pages} {4719} (\bibinfo {year} {2025})}\BibitemShut {NoStop}%
\bibitem [{\citenamefont {Zhang}\ \emph {et~al.}(2003)Zhang, Yuxing and Wang, Zuoqing and Cheeke, J.D.N.}]{zhang_resonant_2003}%
  \BibitemOpen
  \bibfield  {author} {\bibinfo {author} {Zhang, Y.}, \bibinfo {author} {Wang, Z.},\ and\ \bibinfo {author} {Cheeke, J.},\ }\bibfield  {title} {\bibinfo {title} {Resonant spectrum method to characterize piezoelectric films in composite resonators},\ }\href {https://doi.org/10.1109/TUFFC.2003.1193626} {\bibfield  {journal} {\bibinfo  {journal} {IEEE Transactions on Ultrasonics, Ferroelectrics, and Frequency Control}\ }\textbf {\bibinfo {volume} {50}},\ \bibinfo {pages} {321} (\bibinfo {year} {2003})}\BibitemShut {NoStop}%
\bibitem [{\citenamefont {Mason}(1948)Mason, Warren Perry}]{mason_electromechanical_1948}%
  \BibitemOpen
  \bibfield  {author} {\bibinfo {author} {Mason, W.~P.},\ }\href@noop {} {\emph {\bibinfo {title} {Electromechanical transducers and wave filters}}},\ \bibinfo {edition} {2nd}\ ed.\ (\bibinfo  {publisher} {D. Van Nostrand Company},\ \bibinfo {address} {New York, NY, USA},\ \bibinfo {year} {1948})\BibitemShut {NoStop}%
\bibitem [{\citenamefont {Lee}\ \emph {et~al.}(2023{\natexlab{b}})Lee, Seung-Jae and Jeon, Seong Ran and Jung, Sung Hoon and Choi, Young-Jun and Oh, Hae-Gon and Lee, Hae-Yong and Kwon, Min-Ki and Hong, Soon-Ku}]{lee_realization_2023}%
  \BibitemOpen
  \bibfield  {author} {\bibinfo {author} {Lee, S.-J.}, \bibinfo {author} {Jeon, S.~R.}, \bibinfo {author} {Jung, S.~H.}, \emph {et~al.},\ }\bibfield  {title} {\bibinfo {title} {Realization of low dislocation density aln on patterned sapphire substrate by hydride vapor-phase epitaxy for deep ultraviolet light-emitting diodes},\ }\href {https://doi.org/10.1002/pssa.202200835} {\bibfield  {journal} {\bibinfo  {journal} {physica status solidi (a)}\ }\textbf {\bibinfo {volume} {220}},\ \bibinfo {pages} {2200835} (\bibinfo {year} {2023}{\natexlab{b}})}\BibitemShut {NoStop}%
\bibitem [{\citenamefont {Wang}\ \emph {et~al.}(2015)Wang, C. and Axline, C. and Gao, Y. Y. and Brecht, T. and Chu, Y. and Frunzio, L. and Devoret, M. H. and Schoelkopf, R. J.}]{wang_surface_2015}%
  \BibitemOpen
  \bibfield  {author} {\bibinfo {author} {Wang, C.}, \bibinfo {author} {Axline, C.}, \bibinfo {author} {Gao, Y.~Y.}, \emph {et~al.},\ }\bibfield  {title} {\bibinfo {title} {Surface participation and dielectric loss in superconducting qubits},\ }\href {https://doi.org/10.1063/1.4934486} {\bibfield  {journal} {\bibinfo  {journal} {Applied Physics Letters}\ }\textbf {\bibinfo {volume} {107}},\ \bibinfo {pages} {162601} (\bibinfo {year} {2015})}\BibitemShut {NoStop}%
\bibitem [{\citenamefont {Minev}\ \emph {et~al.}(2021)Minev, Zlatko K. and Leghtas, Zaki and Mundhada, Shantanu O. and Christakis, Lysander and Pop, Ioan M. and Devoret, Michel H.}]{minev_energy-participation_2021}%
  \BibitemOpen
  \bibfield  {author} {\bibinfo {author} {Minev, Z.~K.}, \bibinfo {author} {Leghtas, Z.}, \bibinfo {author} {Mundhada, S.~O.}, \emph {et~al.},\ }\bibfield  {title} {\bibinfo {title} {Energy-participation quantization of {Josephson} circuits},\ }\href {https://doi.org/10.1038/s41534-021-00461-8} {\bibfield  {journal} {\bibinfo  {journal} {npj Quantum Information}\ }\textbf {\bibinfo {volume} {7}},\ \bibinfo {pages} {131} (\bibinfo {year} {2021})}\BibitemShut {NoStop}%
\bibitem [{\citenamefont {Braginsky}\ \emph {et~al.}(1986)Braginsky, V. B. and Mitrofanov, V. P. and Panov, V. I.}]{braginsky_systems_1986}%
  \BibitemOpen
  \bibfield  {author} {\bibinfo {author} {Braginsky, V.~B.}, \bibinfo {author} {Mitrofanov, V.~P.},\ and\ \bibinfo {author} {Panov, V.~I.},\ }\href {https://press.uchicago.edu/ucp/books/book/chicago/S/bo5973099.html} {\emph {\bibinfo {title} {Systems with {Small} {Dissipation}}}}\ (\bibinfo  {publisher} {University of Chicago Press},\ \bibinfo {address} {Chicago, IL},\ \bibinfo {year} {1986})\BibitemShut {NoStop}%
\bibitem [{\citenamefont {Hutchison}(1960)Hutchison, Thomas S.}]{hutchison_ultrasonic_1960}%
  \BibitemOpen
  \bibfield  {author} {\bibinfo {author} {Hutchison, T.~S.},\ }\bibfield  {title} {\bibinfo {title} {Ultrasonic absorption in solids},\ }\href {https://doi.org/10.1126/science.132.3428.643} {\bibfield  {journal} {\bibinfo  {journal} {Science}\ }\textbf {\bibinfo {volume} {132}},\ \bibinfo {pages} {643} (\bibinfo {year} {1960})}\BibitemShut {NoStop}%
\bibitem [{\citenamefont {Nowick}\ and\ \citenamefont {Berry}(1972)Nowick, A. S. and Berry, B. S.}]{nowick_chapter_1972-3}%
  \BibitemOpen
  \bibfield  {author} {\bibinfo {author} {Nowick, A.~S.}\ and\ \bibinfo {author} {Berry, B.~S.},\ }\bibfield  {title} {\bibinfo {title} {Boundary relaxation processes and internal friction at high temperatures},\ }in\ \href {https://doi.org/10.1016/B978-0-12-522650-9.50021-5} {\emph {\bibinfo {booktitle} {Anelastic relaxation in crystalline solids}}}\ (\bibinfo  {publisher} {Academic Press},\ \bibinfo {year} {1972})\ pp.\ \bibinfo {pages} {435--462}\BibitemShut {NoStop}%
\bibitem [{\citenamefont {Akhiezer}(1939)Akhiezer, A.}]{akhiezer_absorption_1939}%
  \BibitemOpen
  \bibfield  {author} {\bibinfo {author} {Akhiezer, A.},\ }\bibfield  {title} {\bibinfo {title} {On the absorption of sound in solids},\ }\href@noop {} {\bibfield  {journal} {\bibinfo  {journal} {Journal of Physics (USSR)}\ }\textbf {\bibinfo {volume} {1}},\ \bibinfo {pages} {277} (\bibinfo {year} {1939})}\BibitemShut {NoStop}%
\bibitem [{\citenamefont {Ghaffari}\ \emph {et~al.}(2013)Ghaffari, Shirin and Chandorkar, Saurabh A. and Wang, Shasha and Ng, Eldwin J. and Ahn, Chae H. and Hong, Vu and Yang, Yushi and Kenny, Thomas W.}]{ghaffari_quantum_2013}%
  \BibitemOpen
  \bibfield  {author} {\bibinfo {author} {Ghaffari, S.}, \bibinfo {author} {Chandorkar, S.~A.}, \bibinfo {author} {Wang, S.}, \emph {et~al.},\ }\bibfield  {title} {\bibinfo {title} {Quantum limit of quality factor in silicon micro and nano mechanical resonators},\ }\href {https://doi.org/10.1038/srep03244} {\bibfield  {journal} {\bibinfo  {journal} {Scientific Reports}\ }\textbf {\bibinfo {volume} {3}},\ \bibinfo {pages} {3244} (\bibinfo {year} {2013})}\BibitemShut {NoStop}%
\bibitem [{\citenamefont {Phillips}(1987)Phillips, W. A.}]{phillips_two-level_1987}%
  \BibitemOpen
  \bibfield  {author} {\bibinfo {author} {Phillips, W.~A.},\ }\bibfield  {title} {\bibinfo {title} {Two-level states in glasses},\ }\href {https://doi.org/10.1088/0034-4885/50/12/003} {\bibfield  {journal} {\bibinfo  {journal} {Reports on Progress in Physics}\ }\textbf {\bibinfo {volume} {50}},\ \bibinfo {pages} {1657} (\bibinfo {year} {1987})}\BibitemShut {NoStop}%
\bibitem [{\citenamefont {Hunklinger}\ and\ \citenamefont {Arnold}(1976)Hunklinger, S. and Arnold, W.}]{hunklinger_3_1976}%
  \BibitemOpen
  \bibfield  {author} {\bibinfo {author} {Hunklinger, S.}\ and\ \bibinfo {author} {Arnold, W.},\ }\bibfield  {title} {\bibinfo {title} {Ultrasonic properties of glasses at low temperatures},\ }in\ \href {https://doi.org/10.1016/B978-0-12-477912-9.50008-4} {\emph {\bibinfo {booktitle} {Physical {Acoustics}}}},\ Vol.~\bibinfo {volume} {12}\ (\bibinfo  {publisher} {Academic Press},\ \bibinfo {year} {1976})\ pp.\ \bibinfo {pages} {155--215}\BibitemShut {NoStop}%
\bibitem [{\citenamefont {Emser}\ \emph {et~al.}(2024)Emser, A. L. and Metzger, C. and Rose, B. C. and Lehnert, K. W.}]{emser_thin-film_2024}%
  \BibitemOpen
  \bibfield  {author} {\bibinfo {author} {Emser, A.~L.}, \bibinfo {author} {Metzger, C.}, \bibinfo {author} {Rose, B.~C.},\ and\ \bibinfo {author} {Lehnert, K.~W.},\ }\bibfield  {title} {\bibinfo {title} {Thin-film quartz for high-coherence piezoelectric phononic crystal resonators},\ }\href {https://doi.org/10.1103/PhysRevApplied.22.064032} {\bibfield  {journal} {\bibinfo  {journal} {Physical Review Applied}\ }\textbf {\bibinfo {volume} {22}},\ \bibinfo {pages} {064032} (\bibinfo {year} {2024})}\BibitemShut {NoStop}%
\bibitem [{\citenamefont {Undershute}\ \emph {et~al.}(2025)Undershute, Camryn and Kitzman, Joseph M. and Mikolas, Camille A. and Pollanen, Johannes}]{undershute_decoherence_2025}%
  \BibitemOpen
  \bibfield  {author} {\bibinfo {author} {Undershute, C.}, \bibinfo {author} {Kitzman, J.~M.}, \bibinfo {author} {Mikolas, C.~A.},\ and\ \bibinfo {author} {Pollanen, J.},\ }\bibfield  {title} {\bibinfo {title} {Decoherence of surface phonons in a quantum acoustic system},\ }\href {https://doi.org/10.1103/PhysRevA.111.012615} {\bibfield  {journal} {\bibinfo  {journal} {Physical Review A}\ }\textbf {\bibinfo {volume} {111}},\ \bibinfo {pages} {012615} (\bibinfo {year} {2025})}\BibitemShut {NoStop}%
\bibitem [{\citenamefont {Bolgar}\ \emph {et~al.}(2018)Bolgar, Aleksey N. and Zotova, Julia I. and Kirichenko, Daniil D. and Besedin, Ilia S. and Semenov, Aleksander V. and Shaikhaidarov, Rais S. and Astafiev, Oleg V.}]{bolgar_quantum_2018}%
  \BibitemOpen
  \bibfield  {author} {\bibinfo {author} {Bolgar, A.~N.}, \bibinfo {author} {Zotova, J.~I.}, \bibinfo {author} {Kirichenko, D.~D.}, \emph {et~al.},\ }\bibfield  {title} {\bibinfo {title} {Quantum regime of a two-dimensional phonon cavity},\ }\href {https://doi.org/10.1103/PhysRevLett.120.223603} {\bibfield  {journal} {\bibinfo  {journal} {Physical Review Letters}\ }\textbf {\bibinfo {volume} {120}},\ \bibinfo {pages} {223603} (\bibinfo {year} {2018})}\BibitemShut {NoStop}%
\bibitem [{\citenamefont {Moores}\ \emph {et~al.}(2018)Moores, Bradley A. and Sletten, Lucas R. and Viennot, Jeremie J. and Lehnert, K. W.}]{moores_cavity_2018}%
  \BibitemOpen
  \bibfield  {author} {\bibinfo {author} {Moores, B.~A.}, \bibinfo {author} {Sletten, L.~R.}, \bibinfo {author} {Viennot, J.~J.},\ and\ \bibinfo {author} {Lehnert, K.~W.},\ }\bibfield  {title} {\bibinfo {title} {Cavity quantum acoustic device in the multimode strong coupling regime},\ }\href {https://doi.org/10.1103/PhysRevLett.120.227701} {\bibfield  {journal} {\bibinfo  {journal} {Physical Review Letters}\ }\textbf {\bibinfo {volume} {120}},\ \bibinfo {pages} {227701} (\bibinfo {year} {2018})}\BibitemShut {NoStop}%
\bibitem [{\citenamefont {Bienfait}\ \emph {et~al.}(2019)Bienfait, A. and Satzinger, K. J. and Zhong, Y. P. and Chang, H.-S. and Chou, M.-H. and Conner, C. R. and Dumur, {\'E}. and Grebel, J. and Peairs, G. A. and Povey, R. G. and Cleland, A. N.}]{bienfait_phonon-mediated_2019}%
  \BibitemOpen
  \bibfield  {author} {\bibinfo {author} {Bienfait, A.}, \bibinfo {author} {Satzinger, K.~J.}, \bibinfo {author} {Zhong, Y.~P.}, \emph {et~al.},\ }\bibfield  {title} {\bibinfo {title} {Phonon-mediated quantum state transfer and remote qubit entanglement},\ }\href {https://doi.org/10.1126/science.aaw8415} {\bibfield  {journal} {\bibinfo  {journal} {Science}\ }\textbf {\bibinfo {volume} {364}},\ \bibinfo {pages} {368} (\bibinfo {year} {2019})}\BibitemShut {NoStop}%
\bibitem [{\citenamefont {Crump}\ \emph {et~al.}(2023)Crump, Wayne and V{\"a}limaa, Alpo and Sillanp{\"a}{\"a}, Mika A.}]{crump_coupling_2023}%
  \BibitemOpen
  \bibfield  {author} {\bibinfo {author} {Crump, W.}, \bibinfo {author} {V{\"a}limaa, A.},\ and\ \bibinfo {author} {Sillanp{\"a}{\"a}, M.~A.},\ }\bibfield  {title} {\bibinfo {title} {Coupling high-overtone bulk acoustic wave resonators via superconducting qubits},\ }\href {https://doi.org/10.1063/5.0166924} {\bibfield  {journal} {\bibinfo  {journal} {Applied Physics Letters}\ }\textbf {\bibinfo {volume} {123}},\ \bibinfo {pages} {134004} (\bibinfo {year} {2023})}\BibitemShut {NoStop}%
\bibitem [{\citenamefont {Galliou}\ \emph {et~al.}(2016)Galliou, Serge and Del{\'e}glise, Samuel and Goryachev, Maxim and Neuhaus, Leonhard and Cagnoli, Gianpietro and Zerkani, Salim and Dolique, Vincent and Bon, J{\'e}r{\'e}my and Vacheret, Xavier and Abb{\'e}, Philippe and Pinard, Laurent and Michel, Christophe and Karassouloff, Thibaut and Briant, Tristan and Cohadon, Pierre-Fran{\c c}ois and Heidmann, Antoine and Tobar, Michael E. and Bourquin, Roger}]{galliou_new_2016}%
  \BibitemOpen
  \bibfield  {author} {\bibinfo {author} {Galliou, S.}, \bibinfo {author} {Del{\'e}glise, S.}, \bibinfo {author} {Goryachev, M.}, \emph {et~al.},\ }\bibfield  {title} {\bibinfo {title} {A new method of probing mechanical losses of coatings at cryogenic temperatures},\ }\href {https://doi.org/10.1063/1.4972106} {\bibfield  {journal} {\bibinfo  {journal} {Review of Scientific Instruments}\ }\textbf {\bibinfo {volume} {87}},\ \bibinfo {pages} {123906} (\bibinfo {year} {2016})}\BibitemShut {NoStop}%
\bibitem [{\citenamefont {Zhang}\ \emph {et~al.}(2006)Zhang, Hao and Pang, Wei and Yu, Hongyu and Kim, Eun Sok}]{zhang_high-tone_2006}%
  \BibitemOpen
  \bibfield  {author} {\bibinfo {author} {Zhang, H.}, \bibinfo {author} {Pang, W.}, \bibinfo {author} {Yu, H.},\ and\ \bibinfo {author} {Kim, E.~S.},\ }\bibfield  {title} {\bibinfo {title} {High-tone bulk acoustic resonators on sapphire, crystal quartz, fused silica, and silicon substrates},\ }\href {https://doi.org/10.1063/1.2209029} {\bibfield  {journal} {\bibinfo  {journal} {Journal of Applied Physics}\ }\textbf {\bibinfo {volume} {99}},\ \bibinfo {pages} {124911} (\bibinfo {year} {2006})}\BibitemShut {NoStop}%
\bibitem [{\citenamefont {Tian}\ \emph {et~al.}(2020)Tian, Hao and Liu, Junqiu and Dong, Bin and Skehan, J. Connor and Zervas, Michael and Kippenberg, Tobias J. and Bhave, Sunil A.}]{tian_hybrid_2020}%
  \BibitemOpen
  \bibfield  {author} {\bibinfo {author} {Tian, H.}, \bibinfo {author} {Liu, J.}, \bibinfo {author} {Dong, B.}, \emph {et~al.},\ }\bibfield  {title} {\bibinfo {title} {Hybrid integrated photonics using bulk acoustic resonators},\ }\href {https://doi.org/10.1038/s41467-020-16812-6} {\bibfield  {journal} {\bibinfo  {journal} {Nature Communications}\ }\textbf {\bibinfo {volume} {11}},\ \bibinfo {pages} {3073} (\bibinfo {year} {2020})}\BibitemShut {NoStop}%
\bibitem [{\citenamefont {von L{\"u}pke}(2023)von L{\"u}pke, Uwe}]{von_lupke_quantum_2023}%
  \BibitemOpen
  \bibfield  {author} {\bibinfo {author} {von L{\"u}pke, U.},\ }\emph {\bibinfo {title} {Quantum control of a multimode bulk acoustic wave resonator}},\ \href {https://doi.org/10.3929/ethz-b-000656573} {\bibinfo {type} {Doctoral {Thesis}}},\ \bibinfo  {school} {ETH Zurich} (\bibinfo {year} {2023})\BibitemShut {NoStop}%
\bibitem [{\citenamefont {Gargiulo}\ \emph {et~al.}(2021)Gargiulo, O. and Oleschko, S. and Prat-Camps, J. and Zanner, M. and Kirchmair, G.}]{gargiulo_fast_2021}%
  \BibitemOpen
  \bibfield  {author} {\bibinfo {author} {Gargiulo, O.}, \bibinfo {author} {Oleschko, S.}, \bibinfo {author} {Prat-Camps, J.}, \emph {et~al.},\ }\bibfield  {title} {\bibinfo {title} {Fast flux control of {3D} transmon qubits using a magnetic hose},\ }\href {https://doi.org/10.1063/5.0032615} {\bibfield  {journal} {\bibinfo  {journal} {Applied Physics Letters}\ }\textbf {\bibinfo {volume} {118}},\ \bibinfo {pages} {012601} (\bibinfo {year} {2021})}\BibitemShut {NoStop}%
\bibitem [{\citenamefont {Frattini}\ \emph {et~al.}(2018)Frattini, NE and Sivak, VV and Lingenfelter, A and Shankar, S and Devoret, MH}]{frattini2018optimizing}%
  \BibitemOpen
  \bibfield  {author} {\bibinfo {author} {Frattini, N.}, \bibinfo {author} {Sivak, V.}, \bibinfo {author} {Lingenfelter, A.}, \emph {et~al.},\ }\bibfield  {title} {\bibinfo {title} {Optimizing the nonlinearity and dissipation of a snail parametric amplifier for dynamic range},\ }\href {https://journals.aps.org/prapplied/abstract/10.1103/PhysRevApplied.10.054020} {\bibfield  {journal} {\bibinfo  {journal} {Physical Review Applied}\ }\textbf {\bibinfo {volume} {10}},\ \bibinfo {pages} {054020} (\bibinfo {year} {2018})}\BibitemShut {NoStop}%
\bibitem [{\citenamefont {Rieger}\ \emph {et~al.}(2023)Rieger, D. and G{\"u}nzler, S. and Spiecker, M. and Nambisan, A. and Wernsdorfer, W. and Pop, I.M.}]{rieger_fano_2023}%
  \BibitemOpen
  \bibfield  {author} {\bibinfo {author} {Rieger, D.}, \bibinfo {author} {G{\"u}nzler, S.}, \bibinfo {author} {Spiecker, M.}, \emph {et~al.},\ }\bibfield  {title} {\bibinfo {title} {Fano interference in microwave resonator measurements},\ }\href {https://doi.org/10.1103/PhysRevApplied.20.014059} {\bibfield  {journal} {\bibinfo  {journal} {Physical Review Applied}\ }\textbf {\bibinfo {volume} {20}},\ \bibinfo {pages} {014059} (\bibinfo {year} {2023})}\BibitemShut {NoStop}%
\bibitem [{\citenamefont {Probst}\ \emph {et~al.}(2015)Probst, S. and Song, F. B. and Bushev, P. A. and Ustinov, A. V. and Weides, M.}]{probst_efficient_2015}%
  \BibitemOpen
  \bibfield  {author} {\bibinfo {author} {Probst, S.}, \bibinfo {author} {Song, F.~B.}, \bibinfo {author} {Bushev, P.~A.}, \emph {et~al.},\ }\bibfield  {title} {\bibinfo {title} {Efficient and robust analysis of complex scattering data under noise in microwave resonators},\ }\href {https://doi.org/10.1063/1.4907935} {\bibfield  {journal} {\bibinfo  {journal} {Review of Scientific Instruments}\ }\textbf {\bibinfo {volume} {86}},\ \bibinfo {pages} {024706} (\bibinfo {year} {2015})}\BibitemShut {NoStop}%
\bibitem [{noa()}]{noauthor_steelelab-delftstlab_2025}%
  \BibitemOpen
  \href {https://github.com/steelelab-delft/stlab} {\bibinfo {title} {steelelab-delft/stlab}}\BibitemShut {NoStop}%
\bibitem [{\citenamefont {Pozar}(2021)Pozar, David M}]{pozar2021microwave}%
  \BibitemOpen
  \bibfield  {author} {\bibinfo {author} {Pozar, D.~M.},\ }\href@noop {} {\emph {\bibinfo {title} {Microwave engineering: theory and techniques}}}\ (\bibinfo  {publisher} {John Wiley \& Sons, Ltd},\ \bibinfo {year} {2021})\BibitemShut {NoStop}%
\bibitem [{\citenamefont {Saleh}\ and\ \citenamefont {Teich}(1991)Saleh, Bahaa E. A. and Teich, Melvin Carl}]{saleh_fundamentals_1991}%
  \BibitemOpen
  \bibfield  {author} {\bibinfo {author} {Saleh, B. E.~A.}\ and\ \bibinfo {author} {Teich, M.~C.},\ }\href@noop {} {\emph {\bibinfo {title} {Fundamentals of photonics}}}\ (\bibinfo  {publisher} {John Wiley \& Sons, Ltd},\ \bibinfo {year} {1991})\BibitemShut {NoStop}%
\bibitem [{\citenamefont {Goryachev}\ \emph {et~al.}(2013)Goryachev, Maxim and Creedon, Daniel L. and Galliou, Serge and Tobar, Michael E.}]{goryachev_observation_2013}%
  \BibitemOpen
  \bibfield  {author} {\bibinfo {author} {Goryachev, M.}, \bibinfo {author} {Creedon, D.~L.}, \bibinfo {author} {Galliou, S.},\ and\ \bibinfo {author} {Tobar, M.~E.},\ }\bibfield  {title} {\bibinfo {title} {Observation of {Rayleigh} phonon scattering through excitation of extremely high overtones in low-loss cryogenic acoustic cavities for hybrid quantum systems},\ }\href {https://doi.org/10.1103/PhysRevLett.111.085502} {\bibfield  {journal} {\bibinfo  {journal} {Physical Review Letters}\ }\textbf {\bibinfo {volume} {111}},\ \bibinfo {pages} {085502} (\bibinfo {year} {2013})}\BibitemShut {NoStop}%
\bibitem [{\citenamefont {Cleland}(2013)Cleland, Andrew N}]{cleland2013foundations}%
  \BibitemOpen
  \bibfield  {author} {\bibinfo {author} {Cleland, A.~N.},\ }\href {https://link.springer.com/book/10.1007/978-3-662-05287-7} {\emph {\bibinfo {title} {Foundations of nanomechanics: from solid-state theory to device applications}}}\ (\bibinfo  {publisher} {Springer Science \& Business Media},\ \bibinfo {year} {2013})\BibitemShut {NoStop}%
\bibitem [{\citenamefont {Ziman}(2001)Ziman, J. M.}]{ziman_electrons_2001}%
  \BibitemOpen
  \bibfield  {author} {\bibinfo {author} {Ziman, J.~M.},\ }\href@noop {} {\emph {\bibinfo {title} {Electrons and phonons: The theory of transport phenomena in solids}}},\ Oxford classic texts in the physical sciences\ (\bibinfo  {publisher} {Oxford University Press},\ \bibinfo {address} {Oxford},\ \bibinfo {year} {2001})\BibitemShut {NoStop}%
\bibitem [{\citenamefont {Maznev}(2015)Maznev, A. A.}]{maznev_boundary_2015}%
  \BibitemOpen
  \bibfield  {author} {\bibinfo {author} {Maznev, A.~A.},\ }\bibfield  {title} {\bibinfo {title} {Boundary scattering of phonons: {Specularity} of a randomly rough surface in the small-perturbation limit},\ }\href {https://doi.org/10.1103/PhysRevB.91.134306} {\bibfield  {journal} {\bibinfo  {journal} {Physical Review B}\ }\textbf {\bibinfo {volume} {91}},\ \bibinfo {pages} {134306} (\bibinfo {year} {2015})}\BibitemShut {NoStop}%
\bibitem [{\citenamefont {Stevens}\ and\ \citenamefont {Tiersten}(1986)Stevens, D. S. and Tiersten, H. F.}]{stevens_analysis_1986}%
  \BibitemOpen
  \bibfield  {author} {\bibinfo {author} {Stevens, D.~S.}\ and\ \bibinfo {author} {Tiersten, H.~F.},\ }\bibfield  {title} {\bibinfo {title} {An analysis of doubly rotated quartz resonators utilizing essentially thickness modes with transverse variation},\ }\href {https://doi.org/10.1121/1.393190} {\bibfield  {journal} {\bibinfo  {journal} {The Journal of the Acoustical Society of America}\ }\textbf {\bibinfo {volume} {79}},\ \bibinfo {pages} {1811} (\bibinfo {year} {1986})}\BibitemShut {NoStop}%
\bibitem [{\citenamefont {Gokhale}\ \emph {et~al.}(2021)Gokhale, Vikrant J. and Downey, Brian P. and Katzer, D. Scott and Meyer, David J.}]{gokhale_phonon_2021}%
  \BibitemOpen
  \bibfield  {author} {\bibinfo {author} {Gokhale, V.~J.}, \bibinfo {author} {Downey, B.~P.}, \bibinfo {author} {Katzer, D.~S.},\ and\ \bibinfo {author} {Meyer, D.~J.},\ }\bibfield  {title} {\bibinfo {title} {Phonon diffraction limited performance of {Fabry}-{P{\'e}rot} cavities in piezoelectric epi{\textendash}{H}{B}{A}{R}s},\ }in\ \href {https://doi.org/10.1109/MEMS51782.2021.9375450} {\emph {\bibinfo {booktitle} {2021 {IEEE} 34th {International} {Conference} on {Micro} {Electro} {Mechanical} {Systems} ({MEMS})}}}\ (\bibinfo {year} {2021})\ pp.\ \bibinfo {pages} {206--209}\BibitemShut {NoStop}%
\bibitem [{\citenamefont {Renninger}\ \emph {et~al.}(2018)Renninger, W. H. and Kharel, P. and Behunin, R. O. and Rakich, P. T.}]{renninger_bulk_2018}%
  \BibitemOpen
  \bibfield  {author} {\bibinfo {author} {Renninger, W.~H.}, \bibinfo {author} {Kharel, P.}, \bibinfo {author} {Behunin, R.~O.},\ and\ \bibinfo {author} {Rakich, P.~T.},\ }\bibfield  {title} {\bibinfo {title} {Bulk crystalline optomechanics},\ }\href {https://doi.org/10.1038/s41567-018-0090-3} {\bibfield  {journal} {\bibinfo  {journal} {Nature Physics}\ }\textbf {\bibinfo {volume} {14}},\ \bibinfo {pages} {601} (\bibinfo {year} {2018})}\BibitemShut {NoStop}%
\bibitem [{\citenamefont {Martinis}\ \emph {et~al.}(2005)Martinis, John M. and Cooper, K. B. and McDermott, R. and Steffen, Matthias and Ansmann, Markus and Osborn, K. D. and Cicak, K. and Oh, Seongshik and Pappas, D. P. and Simmonds, R. W. and Yu, Clare C.}]{martinis_decoherence_2005}%
  \BibitemOpen
  \bibfield  {author} {\bibinfo {author} {Martinis, J.~M.}, \bibinfo {author} {Cooper, K.~B.}, \bibinfo {author} {McDermott, R.}, \emph {et~al.},\ }\bibfield  {title} {\bibinfo {title} {Decoherence in {Josephson} qubits from dielectric loss},\ }\href {https://doi.org/10.1103/PhysRevLett.95.210503} {\bibfield  {journal} {\bibinfo  {journal} {Physical Review Letters}\ }\textbf {\bibinfo {volume} {95}},\ \bibinfo {pages} {210503} (\bibinfo {year} {2005})}\BibitemShut {NoStop}%
\bibitem [{\citenamefont {Trif}\ \emph {et~al.}(2018)Trif, Mircea and Dmytruk, Olesia and Bouchiat, H{\'e}l{\`e}ne and Aguado, Ram{\'o}n and Simon, Pascal}]{trif_dynamic_2018}%
  \BibitemOpen
  \bibfield  {author} {\bibinfo {author} {Trif, M.}, \bibinfo {author} {Dmytruk, O.}, \bibinfo {author} {Bouchiat, H.}, \emph {et~al.},\ }\bibfield  {title} {\bibinfo {title} {Dynamic current susceptibility as a probe of {Majorana} bound states in nanowire-based {Josephson} junctions},\ }\href {https://doi.org/10.1103/PhysRevB.97.041415} {\bibfield  {journal} {\bibinfo  {journal} {Physical Review B}\ }\textbf {\bibinfo {volume} {97}},\ \bibinfo {pages} {041415} (\bibinfo {year} {2018})}\BibitemShut {NoStop}%
\bibitem [{\citenamefont {Banderier}\ \emph {et~al.}(2023)Banderier, Hugo and Drimmer, Maxwell and Chu, Yiwen}]{banderier_unified_2023}%
  \BibitemOpen
  \bibfield  {author} {\bibinfo {author} {Banderier, H.}, \bibinfo {author} {Drimmer, M.},\ and\ \bibinfo {author} {Chu, Y.},\ }\bibfield  {title} {\bibinfo {title} {Unified simulation methods for quantum acoustic devices},\ }\href {https://doi.org/10.1103/PhysRevApplied.20.024024} {\bibfield  {journal} {\bibinfo  {journal} {Physical Review Applied}\ }\textbf {\bibinfo {volume} {20}},\ \bibinfo {pages} {024024} (\bibinfo {year} {2023})}\BibitemShut {NoStop}%
\bibitem [{\citenamefont {Reagor}\ \emph {et~al.}(2016)Reagor, Matthew and Pfaff, Wolfgang and Axline, Christopher and Heeres, Reinier W. and Ofek, Nissim and Sliwa, Katrina and Holland, Eric and Wang, Chen and Blumoff, Jacob and Chou, Kevin and Hatridge, Michael J. and Frunzio, Luigi and Devoret, Michel H. and Jiang, Liang and Schoelkopf, Robert J.}]{reagor_quantum_2016}%
  \BibitemOpen
  \bibfield  {author} {\bibinfo {author} {Reagor, M.}, \bibinfo {author} {Pfaff, W.}, \bibinfo {author} {Axline, C.}, \emph {et~al.},\ }\bibfield  {title} {\bibinfo {title} {Quantum memory with millisecond coherence in circuit {QED}},\ }\href {https://doi.org/10.1103/PhysRevB.94.014506} {\bibfield  {journal} {\bibinfo  {journal} {Physical Review B}\ }\textbf {\bibinfo {volume} {94}},\ \bibinfo {pages} {014506} (\bibinfo {year} {2016})}\BibitemShut {NoStop}%
\bibitem [{\citenamefont {Gokhale}\ \emph {et~al.}(2020{\natexlab{b}})Gokhale, Vikrant J. and Downey, Brian P. and Katzer, D. Scott and Meyer, David J.}]{gokhale_temperature_2020}%
  \BibitemOpen
  \bibfield  {author} {\bibinfo {author} {Gokhale, V.~J.}, \bibinfo {author} {Downey, B.~P.}, \bibinfo {author} {Katzer, D.~S.},\ and\ \bibinfo {author} {Meyer, D.~J.},\ }\bibfield  {title} {\bibinfo {title} {Temperature evolution of frequency and anharmonic phonon loss for multi-mode epitaxial {HBARs}},\ }\href {https://doi.org/10.1063/5.0013848} {\bibfield  {journal} {\bibinfo  {journal} {Applied Physics Letters}\ }\textbf {\bibinfo {volume} {117}},\ \bibinfo {pages} {124003} (\bibinfo {year} {2020}{\natexlab{b}})}\BibitemShut {NoStop}%
\bibitem [{\citenamefont {Bild}\ \emph {et~al.}(2023)Bild, Marius and Fadel, Matteo and Yang, Yu and von L{\"u}pke, Uwe and Martin, Phillip and Bruno, Alessandro and Chu, Yiwen}]{bild_schrodinger_2023}%
  \BibitemOpen
  \bibfield  {author} {\bibinfo {author} {Bild, M.}, \bibinfo {author} {Fadel, M.}, \bibinfo {author} {Yang, Y.}, \emph {et~al.},\ }\bibfield  {title} {\bibinfo {title} {Schr{\"o}dinger cat states of a 16-microgram mechanical oscillator},\ }\href {https://doi.org/10.1126/science.adf7553} {\bibfield  {journal} {\bibinfo  {journal} {Science}\ }\textbf {\bibinfo {volume} {380}},\ \bibinfo {pages} {274} (\bibinfo {year} {2023})}\BibitemShut {NoStop}%
\bibitem [{\citenamefont {Yu}\ \emph {et~al.}(2009)Yu, H. and Lee, C.-Y. and Pang, W. and Zhang, H. and Brannon, A. and Kitching, J. and Kim, E.S. Sok}]{yu_hbar-based_2009}%
  \BibitemOpen
  \bibfield  {author} {\bibinfo {author} {Yu, H.}, \bibinfo {author} {Lee, C.-Y.}, \bibinfo {author} {Pang, W.}, \emph {et~al.},\ }\bibfield  {title} {\bibinfo {title} {{HBAR}-{Based} 3.6 {GHz} oscillator with low power consumption and low phase noise},\ }\href {https://doi.org/10.1109/TUFFC.2009.1050} {\bibfield  {journal} {\bibinfo  {journal} {IEEE Transactions on Ultrasonics, Ferroelectrics, and Frequency Control}\ }\textbf {\bibinfo {volume} {56}},\ \bibinfo {pages} {400} (\bibinfo {year} {2009})}\BibitemShut {NoStop}%
\bibitem [{\citenamefont {Baron}\ \emph {et~al.}(2011)Baron, T. and Martin, G. and Lebrasseur, E. and Francois, B. and Ballandras, S. and Lasagne, P.-P. and Reinhardt, A. and Chomeloux, Luc and Lesage, Jean-Marc and Gachon, D.}]{baron_rf_2011}%
  \BibitemOpen
  \bibfield  {author} {\bibinfo {author} {Baron, T.}, \bibinfo {author} {Martin, G.}, \bibinfo {author} {Lebrasseur, E.}, \emph {et~al.},\ }\bibfield  {title} {\bibinfo {title} {{RF} oscillators stabilized by temperature compensated {HBARs} based on {LiNbO3}/{Quartz} combination},\ }in\ \href {https://doi.org/10.1109/FCS.2011.5977814} {\emph {\bibinfo {booktitle} {2011 {Joint} {Conference} of the {IEEE} {International} {Frequency} {Control} and the {European} {Frequency} and {Time} {Forum} ({FCS}) {Proceedings}}}}\ (\bibinfo  {publisher} {IEEE},\ \bibinfo {address} {San Francisco, CA, USA},\ \bibinfo {year} {2011})\ pp.\ \bibinfo {pages} {1--4}\BibitemShut {NoStop}%
\bibitem [{\citenamefont {Capannelli}\ \emph {et~al.}(2025)Capannelli, Kenji and Undseth, Brennan and Fern{\'a}ndez de Fuentes, Irene and Raymenants, Eline and Unseld, Florian K. and Pietx-Casas, Oriol and Philips, Stephan G. J. and M{\k a}dzik, Mateusz T. and Amitonov, Sergey V. and Tryputen, Larysa and Scappucci, Giordano and Vandersypen, Lieven M. K.}]{capannelli_tracking_2025}%
  \BibitemOpen
  \bibfield  {author} {\bibinfo {author} {Capannelli, K.}, \bibinfo {author} {Undseth, B.}, \bibinfo {author} {Fern{\'a}ndez~de Fuentes, I.}, \emph {et~al.},\ }\bibfield  {title} {\bibinfo {title} {Tracking spin qubit frequency variations over 912 days},\ }\href {https://doi.org/10.1038/s41534-025-01134-6} {\bibfield  {journal} {\bibinfo  {journal} {npj Quantum Information}\ }\textbf {\bibinfo {volume} {11}},\ \bibinfo {pages} {192} (\bibinfo {year} {2025})}\BibitemShut {NoStop}%
\bibitem [{\citenamefont {Boudot}\ \emph {et~al.}(2016)Boudot, Rodolphe and Martin, Gilles and Friedt, Jean-Michel and Rubiola, Enrico}]{boudot_frequency_2016}%
  \BibitemOpen
  \bibfield  {author} {\bibinfo {author} {Boudot, R.}, \bibinfo {author} {Martin, G.}, \bibinfo {author} {Friedt, J.-M.},\ and\ \bibinfo {author} {Rubiola, E.},\ }\bibfield  {title} {\bibinfo {title} {Frequency flicker of 2.3 {GHz} {AlN}-sapphire high-overtone bulk acoustic resonators},\ }\href {https://doi.org/10.1063/1.4972102} {\bibfield  {journal} {\bibinfo  {journal} {Journal of Applied Physics}\ }\textbf {\bibinfo {volume} {120}},\ \bibinfo {pages} {224903} (\bibinfo {year} {2016})}\BibitemShut {NoStop}%
\end{thebibliography}
%

\clearpage
\setcounter{equation}{0}
\setcounter{figure}{0}
\setcounter{table}{0}
\setcounter{page}{0}
\setcounter{section}{0}
\renewcommand{\theequation}{S\arabic{equation}}
\renewcommand{\thefigure}{S\arabic{figure}}
\renewcommand{\thetable}{S\arabic{table}}

\onecolumngrid

\begin{center}
	\textsc{\Large{Supplementary Information}}
\end{center}

\section{Loss mechanisms in HBARs}
In this supplementary section, we discuss and derive simple analytical expressions for the limiting quality factors of HBARs and BAW resonators due to different loss mechanisms. For the derivation of mechanical absorption, surface scattering, and diffraction loss, we treat the bulk acoustic wave resonator as a Fabry-Perot cavity for phonons. This enables us to use known expressions from resonator optics to calculate the finesse and the quality factor of the phonon cavity starting from estimates of the intensity attenuation factor $r^2$, or the fraction of energy lost per round trip $\Delta E$. An expression for the Finesse in terms of these quantities is \cite{saleh_fundamentals_1991} 
\begin{equation}\label{eq:Finesse}
    \mathcal{F}=\frac{\pi r^{1/2}}{1-r}\approx \frac{2\pi}{\Delta E}.
\end{equation}
Note that the last approximation assumes that losses are small, so $r\approx r^{1/2}\approx1$, and the fraction of energy lost per round trip is $\Delta E=(1-r^2)\approx2(1-r)$. The finesse can then be used to calculate the quality factor
\begin{equation}\label{eq:QfromFinesse}
    Q=\frac{f_n}{\text{FSR}}\mathcal{F}\approx\frac{2\pi f_n \tau_{RT}}{\Delta E}.
\end{equation}
Here, $f_n$ is the acoustic mode frequency, and $\tau_{RT}=\text{FSR}^{-1}=2L/v$ is the round trip travel time of the acoustic wave, where $L$ is the length of the cavity and $v$ is the speed of sound in the resonator's material at the relevant acoustic wave polarization and propagation direction.
\subsection{Mechanical absorption}
We use mechanical absorption as a general term for the loss arising from the attenuation of an acoustic wave as it propagates through a material. This is distinct from surface scattering loss, which occurs only when an acoustic wave reflects from an interface. As a result, it may originate from several physical mechanisms that collectively give rise to an effective absorption coefficient.

In real materials, mechanical absorption loss can have different origins: from viscous damping associated with the internal friction of atoms, to the cyclic rearrangement of grain boundaris, dislocations, and defects under stress \cite{braginsky_systems_1986}. From a wave perspective, we can also think of the scattering of acoustic waves at impurities and defects \cite{goryachev_observation_2013, gruenke-freudenstein_surface_2025}. To model the effects of mechanical absorption in a bulk acoustic wave resonator, we define an effective absorption coefficient per unit length $\alpha$. 
The absorption coefficient is material-dependent, where in general we expect disordered or polycrystalline materials to have a larger $\alpha$ than more ideal or crystalline materials. Therefore, for a resonator with thickness $L$, the intensity of an acoustic wave after a round trip will be attenuated by a factor 

\begin{equation}\label{eq:rabsorption}
    r^2=e^{-\alpha 2L}.
\end{equation} 
Combining Equations \eqref{eq:Finesse}, \eqref{eq:QfromFinesse}, and \eqref{eq:rabsorption}, and expanding $e^{-\alpha 2L}\approx1-\alpha 2L$, we arrive at a simple expression for the absorption-limited quality factor:
\begin{equation}\label{eq:absorptionQ}
    Q_{\text{mech}}\approx\frac{2\pi f_n}{\alpha v}.
\end{equation}
Note that in a monolithic acoustic resonator, mechanical absorption loss distributed uniformly over the bulk does not depend on the thickness of the acoustic cavity. Alternatively, for a composite BAW or HBAR, we can calculate the effective absorption-limited quality factor by weighting the energy participation ratios of its constituent layers $Q^{-1}_{\text{mech}}=\sum_Xp_XQ^{-1}_{\text{X, mech}}$, as we discuss in the main text. 

From another perspective, the absorption of propagating waves in a medium can also be described by a complex wavevector $\tilde{k}=k+ik^*$, such that the corresponding quality factor is \cite{pozar2021microwave}:
\begin{equation}\label{eq:pozarQ}
    Q_{\text{mech}}=-\frac{k}{2k^*}.
\end{equation}
Matching Eqs. \eqref{eq:absorptionQ} and \eqref{eq:pozarQ}, and using $k=2\pi f_n/v$, we identify $\alpha=-2k^*$. Furthermore, in analogy to loss in dielectrics, we can consider a complex stiffness constant for longitudinally polarized waves $\tilde{c}_{33}=c_{33}+ic_{33}^*$ and an acoustic loss tangent $\tan\delta=c_{33}^*/c_{33}$, such that for small loss:
\begin{equation}\label{eq:wavevectorloss}
    \tilde{k}\approx\frac{2\pi f_n}{v}\left(1-i\frac{c^*_{33}}{2c_{33}}\right).
\end{equation}
To arrive at Eq. \eqref{eq:wavevectorloss}, we used $v=\sqrt{\tilde{c}_{33}/\rho}$ and performed a Taylor expansion. Combining Eqs. \eqref{eq:absorptionQ}-\eqref{eq:wavevectorloss}, we consistently find the relations:
\begin{equation}
    Q_{\text{mech}}=\frac{c_{33}}{c^*_{33}}=\frac{1}{\tan\delta},\quad\text{and}\quad \alpha=\frac{2\pi f_n}{v}\tan\delta.
\end{equation}
The nearly flat frequency dependence of the quality factors (ignoring mode-to-mode fluctuations and acoustic impedance mismatch between AlN and sapphire) that we found in the sputter-deposited HBARs (Samples A and E; see Figs. \ref{fig:antennameas} and \ref{fig:EMPA}) is consistent with a constant $c_{33}^*$ and loss tangent. The mathematical interpretation of this fact is the existence of a constant phase lag between stress and strain \cite{braginsky_systems_1986, cleland2013foundations}. Note that the HVPE-deposited samples are limited by surface scattering, which has a different frequency dependence. 

\subsection{Surface scattering}
\label{sec:scattering}
Surface scattering loss in BAWs and HBARs originates from the diffuse reflection of acoustic waves at surfaces or interfaces. To model this loss, we treat the acoustic resonator as a Fabry-Perot cavity with imperfect mirrors of reduced reflectance \cite{saleh_fundamentals_1991, galliou_extremely_2013, luo_lifetime-limited_2025}. 

The reflectance of a mirror $R_i\in[0,1]$ quantifies the fraction of wave intensity preserved after reflection. For perfectly reflective mirrors, $R_i=1$, while for very lossy mirrors $R_i\approx0$. We distinguish two contributions in the reflectance, namely
\begin{equation}
    R_i=R_{0}p,
\end{equation}
where $R_0$ is the coherent reflection coefficient of a 'leaky' mirror, and $p\in[0,1]$ is the specularity parameter, which quantifies the degree of specular ($p\to1$) versus diffuse ($p\to0$) reflection. At the bulk-vacuum boundary there is a large impedance mismatch, so we can assume that acoustic waves are completely reflected and $R_0=1$. From Ziman's formula \cite{ziman_electrons_2001, maznev_boundary_2015} 
\begin{equation}\label{eq:ziman}
    p=e^{-\left(\frac{4\pi\sigma\cos\theta}{\lambda}\right)^2},
\end{equation}
where $\lambda$ is the acoustic wavelength, $\sigma$ is the RMS surface roughness, and $\theta$ is the angle of incidence. Therefore, the total round-trip intensity attenuation factor of the sound waves from non-specular reflection is \cite{saleh_fundamentals_1991}:
\begin{equation}\label{eq:rscattering}
    r^2=R_{1}R_{2}\approx p_{1}p_{2}=e^{-\left(\frac{4\pi\sigma_{1}}{\lambda}\right)^2}e^{-\left(\frac{4\pi\sigma_{2}}{\lambda}\right)^2}
    \approx 1-\left(\frac{4\pi\sigma_{1}}{\lambda}\right) ^2-\left(\frac{4\pi\sigma_{2}}{\lambda}\right) ^2,
\end{equation}
where we consider normal incidence ($\cos\theta =1$) and small surface roughness compared with the wavelength ($\sigma\ll\lambda$). The subindex $1$ and $2$ refer to the top and bottom surfaces, respectively. Combining Eqs. \eqref{eq:rscattering} and \eqref{eq:QfromFinesse} we find
\begin{equation}
    Q_{\sigma}\approx\frac{Lv}{4\pi f_n(\sigma_1^2+\sigma_2^2)},
\end{equation}
 where we also replaced $\lambda_n=v/f_n$ to express the quality factor only in terms of the acoustic mode frequency $f_n$.
This result is not surprising, it is well known that surface related loss mechanisms scale with the length $L$ of the resonator \cite{luo_lifetime-limited_2025}. Additionally, surface scattering loss is known to have a $Q\times f=const.$ relation, and experimental measurements have confirmed the degradation of the quality factors with surface roughness in FBARs \cite{vorobiev_effect_2011}.
We find that this model describes well the experimentally observed frequency dependence of the HBAR quality factors, see Fig. 2 in the main text, where the RMS roughness of the top and back surfaces and the interface are estimated independently from AFM scans (Fig. 8 in Methods).

\subsection{Diffraction}\label{sec:diffraction}
\begin{figure}[t]
    \centering
    \includegraphics[width = 0.5\linewidth]{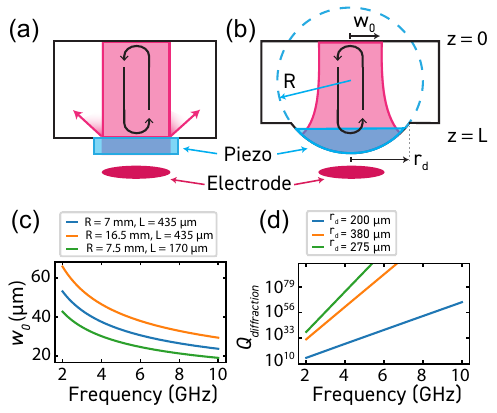}
    \caption{\textbf{Diffraction loss.} (a) Schematic of a flat-flat HBAR illustrating the absence of lateral confinement of the acoustic energy. (b) Schematic of a plano-convex HBAR with a dome-shaped surface of radius of curvature $R$. This geometry supports stable Laguerre-Gaussian (LG) longitudinal modes propagating between the boundaries at $z=0$ and $z=L$, and determines the mode waist $w_0$ of the modes. A dome radius $r_d\gg w_0$ ensures that most of the acoustic energy is refocused after each round trip, thereby suppressing diffraction loss. (c) Calculated mode waist $w_0$ of the fundamental LG$_{00}$ mode as a function of frequency for various radii of curvature $R$ and HBAR thickness $L$. (d) Diffraction-limited quality factor $Q_{\text{diffraction}}$ as a function of frequency for different dome radii $r_d$, obtained using the mode waists from panel (c). }
    \label{fig:DiffractionTheory}
\end{figure}
When an HBAR is driven with an electrode and a piezoelectric transducer, the acoustic waves are launched towards the bulk and will bounce back and forth between the surfaces of the HBAR. Nevertheless, these waves might not be perfectly confined laterally. The acoustic beam might spread sideways as it propagates (just like an optical beam diverges).  Only the waves that remain confined to the region facing the electrode can continue to undergo electromechanical interactions, whereas the rest of the energy leaks away towards the sides, leading to diffraction loss. This is schematically illustrated in Fig. \ref{fig:DiffractionTheory}(a).

A very effective strategy to increase the lateral confinement of the acoustic waves and mitigate diffraction loss is to engineer a plano-convex geometry for the HBAR \cite{stevens_analysis_1986, kharel_ultra-high-_2018, galliou_extremely_2013}, where one of the surfaces is given a dome shape with a certain radius of curvature $R$, see Fig. \ref{fig:DiffractionTheory}(b). Alternatively, for flat-flat HBAR geometries, increasing the lateral size of the transduced acoustic mode reduces the diffraction loss \cite{gokhale_phonon_2021}. 

We first analyze the diffraction loss limit for a plano-convex HBAR following Ref. \cite{kharel_ultra-high-_2018}. If the total thickness of the HBAR is $L\gg\lambda_n$, where $\lambda_n$ is the acoustic wavelength of the $n$th overtone, it is possible to apply the paraxial wave approximation and treat the propagation of acoustic waves with a Gaussian beam optics approach \cite{saleh_fundamentals_1991}. Under this approximation, Christoffels' equation can be simplified to an equivalent paraxial Helmholtz equation \cite{kharel_ultra-high-_2018}, whose solutions are families of Laguerre-Gaussian (LG) or Hermite-Gaussian (HG) modes. Given the radial symmetry of our system, we use cylindrical coordinates ($r,\phi,z$) and LG-type solutions. Assuming that longitudinal acoustic waves propagate in the $\hat{z}$ direction with a displacement field $\mathbf{u}_{npl}(\mathbf{r},t)=U_{npl}(r,\phi)e^{-i(k_nz-2\pi f_nt)}\hat{z}$, we can write the spatial dependence of the modes as
\begin{equation}\label{eq:profile}
    U_{npl}(r,\phi)e^{-ik_nz}=U_0\cdot \text{LG}_{pl}(r,\phi)e^{-i\left(k_nz+k_n\frac{r^2}{R'(z)}-(|l|+2p+1)\psi'(z)\right)}
\end{equation}
where $U_0$ is a normalization constant, $k_n=2\pi/\lambda_n$ is the acoustic wavevector, $R'(z)$ is the radius of curvature of the acoustic beam wavefront, $\psi'(z)$ is the acoustic Guoy phase, and $\text{LG}_{pl}$ is the Laguerre-Gaussian mode with radial index $p$ and azimuthal index $l$
\begin{equation}\label{eq:profileLG}
    \text{LG}_{pl}=\sqrt{\frac{2p!}{\pi(p+|l|)!}}\frac{w_0}{w(z)}\left(\frac{r\sqrt{2}}{w(z)}\right)^{|l|}e^{-\frac{r^2}{w(z)^2}} \cdot L_p^{|l|}\left(\frac{2r^2}{w(z)^2}\right)e^{-il\phi}.
\end{equation}
Here, $L_p^{|l|}$ denotes the generalized Laguerre polynomial and $w(z)=w_0\sqrt{1+(z/z_R)^2}$ is the beam radius at $z$, where $w_0$ is the mode waist. The acoustic Rayleigh length 
\begin{equation}\label{eq:rayleighlength}
    z_R=\frac{\pi w_0^2}{\lambda_n}\chi
\end{equation}
is a characteristic length scale that quantifies the spread of the beam with distance. The anisotropy-parameter $\chi$ defined in the supplementary of \cite{kharel_ultra-high-_2018} is a correction factor introduced to account for the effects of acoustic wave propagation in an anisotropic medium. For sapphire, $\chi\approx1.5$. The beam front radius of curvature also depends on the Rayleigh length
\begin{equation}
    R'(z)=\frac{z}{\chi}\left(1+\frac{z_R^2}{z^2}\right),
\end{equation}
and since at $z=L$ it must match the radius of curvature of the dome $(R'(L)=R)$, we can use this fact to find the mode waist as a function of the HBAR's geometric parameters and acoustic wavelength:
\begin{equation}\label{eq:modewaist}
    w_0=\left(\frac{LR\lambda_n^2}{\chi\pi^2}\right)^{1/4}=\left(\frac{LRv^2}{\chi\pi^2f_n^2}\right)^{1/4}.
\end{equation}
To obtain this simple expression for the mode waist, we assume $R\gg L$, and also use $\lambda_nf_n\approx v$ in the second equality. Note that the mode waist depends on the frequency, with lower frequency modes having a larger $w_0$. This is shown in Fig. \ref{fig:DiffractionTheory}(c), where we calculate the mode waist using the radius of curvature $R$ and the total thickness $L$ of Samples B, C, and D, respectively. Having control over the radius of the beam is important because, for the $\text{LG}_{00}$ mode, $99\%$ of the total beam power is carried within a circle of radius $r=1.5 w(z)$ \cite{saleh_fundamentals_1991}. In fact, assuming that the radius of the beam does not vary significantly throughout the entire length of the HBAR, $w(z)\approx w_0$, the fraction of total energy in the $\text{LG}_{00}$ mode confined under the dome area is  
\begin{equation}
    \frac{E_{\text{dome}}}{E_{\text{tot}}}=1-e^{-\frac{2r_d^2}{w_0^2}},
\end{equation}
where $r_d$ is the radius of the dome, as illustrated in Fig. \ref{fig:DiffractionTheory}(b), and we use that $E_{\text{dome}}=L\int_0^{r_d}2\pi r |U_{n00}(r)|^2dr$. So the fraction of the total energy lost per round trip is $\Delta E=e^{-2r_d^2/w_0^2}$, leading to
\begin{equation}\label{eq:diffractionloss}
    Q_{\text{diffraction}}\approx\frac{4\pi f_nL}{v e^{-2r_d^2/w_0^2}}.
\end{equation}
We use Eq. \eqref{eq:diffractionloss} to calculate the diffraction-limited quality factors of Samples B, C and D shown in Fig. \ref{fig:DiffractionTheory}(d). For the calculation we use the dome radius of these samples, $r_d=200,~380~\text{and }275~\mu$m, respectively, and their respective mode waists shown in Fig. \ref{fig:DiffractionTheory}(c). We observe that diffraction loss in plano-convex HBARs is completely negligible for modes with frequencies above $2~$GHz.

Note that in the case of a flat-flat HBAR, we cannot simply use Eq. \eqref{eq:diffractionloss} to predict the diffraction-limited quality factor because we do not have an analytical expression to predict the beam radius or the mode waist. Instead, we need to rely on numerical simulations to estimate the mode's energy distribution after a round trip. This method was recently described in Ref. \cite{gruenke-freudenstein_surface_2025}. Using numerical simulations based on acoustic beam propagation \cite{von_lupke_quantum_2023, renninger_bulk_2018}, we are able to extract the energy lost per round trip by first computing the overlap $\eta$ between the initial forcing field of the electrode or the antenna, $u_0$, with the acoustic field profile after a single round trip, $u_1$:
\begin{equation}
    \eta=\frac{\int u_0^*u_1 }{\sqrt{\int|u_0|^2\int|u_1|^2}}.
\end{equation}
We perform this simulation for different frequencies to also extract any frequency dependence in $\eta$. The energy lost per round trip is then $\Delta E=1-\eta^2$, which we use to compute the diffraction-limited quality factor from Eq. \eqref{eq:QfromFinesse}
\begin{equation}
    Q_{\text{diffraction}}\approx\frac{4\pi f_n L}{v(1-\eta^2)}.
\end{equation}
 Using this method, we obtain good agreement with the quality factors measured for a flat-flat HBAR (see Supplementary section \ref{sec:diffractionlimited}).

\subsection{TLS}\label{sec:TLSsupplementary}
In this work, we observe that at low temperatures and phonon occupation, spurious coupling to two-level systems (TLS) can become a relevant source of loss in composite high-overtone bulk acoustic wave resonators. This effect is most pronounced for HBAR modes with large participation in the piezoelectric transducer, since polycrystalline or amorphous films host a higher density of defects than crystalline substrates \cite{phillips_two-level_1987}. TLS originate from low-energy defect centers in solids, which behave effectively as quantum two-level systems and couple to strain and electric fields via elastic and electric dipole interactions, respectively \cite{behunin_dimensional_2016}. TLS-induced dissipation has been extensively studied in amorphous solids and microwave circuits \cite{phillips_two-level_1987, hunklinger_3_1976, martinis_decoherence_2005}. Here, we summarize the key TLS concepts relevant to our measurements. This section is not intended to be a rigorous derivation of TLS expressions; rather, it aims to provide an intuition for how these expressions are obtained and to justify our observations.

Early TLS models describe defects as atoms or molecular groups tunneling between the minima of an asymmetric double-well potential \cite{hunklinger_3_1976, phillips_two-level_1987}. A TLS is characterized by an energy splitting 
\begin{equation}
    E=\sqrt{\Delta^2+\Delta_0^2},
\end{equation}
where $\Delta$ is the asymmetry energy and $\Delta_0$ is the tunneling matrix element. In what follows, we do not distinguish between tunneling states (TS) and TLS; we simply assume that each defect can be modeled by a two-level system of energy $E=\hbar\omega_{\text{TLS}}$. A distribution of such defects is described by a density of states (DOS) $P(E)$ (or $P(\omega_{\text{TLS}})$) per unit volume.

Following \cite{behunin_dimensional_2016, emser_thin-film_2024,maccabe_nano-acoustic_2020}, we first discuss the coupling of a phonon mode with an individual TLS. This coupling occurs via the deformation of the potential landscape around the defect by the strain field, leading to the interaction Hamiltonian in second quantization
\begin{equation}\label{eq:TLShamiltonian}
    H_{\text{int, TLS-ph}}=[D\sigma_z+M\sigma_x]s_z(\mathbf{r_0})(a^\dagger+a)=\hbar[g_z(\mathbf{r_0})\sigma_z+g_x(\mathbf{r_0})\sigma_x](a^\dagger+a),
\end{equation}

Here, $s_{z}(\mathbf{r_0})\equiv S_{33}(\mathbf{r_0})$ is the longitudinal (z-polarized) strain component of the phonon mode around the position of the defect $\mathbf{r_0}$, which is the only relevant component for the longitudinal HBAR modes considered here. The parameters $M$ and $D$ are the transverse and longitudinal elastic coupling potentials, respectively, determined by the TLS potential and the projection of the TLS elastic dipole moment onto the strain field. $s_z(\mathbf{r_0})$ is normalized such that 
\begin{equation}
\int_{\text{Tot}} c_{33}|s_z(\mathbf{r_0})|^2dV\equiv\bar{c}_{33}|S_{\text{vac}}|^2V,
\end{equation}
where $S_{\text{vac}}=\sqrt{\hbar\omega_r/(2\bar{c}_{33}V)}$ is the vacuum strain amplitude, $\omega_r$ is the mechanical resonance frequency, $V$ is the resonator volume, $c_{33}$ is the stiffness constant of the corresponding material (in Voigt notation), and $\bar{c}_{33}$ is a volume-averaged stiffness constant (only relevant for a resonator composed of multiple materials with different stiffnesses). We define the longitudinal and transverse coupling strenghts:
\begin{align}
    g_z(\mathbf{r_0})=\frac{Ds_z(\mathbf{r_0})}{\hbar},\\
    g_x(\mathbf{r_0})=\frac{Ms_z(\mathbf{r_0})}{\hbar}. \label{eq:transversecoupling}
\end{align}
The transverse term ($\propto\sigma_x$) is associated with 'resonant' TLS effects, while the longitudinal term ($\propto\sigma_z$) leads to 'relaxation' TLS effects.

We first consider resonant TLS, where the transverse coupling between the phonon mode and the TLS produces both a frequency shift of the phonon mode $\delta\omega_r$ and a contribution to the damping rate $\kappa_r$ \cite{emser_thin-film_2024}:
\begin{align}
    &\delta\omega_r=-\frac{S_{\text{vac}}^2}{\hbar}\text{Re}[\chi(\omega_r)],\label{eq:freqshiftsusceptibility}\\
    &\kappa_r=\frac{2S_{\text{vac}}^2}{\hbar}\text{Im}[\chi(\omega_r)],\label{eq:dampingsusceptibility}
\end{align}
where $\chi(\omega_r)$ is the acoustic susceptibility, which quantifies the response of the TLS ensemble to a time-varying strain field. There are different techniques to calculate $\chi(\omega_r)$ for a single TLS \cite{emser_thin-film_2024, trif_dynamic_2018}. Importantly, the susceptibility of the full TLS bath is obtained by summing the susceptibilities of individual TLS, which in the continuum limit becomes $\sum_{\text{TLS}}\to\hbar\int dV\int P(\omega_{\text{TLS}})d\omega_{\text{TLS}}$, including all TLS in the resonator volume. The acoustic susceptibility due to transverse coupling can then be written \cite{emser_thin-film_2024, maccabe_nano-acoustic_2020}:
\begin{equation}\label{eq:originalsusceptibility}
    \chi(\omega)=\frac{\hbar^2}{S_{\text{vac}}^2}\int dV\int_0^{\omega_{\text{max}}} d\omega_{\text{TLS}}P(\omega_{\text{TLS}})|g_x(\mathbf{r})|^2\tanh\left(\frac{\hbar\omega_{\text{TLS}}}{2k_BT}\right)\left(\frac{1}{\omega_{\text{TLS}}-(\omega+i\Gamma_2^{\text{TLS}})}+\frac{1}{\omega_{\text{TLS}}+\omega+i\Gamma_2^{\text{TLS}}}\right),
\end{equation}
where $\Gamma_2^{\text{TLS}}$ is the TLS relaxation rate, and $\omega_{\text{max}}$ is the maximum transition frequency in the TLS ensemble. 
Typically, Eq. \eqref{eq:originalsusceptibility} is integrated over the volume hosting TLS, $V_h$. To be more general and accommodate the possibility of having multiple materials or volumes hosting different densities of TLS, we partition the resonator volume into smaller regions $X$, each with constant stiffness $c_{33,X}$ and uniform TLS density. We also introduce a factor $c_{33}/c_{33}$ in the integral that allows us to establish a connection with the energy participation ratios:
\begin{equation}\label{eq:susceptibilityintegral}
    \chi(\omega)=\frac{1}{S_{\text{vac}}^2}\sum_X\int_X dVc_{33,X}|s_z(\mathbf{r})|^2\bar{P}_X\frac{\bar{M}_X^2}{c_{33,X}}\int_0^{\omega_{\text{max}}} d\omega_{\text{TLS}}\tanh\left(\frac{\hbar\omega_{\text{TLS}}}{2k_BT}\right)\left(\frac{1}{\omega_{\text{TLS}}-(\omega+i\Gamma_2^{\text{TLS}})}+\frac{1}{\omega_{\text{TLS}}+\omega+i\Gamma_2^{\text{TLS}}}\right)
\end{equation}
Here, we have replaced the transverse coupling constant $g_x$ using its definition in Eq. \eqref{eq:transversecoupling}. In addition, we assume a uniform TLS DOS distribution with a mean value $\bar{P}_X$ and an averaged transverse elastic coupling potential $\bar{M}_X$ in each region. The integral over $d\omega_{\text{TLS}}$ can be evaluated for $\hbar\omega_{\text{max}}\gg k_BT$ using the digamma function $\Psi$, and the volume integral evaluates to $\int_Xc_{33,X}|s_z(\mathbf{r})|^2dV= c_{33,X}|S_{\text{vac}}|^2V_X$, where $V_X$ is an effective volume. We obtain \cite{emser_thin-film_2024}:
\begin{equation}\label{eq:susceptibilityintegrated}
    \chi(\omega_r)\approx-2\sum_X\frac{\bar{P}_X\bar{M}_X^2}{c_{33,X}}V_Xc_{33,X}\left(\Psi\left(\frac{1}{2}+\frac{\hbar\Gamma_2^{\text{TLS}}}{2\pi k_BT}+\frac{\hbar \omega_r}{ 2\pi i  k_BT}\right)-\ln\left(\frac{\hbar\omega_{\text{max}}}{ 2\pi k_BT}\right)\right)
\end{equation}
Combining Eqs. \eqref{eq:susceptibilityintegrated} and \eqref{eq:freqshiftsusceptibility}, and neglecting the $\Gamma_2^{\text{TLS}}$ term, we arrive at \cite{emser_thin-film_2024, hunklinger_3_1976}:
\begin{equation}\label{eq:TLStemp_appendix}
    \frac{\Delta \omega_r}{\omega_r}=\frac{\sum_Xp^{S}_X\tan\delta_X}{\pi}\left(\Re\left\{\Psi\left(\frac{1}{2}+\frac{\hbar \omega_r}{ 2\pi i  k_BT}\right)\right\}-\ln\left(\frac{\hbar\omega_r}{ 2\pi k_BT}\right)\right)
\end{equation}
where $\Delta\omega_r/\omega_r\equiv\frac{\omega_r(T)-\omega_r(0)}{\omega_r(\infty)}$ and
\begin{equation}\label{eq:TLSlosstangent}
    p^S_X=\frac{V_Xc_{33,X}}{V\bar{c}_{33}}=\frac{\int_Xc_{33,X}|s_z|^2dV}{\int_{\text{Tot}}c_{33}|s_z|^2dV},\qquad \tan \delta_X=\frac{1}{Q_{\text{X,TLS}}}=\frac{\pi \bar{P}_X\bar{M}_X^2}{c_{33,X}}.
\end{equation}
The sum $\sum_Xp^{S}_X \tan \delta_X$ is the quantity defined in Eq. (4) of the main text as $Q_{\text{TLS}}^{-1}$ (later in this section we refer to it as reactive TLS loss tangent $Q_{\text{TLS,reac}}^{-1}$), and $\tan\delta_X=Q_{\text{X,TLS}}^{-1}$ is the intrinsic TLS loss tangent at zero temperature of region $X$. 
For an arbitrary mechanical resonator, $p^S_{X}$ can be obtained from finite-element simulations by integrating the z-polarized strain field over region $X$ and over the full volume \cite{banderier_unified_2023}. This calculation can be extended to other strain and stiffness tensor components to include other types of mechanical motion. Using the stress-strain relation $T_z=c_{33}s_z$, and the definition of potential acoustic energy $\mathcal{E}_{\text{pot}}=\int_{\text{Tot}}\frac{c_{33}}{2}|s_z|^2dV$, we may write
\begin{equation}\label{eq:participationintegral}
    p^S_{X}=\frac{\int_X\frac{1}{2}\frac{|T_z|^2}{c_{33,X}}dV}{\mathcal{E}_{\text{pot}}}
\end{equation}
For HBARs, where the cross-sectional area of the mode is constant across all layers, we can simplify the volume integral in Eq. \eqref{eq:participationintegral} to a one dimensional problem in the longitudinal direction. Using $Z_Xv_X=c_{33}$, where $Z_X$ and $v_X$ are the acoustic impedance and the longitudinal speed of sound in material $X$, respectively, we retrieve the participation expressions in the Methods section ($p^{\text{pot}}_X\approx p_{X}^S$). The advantage of our 1D model is that it allows us to compute the energy participation ratio analytically, making the analysis numerically efficient and more revealing of the underlying relationships between the participation ratios and the system parameters. 

\begin{figure*}
    \centering
    \includegraphics[width=0.85\linewidth]{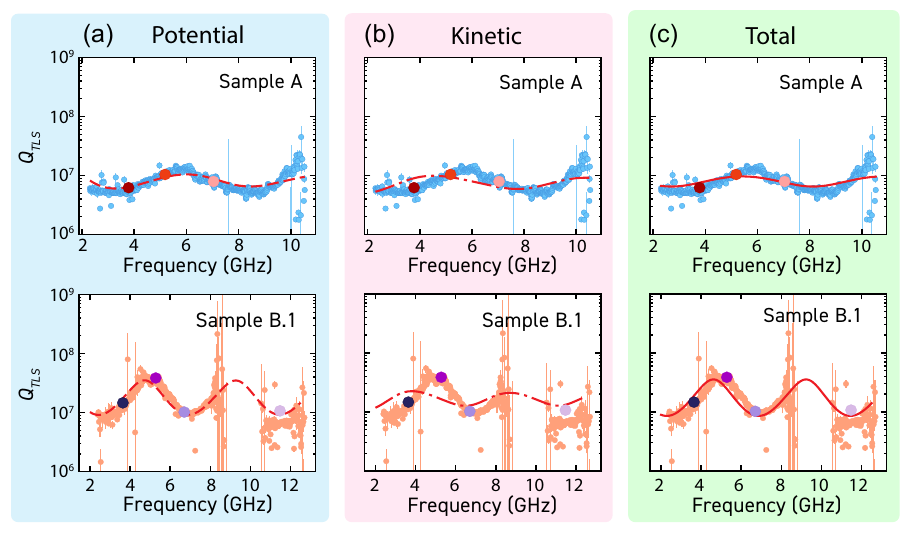}
    \caption{\textbf{Reactive TLS loss tangent and acoustic energy participation.} Best least-squares fits to the model $Q_{\text{TLS}}^{-1}=\sum_Xp^*_XQ_{\text{X,TLS}}^{-1}$ using the potential-energy participation (a), kinetic-energy participation (b) and total (kinetic + potential) energy participation (c). The respective participations are plotted in the Methods Fig. \ref{fig:fig11particips}. For both samples A and B.1, the TLS loss tangent of the bulk layer is assumed to be zero.}
    \label{fig:TLSpotential}
\end{figure*}

In the HBARs studied in this work, both the piezoelectric ($P$) and defect ($D$) layer thicknesses are sub-wavelength in the frequency range considered. As a result, the potential and kinetic energy participation ratios of the mechanical modes in this range can differ substantially; see Fig. \ref{fig:fig11particips} in Methods. In particular, we observe a phase offset in the oscillations of both participation ratios as a function of mode frequency. In our measurements of $Q_{\text{TLS}}$ shown in Fig. \ref{fig:TLS}(c) of the main text, we also observe clear oscillations. Fitting these with the potential-energy participation model yields the best agreement with the data (Fig. \ref{fig:TLSpotential}(a)), compared with fits using the kinetic or total energy participations (Figs. \ref{fig:TLSpotential}(b,c)). These results confirm that TLS loss in mechanical resonators is mediated by the strain, and demonstrate that this loss can be modeled using energy-participation methods \cite{gruenke-freudenstein_surface_2025}, analogously to dielectric loss in superconducting circuits \cite{minev_energy-participation_2021, wang_surface_2015}.

We now consider the effect of resonant TLS on the resonator damping rate $\kappa_r$. At large phonon occupation number $\bar{n}_p$, the TLS are driven by the resonator field. Therefore, we need to take into account their driven steady state and coherence when computing the susceptibility. This yields the well known expression for the inverse quality factor of a mechanical resonator due to TLS:
\begin{equation}
    Q^{-1}=\frac{\kappa_r}{\omega_r}=Q_{\text{TLS}}^{-1}\frac{\tanh\left(\frac{\hbar\omega_r}{2k_BT}\right)}{\sqrt{1+\bar{n}_p/n_c}},
    \end{equation}
where mostly TLS within the linewidth of the resonator contribute to the loss \cite{emser_thin-film_2024, phillips_two-level_1987}, i.e.  $|\omega_{\text{TLS}}-\omega_r|\lesssim\kappa_r$. Driving strongly the TLS reduces the damping by a factor $\sqrt{1+\bar{n}_p/n_c}$, where  
\begin{equation}
    n_c=\frac{\hbar \bar{c}_{33}V}{2\omega_r\bar{M}^2T_1T_2}
\end{equation}
is the critical phonon number at which TLS saturation begins, with $T_1$ and $T_2$ the TLS transverse and longitudinal relaxation times, respectively, and $\bar{M}$ an averaged transverse coupling potential. The zero-temperature TLS loss tangent is again given by $Q_{\text{TLS}}^{-1}=\sum_Xp_X^S\tan\delta_X$, which we refer to as the dissipative TLS loss tangent, $Q_{\text{TLS, diss}}^{-1}$, in contrast to the reactive loss tangent in Eq. \eqref{eq:TLStemp_appendix}. Note that $\tan\delta_X$ has the same expression as in Eq. \eqref{eq:TLSlosstangent}, however, if the TLS DOS $P$ and coupling $M$ depend on energy, the effective averages $\bar{M}_X$ and $\bar{P}_X$ may differ between reactive and dissipative measurements: the dissipative term is dominated by near-resonant TLS, whereas the reactive term is dominated by TLS with $\hbar\omega_{\text{TLS}}\sim k_BT$ \cite{phillips_two-level_1987, emser_thin-film_2024}. Consequently, an inhomogeneous TLS frequency distribution leads to a larger variance in the dissipative TLS loss tangent than in the reactive one, as derived in Ref. \cite{emser_thin-film_2024}:
\begin{equation}\label{eq:variance}
    \frac{\text{Var}[Q^{-1}_{\text{TLS, diss}}]}{\text{Var}[Q^{-1}_{\text{TLS, reac}}]}\approx\frac{4}{\pi^2}\log\left(\frac{\omega_{\text{max}}}{\omega_r}\right)^2.
\end{equation}

\begin{figure}
    \centering
    \includegraphics[width=0.5\linewidth]{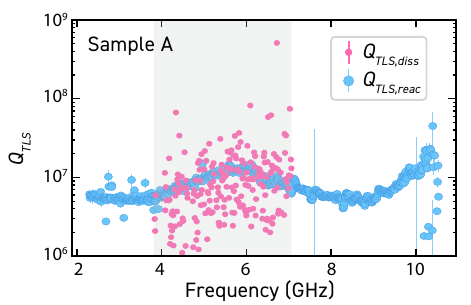}
    \caption{\textbf{Dissipative and reactive resonant TLS loss in sample A}. Dissipative and reactive TLS quality factors measured for several modes of sample A. The reactive TLS quality factors $Q_{\text{TLS,reac}}$ are calculated from the fractional frequency shifts of the modes between $10~$mK and $4~$K. The dissipative TLS quality factors $Q_{\text{TLS,diss}}$ are calculated using Eq. \eqref{eq:supplementaryTLSdiss}. In the text, we compare the variance of both measurements in the frequency range highlighted.}
    \label{fig:TLSreacdiss}
\end{figure}

In this work, we extract the TLS loss tangent using both temperature-dependent frequency measurements (yielding the reactive TLS loss tangent via Eq. \eqref{eq:TLStemp_appendix}), and high- vs. low-power measurements of the quality factors (yielding the dissipative loss tangent). For sample A, these measurements were performed for hundreds of phonon modes spanning a few GHz frequency, see Fig. \ref{fig:TLSreacdiss}. We obtain $Q_{\text{TLS,diss}}^{-1}$ for each mode from the difference between the inverse quality factor measured with high-power spectroscopy and that measured using the qubit to observe the decay of a single phonon excitation:
\begin{equation}\label{eq:supplementaryTLSdiss}
    Q_{\text{TLS,diss}}^{-1}=Q_{\text{i,(qubit)}}^{-1}-Q_{\text{i,(spec.)}}^{-1},
\end{equation}
assuming $\tanh(\hbar\omega_r/2k_BT)/\sqrt{1+\bar{n}_p/n_c}\approx1$ at millikelvin temperatures and $\bar{n}_p=1$. Evaluating the variance of both measurements over the highlighted frequency range in Fig. \ref{fig:TLSreacdiss} (after removing the underlying trend) we obtain $\text{Var}[Q^{-1}_{\text{TLS, diss}}]/\text{Var}[Q^{-1}_{\text{TLS, reac}}]\approx550$. Using Eq. \eqref{eq:variance} and taking a representative value $\omega_r/2\pi\sim5.5~$GHz, we find $\omega_{\text{max}}/2\pi\sim10^{25}$ Hz. For reference, this estimate is several orders of magnitude larger than the Debye frequencies of AlN and sapphire ($\sim10^{13}-10^{14}~$Hz). Although a detailed assessment of the reliability of the extracted $\omega_{\text{max}}$ is beyond the scope of this work, we note that the multimode nature of HBARs provides a useful platform for systematically investigating the properties of the TLS ensemble. 

Finally, the longitudinal coupling term ($\propto\sigma_z$) in Eq. \eqref{eq:TLShamiltonian} leads to relaxation TLS frequency shifts and dissipation in the mechanical resonator. In this case, the strain field modulates the TLS energy splitting. Computing the corresponding TLS susceptibility due to the longitudinal coupling term and inserting it into Eq. \eqref{eq:dampingsusceptibility}, yields a phonon mode decay rate with a power-law temperature dependence $\kappa_r\propto T^d$, where $d$ is the dimension of the phonon bath that the TLS couple to \cite{emser_thin-film_2024, behunin_dimensional_2016}. For a 3D bulk system such as an HBAR, one expects $d\approx3$. Relaxation TLS also produces a frequency shift on the phonon mode that depends on the temperature, this is given by \cite{emser_thin-film_2024}:
\begin{equation}
    \frac{\Delta\omega_r}{\omega_r}=-\frac{8\pi^3}{21}\frac{Q_{\text{TLS}}^{-1}}{(\rho\hbar\omega_r)^2}\frac{\bar{M}^2\bar{D}^2}{v^{10}}\left(\frac{k_BT}{\hbar}\right)^6.
\end{equation}
Taking typical values for sapphire $\rho=3980~$kg/m$^3$, $v=11000~$m/s, $\omega_r=2\pi\times5.5~$GHz, and assuming $\bar{M}\sim\bar{D}\sim1~$eV, we find that $\Delta\omega_r/\omega_r\sim10^{-15}Q_{\text{TLS}}$ at $T=100~$mK, and $\Delta\omega_r/\omega_r\sim10^{-5}Q_{\text{TLS}}$ at $T=4~$K, thus we expect the effects of relaxation TLS to be negligible compared to the resonant TLS contribution.

\subsection{Inverse Purcell effect}\label{sec:purcell}
In a cQAD device, coupling between the mechanical resonator and the qubit induces an additional decay channel for the resonator, an effect known as the inverse Purcell effect \cite{yang_mechanical_2024, reagor_quantum_2016}. To ensure that our measurements yield accurate estimates of the TLS loss tangents when comparing the quality factors obtained with the qubit and high power spectroscopy, we subtract the loss contribution from this decay channel on the phonon modes. The induced loss rate is
\begin{equation}
    \kappa_{\text{Purcell}}\approx \frac{g_0^2}{\Delta^2}\kappa_{\text{qubit}}
\end{equation}
where $\kappa_{\text{qubit}}$ is the qubit decay rate, $\Delta$ is the detuning between the HBAR mode and the qubit idle frequency, and $g_0$ is the qubit-phonon coupling strength. Using the measured qubit decay rates ($\kappa_{\text{qubit1}}\sim30-320~$kHz and $\kappa_{\text{qubit2}}/2\pi=5~$kHz), the couplings listed in Table \ref{tab:qubitparameters} for qubits 1 and 2, and typical detunings $\Delta/2\pi$ ranging from $6~$ to  $60~$MHz for the first five phonon modes reachable with the AC-Stark shift of the qubit, the resulting inverse Purcell decay rate on the phonon modes lies approximately between $0.1~$Hz in the best case and $2~$kHz in the worst case. In our qubit measurements of sample A, this contribution has been precisely characterized and explicitly accounted for.

\newpage

\section{Additional measurements}\label{sec:additionalmeasurements}
In this section, we compile additional measurements on HBAR samples with different geometries and materials.
\subsection{Temperature sweeps}
To investigate the influence of different loss mechanisms in HBARs, we perform temperature sweeps using the spectroscopy measurement setup, covering a temperature range from $10$ mK to $\sim80$ K. These measurements are performed at high drive powers, corresponding to mean phonon populations between $10^{6}$ and $10^{9}$. In addition, the temperature-dependent spectroscopy allows us to extract relevant material properties, such as the temperature coefficient of frequency (TCF), defined as the fractional rate of change of the resonance frequency with temperature \cite{gokhale_temperature_2020}.

\begin{figure}[h!]
    \centering
    \includegraphics[width=0.95\linewidth]{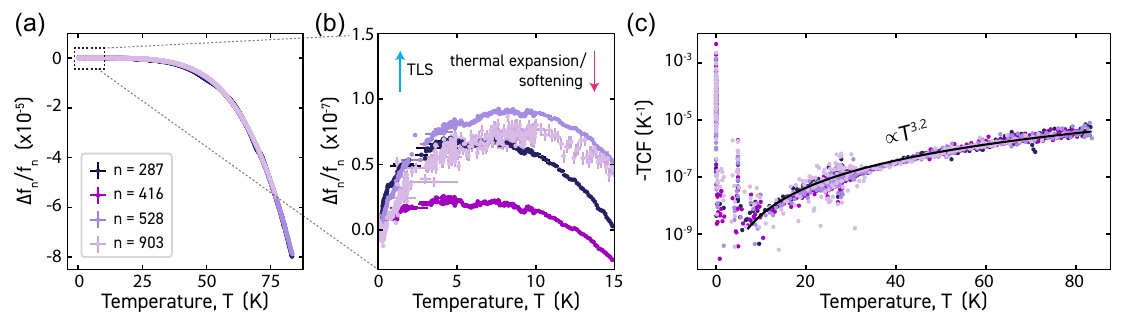}
    \caption{\textbf{Frequency dependence of the HBAR modes with temperature.} (a) Relative frequency shifts of four overtones in Sample B.1 measured from $10$ mK to $\sim80$ K. (b) Zoom-in showing the anomalous low-temperature relative frequency increase attributed to saturation of dispersively coupled TLS. For temperatures above $\sim10$ K the relative frequency shifts decrease, consistent with thermal expansion or softening. (c) Extracted negative temperature coefficient of frequency (TCF) as a function of temperature. The solid black line is a fit to a power-law model $cT^\alpha$ with $\alpha=3.18\pm0.01$. The TCF includes the combined contributions of linear thermal expansion and temperature-dependent elasticity of the substrate. }
    \label{fig:tempfreq}
\end{figure}

We first discuss the relative frequency shifts of four overtones in Sample B.1, shown in Fig. \ref{fig:tempfreq}(a). At temperatures $\gtrsim 15$ K, all modes exhibit a correlated decrease in frequency, consistent with the combined effects of thermal expansion and elastic softening of the substrate. Interestingly, below $10$ K we observe an anomalous upward shift in frequency with increasing temperature, see Fig. \ref{fig:tempfreq}(b). As discussed in the main text, this behavior can be attributed to the saturation of two-level systems (TLS) interacting dispersively with the acoustic modes, and can be modeled by Eq. \eqref{eq:TLStemp_appendix} \cite{phillips_two-level_1987, hunklinger_3_1976}. Because different overtones couple differently to TLS depending on their energy participation ratios on the piezo and bulk, the magnitude of this relative frequency shift varies between modes. 

From the relative frequency shifts, we extract the TCF, as shown in Fig. \ref{fig:tempfreq}(c). The data is well described by a power-law model, $cT^\alpha$, yielding $\alpha=3.18\pm0.01$. Neglecting contributions from the piezoelectric transducer or anisotropy in the crystal, the temperature dependence of the $n$th overtone can be expressed as \cite{gokhale_temperature_2020}
\begin{equation}
    f_n(T)=n\frac{v_B(T)}{2t_B(T)}=\frac{n}{2t_B(T)}\sqrt{\frac{c_{33}(T)}{\rho_B}}.
\end{equation}
where $t_B(T)$ is the temperature-dependent bulk thickness, $c_{33}(T)$ the temperature-dependent longitudinal stiffness coefficient of the bulk, and $\rho_B$ the mass density of the crystal. Differentiating, the TCF separates naturally into two contributions:
\begin{equation}\label{eq:TCF}
    \text{TCF}\equiv\frac{1}{f_n}\frac{\partial f_n}{\partial T}=-\frac{1}{t_B}\frac{\partial t_B}{\partial T}+\frac{1}{2c_{33}}\frac{\partial c_{33}}{\partial T}.
\end{equation}
The first term in Eq. \eqref{eq:TCF} is the linear thermal expansion coefficient (LTE), while the second term is the temperature coefficient of elasticity (TCE). Our measurements cannot distinguish between these contributions directly, so we attribute the observed TCF to the combined effect of both thermal expansion and elastic softening. Notably, the LTE of crystalline solids is generally expected to scale as $T^3$, consistent with the power-law behavior observed in our data. This is a well-known consequence of the Debye model, which predicts a specific heat $C_v\propto T^3$ for crystalline solids, and via the Grüneisen relation, it follows that LTE $\propto C_v\propto T^3$. We note that for $T\to0$, we observe an anomalous enhancement of the TCF, we attribute this behavior to frequency instabilities induced by the strong driving of the HBAR (see Supplementary Information \ref{sec:freqstability} for a further discussion).

\begin{figure*}[h!]
    \centering
    \includegraphics[width=0.85\linewidth]{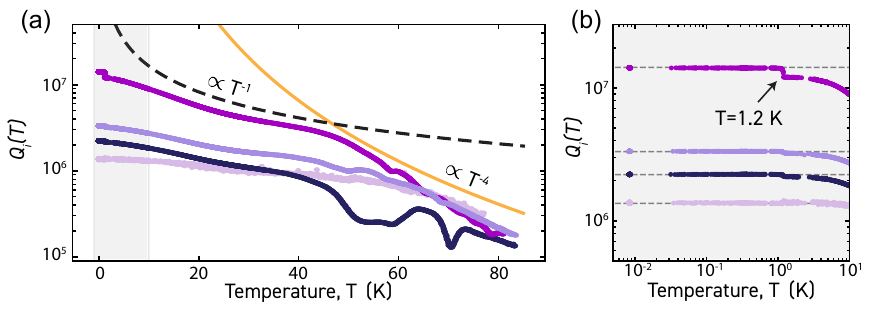}
    \caption{\textbf{Qualify factor dependence of the HBAR modes with temperature.}   (a) Internal quality factors as a function of temperature for four overtone modes in Sample B.1 (same modes as in Fig. \ref{fig:tempfreq}). The black dashed line indicates a $T^{-1}$ dependence, and the solid orange line a $T^{-4}$ dependence; both are shown as guides to the eye. (b) Zoom-in to the $10~$mK -- $10~$K temperature range. The dashed gray lines model the combined effects of temperature-independent loss and resonant TLS loss, described by Eq. (3) in the main text.} 
    \label{fig:tempQ}
\end{figure*}

From the spectroscopy measurements, we also extract the internal quality factors of the  HBAR overtones as a function of temperature $Q_{\text{i}}(T)$, shown in Fig. \ref{fig:tempQ}(a). At low temperatures ($T\to0$), the quality factors plateau due to temperature-independent losses, $Q_{\text{i,0}}^{-1}$, which include contributions from mechanical absorption and surface scattering. As shown in Fig. \ref{fig:tempQ}(b), in this lower temperature range, we observe no significant increase in the quality factors due to the saturation of resonant TLS. This is consistent with Eq. (3) in the main text, where we expect the low TLS density in this sample ($Q_{\text{TLS}}^{-1}\lesssim10^{-7}$) and the large mean phonon occupation of the HBAR modes ($\bar{n}_p>10^6$) in these measurements to mask any TLS saturation contribution.

Above $T\sim1.2$ K, we observe a sudden drop in the quality factors (Fig. \ref{fig:tempQ}(b)), which we attribute to ohmic loss, $Q_{\text{Ohm}}^{-1}$, arising from the aluminum antennas losing superconductivity. We estimate $Q_{\text{Ohm}}^{-1}$ to be in the range $0.5-1\times10^{-8}$. A similar behavior has been observed in quartz BAW resonators with aluminum electrodes deposited on top \cite{valimaa_electrode_2018}. In our case, the antennas are fabricated on a separate substrate, so the aluminum is not directly in contact with the HBAR; nevertheless, the electric field from the antenna can still have a small contribution to the total energy of the HBAR modes.

At higher temperatures ($T\gtrsim10$ K), we observe a rapid decrease in the quality factors with temperature, initially proportional to $T^{-1}$ and then  to $T^{-4}$, as shown in Fig. \ref{fig:tempQ}(a). This behavior is consistent with two distinct regimes of phonon-phonon scattering loss, $Q_{\text{ph-ph}}^{-1}$, in which coherent phonons of the mechanical modes scatter with thermal phonons in the substrate lattice. In the first regime ($T\lesssim50$ K), the $Q_{\text{i}}\propto T^{-1}$ dependence suggests Akhiezer damping \cite{akhiezer_absorption_1939, braginsky_systems_1986}, described by 
\begin{equation}\label{eq:akhiezer}
    Q^{-1}_{\text{Akhiezer}}=\frac{C_v\gamma^2\tau_{th}2\pi f_n 
    T}{\rho v^2},
\end{equation}
where $\rho$ is the density, $v$ is the speed of sound, $C_v$ is the volumetric heat capacity, $\gamma$ is the Grüneisen parameter, $\tau_{th}$ is the thermal phonon lifetime, and $f_n$ is the frequency of the phonon mode. While we are confident in the values of $f_n$, $T$, $\rho$ and $v$ for sapphire, some of the other parameters are difficult to estimate precisely at cryogenic temperatures. Therefore, we treat the product $C_v\gamma^2\tau_{th}$ as a fitting parameter, and find that $C_v\gamma^2\tau_{th}\sim10^{-7}$ reproduces well our data. 

In the higher temperature regime ($T\gtrsim50$ K), the $Q_{\text{i}}\propto T^{-4}$ dependence indicates a Landau-Rumer scattering process \cite{braginsky_systems_1986, gokhale_epitaxial_2020}, described by
\begin{equation}
    Q_{\text{Landau-Rumer}}^{-1}=\frac{\pi^5\gamma^2k_B^4T^4}{15\rho v^5h^3},
\end{equation}
where $k_B$ is the Boltzmann constant and $h$ is Planck's constant. Treating $\gamma$ as an effective fitting parameter, we find that $\gamma\sim5$ reproduces well the observed temperature dependence. This is higher than the typical value for sapphire ($\gamma\sim1.3$), suggesting enhanced lattice anharmonicity in this HBAR, potentially due to the strong drive powers applied. 

Combining all contributions, the total intrinsic loss can be written as
\begin{equation}
    Q_{\text{i}}(T)^{-1}=Q_{\text{i},0}^{-1}+Q_{\text{TLS}}^{-1}(T)+Q_{\text{Ohm}}^{-1}(T)+Q_{\text{ph-ph}}^{-1}(T).
\end{equation}

For completeness, we briefly discuss other temperature-dependent loss mechanisms, such as thermoelastic damping \cite{galliou_extremely_2013}, and relaxation TLS loss \cite{behunin_dimensional_2016, emser_thin-film_2024}. Thermoelastic damping is also expected to have a $T^1$ dependence, but it is negligible for GHz-frequency HBAR modes at cryogenic temperatures \cite{gokhale_epitaxial_2020}. Relaxation TLS loss scales as $T^d$, where $d$ is the effective dimension of the phonon bath to which the TLS couple. For a 3D resonator such as an HBAR we would expect $d\sim3$, however, this trend is not observed in our data.

\subsection{Frequency stability}\label{sec:freqstability}
Our results presented in the main text reveal lifetime-limited coherence times for HBARs, indicating a negligible dephasing rate in these mechanical resonators \cite{luo_lifetime-limited_2025}. This can be attributed to the stable frequencies of the HBAR modes over different time scales, which we investigate in this section. 
\begin{figure*}[b!]
    \centering
    \includegraphics[width=0.85\linewidth]{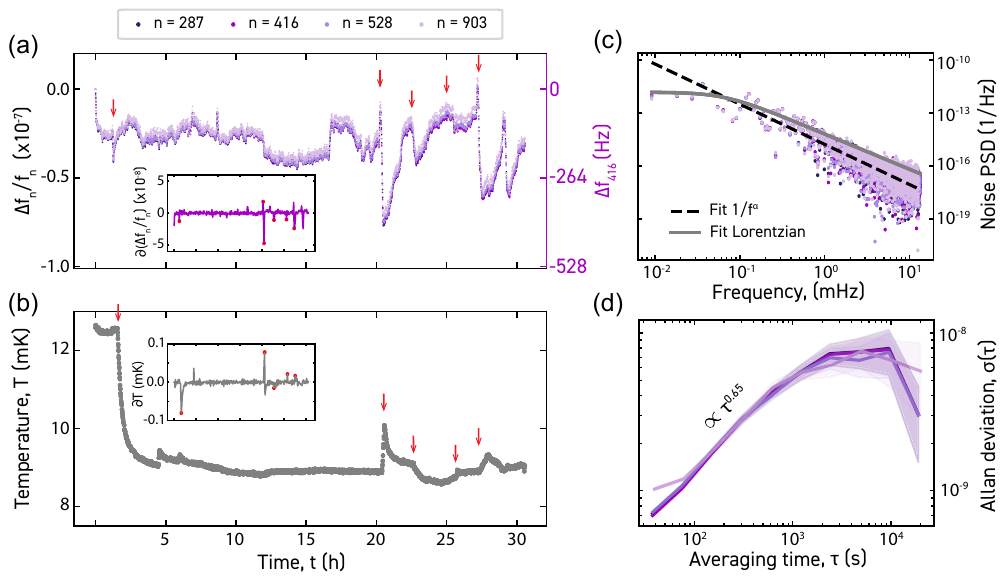}
    \caption{\textbf{Frequency stability of the HBAR modes.} (a) Fractional frequency shift of four overtone modes in Sample B.1 as a function of time. Red arrows mark correlations between sudden frequency jumps and temperature variations in the dilution refrigerator shown in (b). The insets in (a) and (b) display variations of the fractional frequency shift for mode $n=416$ and the corresponding fridge temperature, respectively, obtained by undersampling the data every ten samples and computing consecutive differences; red points highlight the correlated jumps. (c) Noise PSD of the fractional frequency shifts. The dashed black line is a fit to a $1/f^\alpha$ model with $\alpha=2.26(3)$, while the solid gray line shows a fit to a Lorentzian model. (d) Allan deviation of the fractional frequency shifts.}
    \label{fig:freqshift_highpower}
\end{figure*}

First, we measure the relative frequency shifts $\Delta f_n/f_n$ of four HBAR modes in sample B.1 as a function of time for a total of 30 hours at a temperature of $\sim10$~mK; see Fig. \ref{fig:freqshift_highpower}(a). The reflection frequency response of each mode is sampled approximately every 37~s using the spectroscopy measurement setup. We fit the acquired data to Eq. (6) in Methods to extract the resonance frequencies. These measurements are performed at high input powers to improve the signal-to-noise ratio (SNR) and reduce the acquisition time per mode, resulting in large mean phonon populations $\bar{n}_p\sim 10^6-10^9$. We find that the relative frequency shifts of all the measured modes exhibit substantial overlap, as shown in Fig. \ref{fig:freqshift_highpower}(a). This suggests that the observed fluctuations arise from global effects acting on the entire crystal, such as spontaneous thermal expansion/contraction or softening/stiffening, which shift all modes in a correlated manner, and discards the influence of strongly coupled TLS. The relative frequency shifts during the measurement period are bound below $10^{-7}$, corresponding to absolute frequency shifts below the kHz range for the measured modes. Interestingly, we find correlations between sudden frequency shifts and variations in the mixing chamber temperature of the dilution refrigerator to which the HBAR is thermalized, see Fig. \ref{fig:freqshift_highpower}(b). In total, we can identify five such correlated events by examining the consecutive frequency and temperature increments of the data undersampled by a factor ten (insets in Fig. \ref{fig:freqshift_highpower}(a,b)).

To quantify the frequency stability of the HBAR modes, we evaluate both the noise power spectral density (PSD) and the Allan deviation $\sigma(\tau)$ of the relative frequency shifts, shown in Fig. \ref{fig:freqshift_highpower}(c,d), respectively. The PSD is well described by a $1/f^\alpha$ dependence with $\alpha=2.26\pm0.03$, close to Brownian-type noise arising from the cumulative nature of the frequency drifts. Occasional abrupt jumps in frequency suggest the presence of a telegraph-like process, in which the HBAR switches between metastable states activated by thermal stress. Such process would yield a Lorentzian-shaped PSD. Importantly, the noise PSD of the HBAR modes is approximately two to three orders of magnitude lower than that recently reported for other mechanical resonators such as PNCs \cite{bozkurt_mechanical_2025, maksymowych_frequency_2025}. This enhanced stability can be attributed to the larger effective mass of HBAR modes ($\sim16~\mu$g \cite{bild_schrodinger_2023}), which averages out the effects of microscopic fluctuators such as TLS. The Allan deviation further supports this picture: at short averaging times, $\sigma(\tau)$ follows a $\tau^{0.65(2)}$ scaling, close to Brownian-type noise. Moreover, the measured Allan deviation is comparable to that reported for HBARs developed for clocking applications \cite{yu_hbar-based_2009, baron_rf_2011}.
\begin{figure}[b!]
    \centering
    \includegraphics[width=0.5\linewidth]{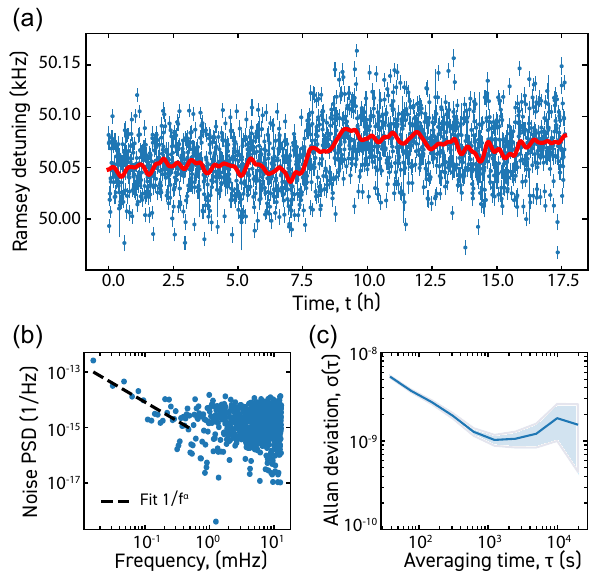}
    \caption{\textbf{Frequency stability of an HBAR mode in the quantum regime.} (a) Ramsey detuning between an HBAR mode in sample B.2 and a microwave drive, measured as a function of time. Each point corresponds to a sampling interval of 38.5~s. The red line shows a Gaussian moving average with a ten-point window. (b) Noise power spectral density of the relative frequency shifts from panel (a). The dashed black line is a fit to a $1/f^{\alpha}$ model with $\alpha = 1.4(2)$. (c) Allan deviation of the relative frequency shifts.}
    \label{fig:freqshift_lowpower}
\end{figure}

Next, we investigate the frequency noise of an HBAR mode in the low-phonon number regime ($\bar{n}_p\sim1$) using a superconducting qubit as a probe. For this purpose, we track the frequency fluctuations of the $n=401$ overtone in Sample B.2, with a mean frequency of $f_{401}=5.0889~$GHz. Following the same method used in \cite{bozkurt_mechanical_2025, capannelli_tracking_2025}, the detuning between a drive tone and the mechanical frequency is extracted from the fringes of Ramsey measurements performed on the phonon mode, see Fig. 4(b) in the main text. We repeat the Ramsey sequence every 38.5~s for a total duration of 17.5~h. The resulting detunings are shown in Fig. \ref{fig:freqshift_lowpower}(a), where the solid red line represents a Gaussian moving average with a ten-point window. Over the full measurement period, the average absolute frequency shifts are on the order of $50$~Hz, comparable to the detection noise floor obtained from the standard deviation of the detrended data. 

To quantify the noise, we compute the PSD and Allan deviation of the relative frequency shifts, shown in Fig. \ref{fig:freqshift_lowpower}(b,c), respectively. The PSD values are consistent with those obtained at high phonon numbers (Fig. \ref{fig:freqshift_lowpower}), and can be fit to a $1/f^\alpha$ model with $\alpha = 1.4\pm0.2$, lying between white and Browninan noise. At Fourier frequencies $>0.1$ mHz the PSD plateaus, which corresponds to the detection noise limit in this measurement. The Allan deviation shows slightly improved stability at long averaging times ($\tau>100$ s) compared to high-power measurements, while for $\tau<100$ s it scales as $\tau^{-0.51(2)}$, again dominated by detection noise. Importantly, unlike in the high-power regime, we do not observe sudden frequency jumps in these qubit-based measurements. This suggests that strong driving of the HBAR could induce additional frequency instabilities \cite{boudot_frequency_2016}, likely of thermal stress origin leading to an enhancement of the TCF, that are absent when probing the resonator at the single-phonon level with the qubit. Interestingly, this is also the regime in which frequency shifts from coupling to unsaturated TLS would be expected; however, such shifts appear to be negligible. This observation is consistent with the lifetime-limited coherence times measured for the HBAR modes in the quantum regime.

\subsection{Diffraction-limited HBAR}\label{sec:diffractionlimited}
To benchmark the effect of the plano-convex geometry in suppressing diffraction loss, we fabricated and characterized an AlN-sapphire HBAR with a flat-flat geometry (Sample F). In this device, the resist reflow step used to define the dome is omitted, and the AlN layer is simply etched to a thickness $t_P=(1.0\pm0.1)~\mu$m to optimize the EM coupling near $\sim5$ GHz. The wafer used to fabricate this sample comes from the same vendor and batch as Samples B and C, with high-crystallinity HVPE-grown AlN. The thickness of the sapphire substrate is $t_B = 435~\mu$m. 
\begin{figure}[b!]
    \centering
    \includegraphics[width=0.5\linewidth]{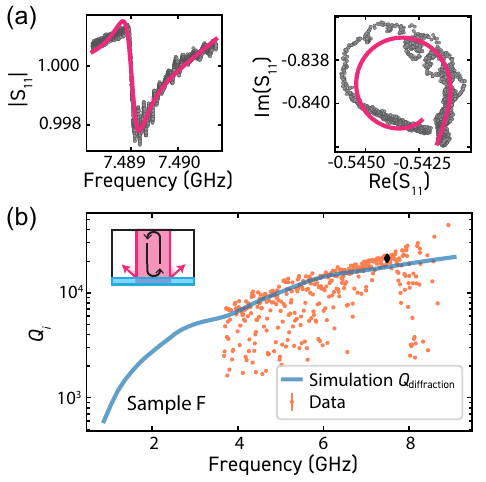}
    \caption{\textbf{Diffraction-limited quality factors of a flat-flat HBAR.} (a) Frequency response of a mechanical mode in a flat HBAR (black dots). The solid red line is a fit to Eq. (6) in Methods to extract the internal quality factor. (b) Internal quality factor $Q_{\text{i}}$ measured for different overtones in this HBAR. The quality factors improve with increasing frequency, a signature of diffraction-limited loss (inset). The solid blue line shows the prediction for the diffraction-limited quality factors as a function of frequency obtained from beam propagation simulations. }
    \label{fig:flatHBAR}
\end{figure}
We flip-chip bond Sample F to a circular coplanar waveguide antenna with a radius of $30~\mu$m (see. Fig. \ref{fig:bigantenna}(b)), and measure the frequency response of the HBAR overtones through the reflected signal from the antenna. The resulting spectrum shows significantly broadened resonances, with the antenna coupling to multiple high-order transverse modes simultaneously rather than selectively exciting the LG$_{00}$ overtone, as can be seen in Fig. \ref{fig:flatHBAR} (a). Fitting the spectroscopy data to the reflection model from Eq. (6) in Methods (solid red line) yields the internal quality factors shown in Fig. \ref{fig:flatHBAR}(b), which are two to three orders of magnitude lower than those measured in the plano-convex HBARs. The observed increase of $Q_{\text{i}}$ with frequency is consistent with diffraction-limited loss. The spread in the data is possibly due to the non-Lorentizan shape of the resonances and the low signal-to-noise ratio due to the reduced external coupling, which complicates the fitting.

To verify this interpretation, we compare the measured quality factors with estimates of the diffraction-limited quality factors ($Q_{\text{diffraction}}$) obtained from beam propagation simulations, following the procedure described in Supplementary Section \ref{sec:diffraction} and Ref. \cite{gruenke-freudenstein_surface_2025}. The initial forcing field $u_0$ in this simulation is taken as the out-of plane ($z$-component) electric field generated by the antenna on the piezoelectric layer, obtained from finite-element electrostatic simulations (Fig. \ref{fig:bigantenna}(b)). The simulated diffraction-limited quality factors for the given antenna and flat-flat HBAR geometry agree well with the experimental data, as shown in Fig. \ref{fig:flatHBAR}(b). These results demonstrate that the plano-convex HBAR design effectively suppresses diffraction losses, enabling substantially higher quality factors.


\subsection{Antenna size and high-order LG modes}
\begin{figure*}[b!]
    \centering
    \includegraphics[width=1\linewidth]{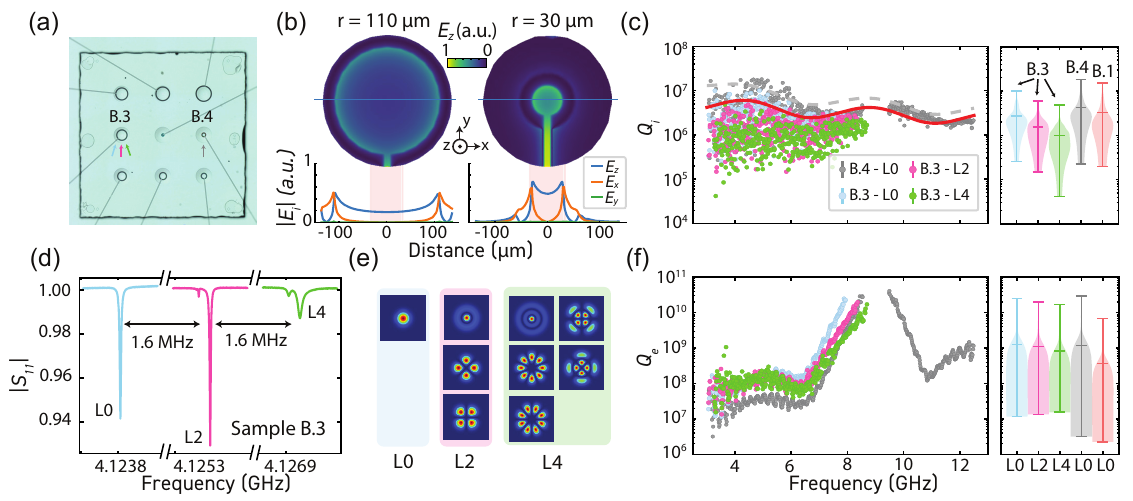}
    \caption{\textbf{Influence of microwave antenna size.} (a) Coplanar waveguide antenna chip with nine antennas of varying radius size. An HBAR chip (transparent) is bonded on top, comprising nine identical AlN-sapphire HBAR domes aligned to the antennas.  We characterize only two of these HBARs, labeled B.3 and B.4, measured with antennas of radii $r=110~\mu$m and $r=30~\mu$m, respectively.  (b) Electrostatic finite-element method simulation of the electric field of the antennas inside the piezoelectric HBAR dome. (c,f) Internal and external quality factors of the L0 longitudinal overtones in samples B.3 and B.4, and high order (L2, L4) transverse overtones in sample B.3. The gray dashed line indicates the surface scattering loss limit, and the solid red line is a fit to the participation model. The violin plots on the right show the distribution of the data, including sample B.1 characterized in the main text. (d) Frequency response of the $n=325$ fundamental (L0) longitudinal overtone in HBAR B.3 and two of its higher order modes (L2, L4). The FSR between the $n$ and $n+1$ overtones is $\sim12.7$ MHz, while the frequency spacing between the radially symmetric high order transverse modes is $\sim1.6\pm0.1$ MHz. (e) Intensity profiles of the L$N$ phonon mode families with even $N=|l|+2p$ supported by the plano-convex HBAR geometry.}
    \label{fig:bigantenna}
\end{figure*}
While in flat-flat HBAR geometries the shape and size of the driving electrode determines the profile of the acoustic modes and is related to the diffraction loss limit \cite{gokhale_phonon_2021}, in plano-convex HBARs the mode profiles are determined by the geometry of the phonon cavity itself, as discussed in Supplementary section \ref{sec:diffraction}. In this case, we expect the shape and size of the antenna to have little or no influence on the internal losses of the acoustic modes, but should instead affect their electromechanical coupling strengths. To test this hypothesis, we fabricate a chip with nine CPW antennas of varying radius size, shown in Fig. \ref{fig:bigantenna}(a). An HBAR chip is flip-chip bonded on top with nine identical AlN-sapphire HBAR domes aligned to the antennas. We measure only two of these HBARs, which we label B.3 and B.4 (see Fig. \ref{fig:bigantenna}(a)), coupled to antennas of radii $r=110~\mu$m and $r=30~\mu$m, respectively. These HBARs are fabricated on the same wafer and with geometry parameters identical to samples B.1 and B.2 characterized in the main text. 

First, we measure the frequency response of Sample B.4, and observe prominent dips separated by an FSR of $\sim12.7$~MHz, as expected for the overtone modes in this HBAR. We fit the frequency response using Eq. (6) in Methods to extract the internal and external quality factors of the modes, shown as gray dots in Fig. \ref{fig:bigantenna}(c,f). The internal quality factors show an oscillatory behavior and frequency dependence similar to those of Sample B.1. We fit this trend using the energy participation model with parameters summarized in Table \ref{tab:parameters} to account for surface sacttering and mechanical absorption loss in the AlN layer and the lossy interface (solid red line in Fig. \ref{fig:bigantenna}(c)).

Next, we measure the frequency response of Sample B.3. In addition to modes separated by the overtone FSR of $\sim12.7$ MHz, we observe additional dips separated by $\sim1.6$ MHz, see Fig. \ref{fig:bigantenna}(d). We identify these modes with families of high-order Laguerre-Gaussian modes, shown in Fig. \ref{fig:bigantenna}(e). By fitting the dips to the reflection model from Eq. (6) in Methods we extract the internal and external quality factors of the different families of modes, as shown in Fig. \ref{fig:bigantenna}(c,f) plotted with different colors. We note that, in general, the modes in Sample B.3 (bigger antenna) show nearly an order of magnitude lower electromechanical coupling than the modes in Sample B.4 (smaller antenna), quantified by the higher $Q_{\text{e}}$ near the center of the transduction envelope. In addition, the different LG modes in Sample B.3 have similar coupling among each other, while in Sample B.4 the high-order modes were not visible, indicating negligible coupling. In terms of the losses, we observe that the fundamental LG mode in samples B.3, B.4 and B.1 are compatible with each other within the spread and sample-to-sample variations. Note that sample B.1 is also measured with an antenna of radius $30~\mu$m and presents slighly lower $Q_{\text{i}}$ than sample B.4. Therefore, the size of the antenna does not seem to have a big influence on the losses, as expected. On the other hand, we note that the internal quality factors seem to decrease with increasing LG mode order.

To understand this different behavior and what these families of modes are, we inspect the piezoelectric interaction expression that determines the electromechanical coupling between the electric field from the antenna, $E$, and the strain field $S$ at the piezoelectric transducer.  The energy density due to the piezoelectric effect is $\varepsilon_{\text{piezo}}=-E\cdot e\cdot S$, where $e=c\cdot d$ is the product of the stiffness tensor $c$ , with the piezoelectric tensor $d$ of the AlN. Since the dominant piezoelectric tensor component of AlN is $d_{33}$ (in Voigt notation),  the piezoelectric coupling occurs mostly between the out-of-plane (z component) electric field of the antenna and the longitudinally polarized acoustic modes, yielding $\varepsilon_{\text{piezo}}\approx-c_{33}d_{33}E_zs_z$. Therefore, the electro-mechanical coupling rate is proportional to the following overlap integral defined over the volume of the piezoelectric transducer:
\begin{equation}\label{eq:couplingEM}
    g_{\text{EM}}\propto-c_{33}d_{33}\iint_{A_P} U_{npl} (r,\phi)E_z(r,\phi)d\phi dr\int_0^{t_P}\sin\left(k_nz\right) dz.
\end{equation} 
 Eq. \eqref{eq:couplingEM} makes several assumptions to ease its interpretation. For example, it assumes that the electric field is homogeneous in the z direction since the electromagnetic wavelength is much larger than the thickness of the piezoelectric layer for frequencies in the GHz range. Strictly speaking, we should have written the strain field profile ($S_{npl}(r,\phi)$) instead of the displacement field profile $U_{npl}(r,\phi)$, nevertheless, note that these two coincide except for normalization factors since $s_z=du_z/dz$ , where the subindex z denotes the z-polarized component. Finally, we assume that the piezoelectric transducer is cylindrical instead of dome-shaped to simplify the integration limits. From the sine integral over the thickness of the piezo in Eq. \eqref{eq:couplingEM}, we see that coupling is maximized when $t_P$ is an integer multiple of half acoustic wavelength. Similarly, from the integral over the area of the piezo $A_P$, we see that the coupling depends on the overlap between the acoustic mode profile (as defined in Eqs. \eqref{eq:profile} and \eqref{eq:profileLG} of this Supplementary) and the out-of-plane electric field of the antenna. Therefore, for maximum selective coupling to the LG$_{00}$ mode, the antenna radius should approximately match the mode waist. Given the radius of curvature $R=6.9$~mm of the domes, and the thickness $L\approx435~\mu$m in samples B.3-4, the mode waist $w_0$ is in the range $25-37~\mu$m for modes between 4 and 9~GHz, calculated using Eq. \eqref{eq:modewaist}. Therefore, it is clear that the antenna with $30~\mu$m radius will have higher and more selective coupling than the bigger antenna with $r=110~\mu$m, in agreement with our observations.

To verify which are the families of high-order modes that the $110~\mu$m radius antenna couples to, we calculate the resonance condition for these modes by imposing that the phase shift in Eq. \eqref{eq:profile} along the cavity axis ($r=0$) in a round trip must be an integer multiple of $2\pi$ \cite{kharel_ultra-high-_2018}, so for half round trip:
\begin{equation}
    -k_{n(pl)}L+(|l|+2p+1)(\psi'(L)-\psi'(0))=n\pi.
\end{equation}
We can define the mode order $N=|l|+2p$, since Laguerre-Gaussian modes with the same order $N$ will be degenerate. Then the free-spectral range between high-order mode families L$N$ is
\begin{equation}\label{eq:highFSR}
    \Delta_N=f_{nN+1}-f_{nN}=\frac{v}{2\pi L}\arctan\left(\frac{L}{z_R}\right).
\end{equation}
Here, $z_R$ is the acoustic Rayleigh length defined in \eqref{eq:rayleighlength}, and to arrive at this expression we used $k_{n(pl)}=2\pi f_{nN}/v$ and the definition of the Guoy phase $\psi'(z)=\arctan(z/z_R)$. Note that we are ignoring the effects of the piezoelectric transducer in this calculation, so $v$ is just the longitudinal speed of sound in the bulk. Using the estimates $R=6.9~$mm, $L=435~\mu$m and $v=11.06~$km/s for sample B.3, the Rayleigh length using Eqs. \eqref{eq:rayleighlength} and \eqref{eq:modewaist} is $z_R=1.9$~mm, so the high-order frequency mode spacing using Eq. \eqref{eq:highFSR} is $\Delta_N=0.9$ MHz, and the frequency spacing between even $N$ mode families is $1.8$ MHz, close to our measured value ($\sim1.6$ MHz). The small discrepancy could be induced by the piezoelectric layer or by variations in the radius of curvature. 

Finally, notice that due to the radial symmetry of the electric field generated by the antenna, the azimuthal angle phase dependence of the LG$_{pl}$ modes (Eq. \eqref{eq:profileLG}) will make the integral in Eq. \eqref{eq:couplingEM} vanish for $l\neq0$ when integrating the azimuthal angle between $0$ and $2\pi$. Therefore, a radially symmetric antenna, like ours, couples selectively to LG modes with $l=0$, which belong to L$N$ families with even mode order $N$. Nevertheless, asymmetries in the antenna or misalignment with the dome can lead to residual coupling to modes with $l\neq0$ within the same family. The frequency degeneracy of these modes could explain the lower internal quality factor of the higher-order modes, owing to energy conversion loss.

\subsection{Thin HBAR}
\begin{figure}[h!]
    \centering
    \includegraphics[width=0.85\linewidth]{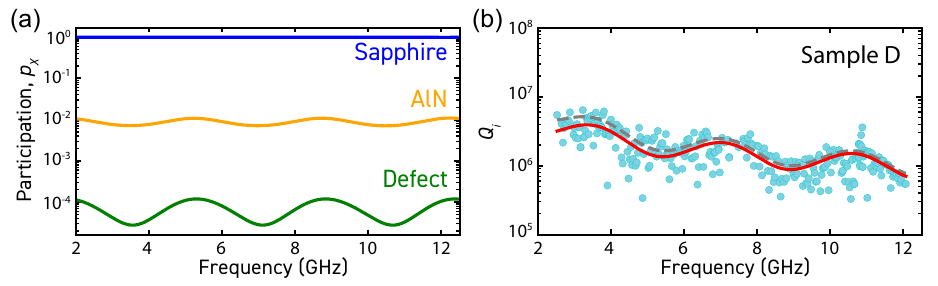}
    \caption{\textbf{Mechanical loss in a thinner HBAR ($t_B=170~\mu$m).} (a) Acoustic energy participation ratios ($p_X$) of the sapphire, AlN and defect layers as a function of frequency in Sample D. The parameters used to calculate these participations are summarized in Table \ref{tab:parameters}. (b) Internal quality factors of the phonon modes. The gray dashed line is the estimated surface scattering loss limit, and the solid red line is a fit to the participation model used to extract the mechanical absorption loss of each layer, as described in the main text.}
    \label{fig:thinHBAR}
\end{figure}

As another point of comparison, we test our energy participation ratio model on an AlN-sapphire HBAR sample with reduced substrate thickness ($t_B\approx170~\mu$m), while keeping a similar piezoelectric layer thickness ($t_P\approx1.5~\mu$m) to the samples characterized in the main text. This reduction in substrate thickness increases the acoustic energy participation ratio of the AlN by roughly a factor of three, as shown in Fig. \ref{fig:thinHBAR}(a). In this HBAR, which we label as sample D, the AlN film was grown on a c-plane sapphire substrate using the same HVPE process as samples B.1-4, C and F. Therefore, the AlN-sapphire interface also exhibits pits and defects, which we model as a $\sim10~$nm-thick defect layer.

We perform spectroscopy measurements on this sample following the same procedure used for the main samples. Namely, the HBAR is flip-chip bonded to a coplanar waveguide antenna chip, and we measure its frequency response in reflection at cryogenic temperatures ($10~$mK). We then fit the internal quality factor of the visible HBAR overtone resonances. The results are shown in Fig. \ref{fig:thinHBAR}(b). The quality factors exhibit an oscillatory behavior as a function of frequency, together with an overall decreasing trend. These oscillations are captured by the acoustic energy participation model, where the frequencies with higher AlN and defect layer participation show lower internal quality factors. 

Following the procedure explained in the main text and Methods, we estimate the limiting quality factor due to surface scattering loss using the measured RMS surface roughness of the AlN top surface ($\sigma_{\text{top}}=0.4~$nm), the fully etched AlN interface ($\sigma_{\text{interface}}=2.4~$nm) and the backside ($\sigma_{\text{back}}=0.5~$nm). This estimate, plotted as a dashed gray line in Fig. \ref{fig:thinHBAR}, reproduces well the observed frequency dependence of the quality factors. We attribute the remaining losses to mechanical absorption in the AlN film and the defect layer. From the energy participation analysis, we extract $Q_{\text{P,mech}}^{-1}=(6\pm3)\times10^{-6}$ and $Q_{\text{D,mech}}^{-1}=(6\pm4)\times10^{-4}$, assuming no absorption loss in the bulk. The corresponding fit is shown as a solid red line in Fig. \ref{fig:thinHBAR}. 

The highest internal quality factors in this thinner sample reach only $6.4\times10^6$ within the measured frequency range, compared to over $2\times10^7$ for the thicker HBARs studied in the main text. This roughly threefold reduction in the best $Q_{\text{i}}$ is  consistent with the decreased bulk thickness and the corresponding increase in the energy participation of the AlN and defect layers.

\subsection{HiPIMS AlN HBAR}\label{sec:hipims}
Finally, we apply the methodology developed in this work to evaluate the quality of AlN films grown by HiPIMS \cite{patidar_low_2025}, an alternative growth method to HVPE or pulsed-DC sputtering. In sample E, the HiPIMS AlN film ($t_P\sim1~\mu$m) is deposited on a $t_B\sim430~\mu$m-thick c-plane sapphire substrate. We fabricate a plano-convex HBAR following our standard process and flip-chip bond this device to a coplanar waveguide antenna.

\begin{figure}[h!]
    \centering
    \includegraphics[width=0.95\linewidth]{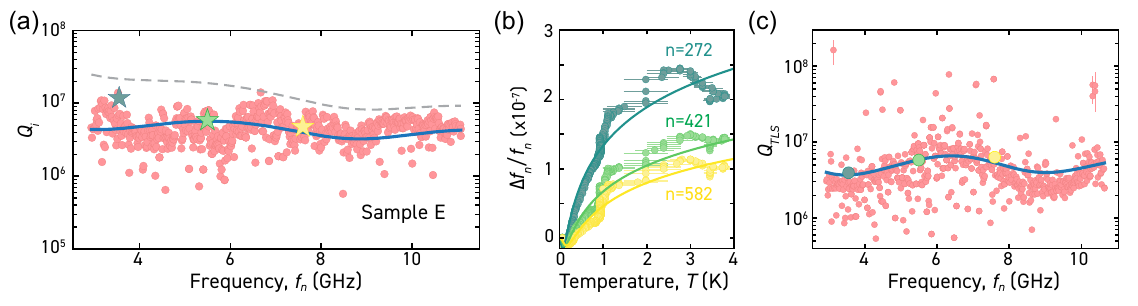}
    \caption{\textbf{Quality factors in an HBAR with HiPIMS AlN.} (a) Internal quality factors of the phonon modes measured in Sample E. The gray dashed line indicates the estimated quality factor limit set by diffuse surface scattering, while the solid blue line shows a fit to the participation model used to extract the mechanical absorption loss of the AlN. The starred phonon modes with overtone numbers $n=272,~421,~582$ are used to investigate TLS loss. (b) Fractional frequency shifts between $10~$mK and $4~$K for the selected modes. The solid line is a fit to the reactive TLS model described by Eq. \eqref{eq:TLStemp} in the main text. (c) Inverse reactive TLS loss tangent extracted from the fractional frequency shift of all phonon modes between $10~$mK and $4~$K. The solid line is a fit to the potential energy participation model discussed in the main text.}
    \label{fig:EMPA}
\end{figure}

We perform spectroscopy measurements at $10~$mK to extract the internal quality factors and resonance frequencies of the HBAR modes. The results are shown in Fig. \ref{fig:EMPA}(a). We calculate the surface scattering loss limit (dashed gray line) from the measured RMS surface roughness $\sigma_{\text{top}}=1.8~$nm, $\sigma_{\text{interface}}=0.5~$nm and $\sigma_{\text{back}}=0.6~$nm. As in sample A of the main text, we find that this HBAR is not limited by surface scattering, since the data do not show a $1/f$ trend. We attribute the remaining loss to mechanical absorption in the AlN layer and extract $Q_{\text{P,mechanical}}^{-1}=7.0(2)\times10^{-5}$ using the acoustic energy participation model. The fit to this model is shown as a solid blue line in Fig. \ref{fig:EMPA}(a), and the parameters used to calculate the energy participation of each layer are summarized in Table \ref{tab:parameters}.

To study TLS loss, we then select the three starred modes in Fig. \ref{fig:EMPA}(a) and measure their fractional frequency shift as a function of temperature. The mode frequencies increase between $10~$mK and $4~$K, consistent with TLS saturation. We fit the frequency shifts using Eq. (4) from the main text to extract $Q_{\text{TLS}}$. The fits are shown as solid lines in Fig. \ref{fig:EMPA}(b), and the fitted $Q_{\text{TLS}}$ are the color-coded points in Fig. \ref{fig:EMPA}(c). We then extend this analysis to all visible modes,  observing an oscillatory trend in  $Q_{\text{TLS}}$ that the potential energy participation model (solid blue line) reproduces well. From this fit, we extract a TLS loss tangent of $Q_{\text{TLS,P}}^{-1}=8.8(2)\times10^{-5}$ for the HiPIMS AlN.

Note that both the mechanical absorption and TLS loss tangent of the HiPIMS AlN closely match those of pulsed-DC-sputtered AlN (Sample A in the main text). This similarity agrees with the comparable AlN rocking curves measured for the two growth methods (see Methods section).

\subsection{HiPIMS AlN growth} 
The AlN thin film in sample E was deposited using HiPIMS in a custom-built AJA ATC-1800 chamber. A base pressure of $<1\times10^{-6}$~Pa was achieved before the deposition to ensure minimal contamination from the chamber during the deposition. The type-2 unbalanced magnetron in the chamber was equipped with a 2-inch Al target (Kurt J. Lesker Company; purity: 99.999 at.$\%$). The film was grown using sputter-up coplanar geometry. The sapphire wafer was placed at a working distance of 12~cm and was heated from the back using five halogen lamps, ensuring uniform heat distribution across the whole sample holder. The deposition was performed at a substrate temperature of 500$^\circ$C. The substrate was baked out at 500$^\circ$C for 1 hour in the deposition chamber prior to the deposition. The substrate was rotated at approximately 15~rpm to promote uniform film growth. The deposition was performed at a pressure of 0.36~Pa, with the flow rates of Ar and N$_2$ in the chamber maintained at 20/12 (sccm/sccm). Power to the sputtering target was supplied by an Ionautics HiPSTER 1 power supply. The film was deposited at a time-averaged power of 100~W, with a pulsing frequency of 7500~Hz and 10~$\mu$s pulse width. These settings resulted in a peak current density of 0.6~A/cm$^2$. The substrate holder was electrically floating throughout the deposition.

\subsection{Pulsed-DC sputtering AlN growth}
The AlN thin film deposition in sample A was performed in a Pfeiffer-Vacuum Spider600 deposition cluster, with a 200~mm-wide, 6~mm-thick Al target as the source and a target-substrate distance of 42.6~mm. AlN sputtering was performed using a pulsed-DC (PDC) generator at 20~kHz. Prior to the AlN deposition, the chamber was heated to 300$^\circ$C in a 90-minute thermalization ramp and was followed by a target cleaning and poisoning routine: first a 5~min target cleaning sputtering in pure Ar atmosphere, with 15~sccm of gas flow to remove the native oxide on the Al surface; then a 10~min chamber conditioning in a pure N$_2$ atmosphere (50~sccm of gas flow) to poison the Al target. For both cleaning and conditioning steps, the sputtering power was set at 1500~W, corresponding to a target power density of 4.77~W/cm$^2$.

The AlN was then deposited on the sapphire wafer with a 1500~W PDC generator, at a frequency of 25~kHz, $35~\mu$s pulse duration, with gas flows of 40~sscm of N$_2$ and 10~sccm of Ar. To enhance ion mobility and promote film nucleation, the substrate was biased with a 13.56~MHz RF generator at 6~W. To deposit $1~\mu$m of AlN, the depostion time was 17~minutes. During deposition, the chamber pressure was $5.6\times 10^{-3}~$mbar, resulting in partial pressures of $4.48\times 10^{-3}~$mbar for N$_2$ and $1.12\times 10^{-3}~$mbar for Ar.

\subsection{EDS analysis of the HVPE AlN-sapphire interface}\label{sec:EDS}
\begin{figure}[h!]
    \centering
    \includegraphics[width=0.8\linewidth]{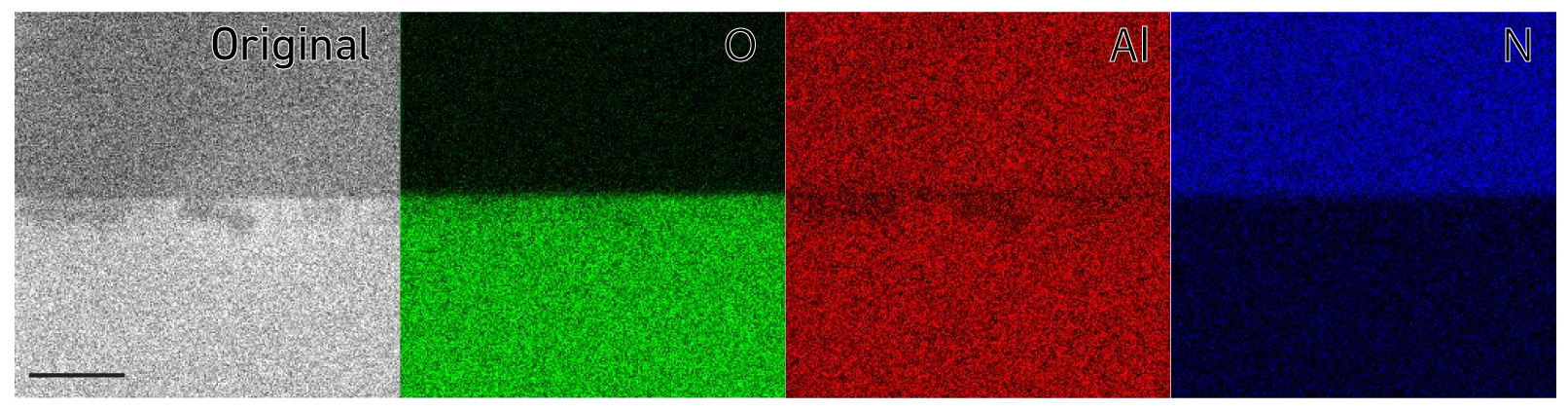}
    \caption{\textbf{EDS analysis.} EDS elemental maps of the HVPE AlN–sapphire interface showing the intensity of the O, Al, and N signals. The scale bar represents 20~nm.}
    \label{fig:EDS}
\end{figure}

During acquisition of the STEM images described in the Methods section, energy-dispersive X-ray spectroscopy (EDS) was performed across the HVPE AlN–sapphire interface (Fig. \ref{fig:EDS}). The EDS maps show a reduced aluminum signal in the damaged interface region.

\end{document}